\documentclass[a4paper]{jpconf}
\usepackage{graphicx,amsmath,amssymb}

\newcommand{\cl}{{\mt{C}}\ell}
\newcommand{\hko}{\hookrightarrow}

\newcommand{\n}{\nonumber\\}

\newcommand{\bu}{\bullet}

\newcommand{\cv}{\circ}

\newcommand{\bec}{\begin{center}}
\newcommand{\eec}{\end{center}}

\newcommand{\bea}{\begin{array}}
\newcommand{\ear}{\end{array}}

\newcommand{\bfr}{\begin{flushright}}

\newcommand{\efr}{\end{flushright}}
\newcommand{\noi}{\noindent}

\newcommand{\RR}{\mathbb{R}}\newcommand{\op}{\oplus}
\newcommand{\HH}{\mathbb{H}}

\newcommand{\la}{\Lambda}
\newcommand{\bege}{\begin{equation}}
\newcommand{\enge}{\end{equation}}
\newcommand{\w}{\wedge}

\newcommand{\beq}{\begin{eqnarray}}\newcommand{\benu}{\begin{enumerate}}\newcommand{\enu}{\end{enumerate}}
\newcommand{\eeq}{\end{eqnarray}}
\newcommand{\mt}{\mathcal}

\newcommand{\vv}{{\bf v}}
\newcommand{\ee}{{\bf e}}
\newcommand{\uu}{{\bf u}}

\newcommand{\ww}{{\bf w}}

\newcommand{\mk}{\mathfrak}

\newcommand{\bx}{\begin{pmatrix}}
\newcommand{\ex}{\end{pmatrix}}

\newcommand{\beo}{\begin{obs}}
\newcommand{\eeo}{\end{obs}}

\begin{document}

\title{Generalized non-associative structures on the 7-sphere}

\author{Rold\~ao da Rocha$^1$ and M\'arcio A Traesel$^2$}
\address{$ˆ1$ Centro de Matem\'atica, Computa\c c\~ao e Cogni\c c\~ao, 
Universidade Federal do ABC, 09210-170 Santo Andr\'e, SP, Brazil}
\address{$ˆ2$ Instituto Federal de Educa\c c\~ao, Ci\^encia e Tecnologia de S\~ao Paulo, Campus Caraguatatuba, 11665-310 Caraguatatuba, SP, Brazil}
\ead{roldao.rocha@ufabc.edu.br,marcio.traesel@ifsc.edu.br}

\begin{abstract}
In this paper we provide a more general class of non-associative products  using the exterior and Clifford bundles on the 7-sphere $S^7$. Some additional properties encompass the previous formalisms presented in \cite{eu1,dix} in the Clifford algebra context, and wider classes
of non-associative structures on $S^7$ are investigated, evinced by the directional non-associative products  and the  mixed composition of generalized non-associative products between Clifford algebra multivectors. These non-associative products are further generalized by considering the non-associative shear of arbitrary Clifford bundle $\cl_{0,7}$ elements into octonions. We assert new properties inherited from the non-associative structure introduced in the whole Clifford bundle on $S^7$, which naturally induce involutions on the Clifford bundle and provide immediate generalizations concerning well-established formal results in, e.g., \cite{mart, beng, roo, ced, brink} and potential applications in physics as pioneered by, e. g., \cite{dix, mart, beng, roo, ced, brink,Toppan,top,toplu,Guna2,guna3,4a,Daboul}.
\end{abstract}

\section{Introduction}
This paper aims to provide a comprehensive investigation concerning a more general class of non-associative structures on $S^7$, in the Clifford algebra formulation of octonionic generalized products. As the octonionic product can be defined from the Clifford algebra structure, such formalism is closely related to the algebraic and geometric structures associated with the sphere $S^7$ \cite{aht,bot,miln}. Generalized octonionic  algebras and Moufang-like identities can be accomplished in this formalism, addressing their possible generalizations and additional properties, as well as prominent applications in physics. 

The triality principle introduced by Cartan \cite{car1, che} asserts  that in an 8-dimensional vector space $V$ there exists an order three automorphism, which  cyclically permutes vectors and semispinors --- also seen as  minimal left ideals of the Clifford algebra $\cl_{8,0}$, or $\cl_{0,8}$, or $\cl_{4,4}$, or their equivalent $\cl_8(\mathbb{C})$ complex case --- that carry non-equivalent representations of the Spin(8) group.  
In \cite{eu1} it was proved that new deformed octonionic units with respect to the non-associative product between Clifford bundle sections and octonionic fields can be introduced, in order to better investigate the generalization of Moufang identities in this context. Using the formalism presented and its generalization \cite{eu2,cru0}, the Poincar\'e superalgebra is obtained from the Clifford orthosymplectic algebra \cite{cru0}. It was also shown in \cite{eu2} --- following \cite{cru0} --- that
 the Chevalley product, an order three automorphism on the vector space constructed as the direct sum of maximal index vector spaces
 and their semispinor associated spaces, induces triality like morphisms on some subspaces of the associated complexified exterior algebra. 
See \cite{baez} for details including historical notes. Some interesting applications may be found in \cite{Toppan,top,toplu,Guna2,guna3,Schray,Daboul,Schafer,Duff1,Manogue,alb}. 
 
A similar question can be formulated, and it motivates the main aim of this paper: to investigate 
in what extent the well known results, concerning the octonionic product deformations on the tangent bundle on $S^7$, can be 
completely generalized to the whole exterior and Clifford bundles on $S^7$. Furthermore, although there is a plethora of new products that can be defined in such scenario, we restrict our formalism to the prominent products that are immediate generalization of the results in, i. e., Cederwall, Bengtsson, Rooman, Preitschopf, Brink
\cite{dix, mart, beng, roo, ced, brink}, and some potential applications regarding the same references directly -- as well as other ones as \cite{Toppan, top, toplu, Guna2, guna3,4a}.

First, the original non-associative deformed products  
between octonions are reviewed \cite{eu1,dix,mart,beng,roo}, together with the extended octonionic products between octonions and Clifford
multivectors, and also the extended generalized non-associative products between Clifford multivectors, in the light 
of \cite{eu1,dix}. Our results are immediately led to the formalism in \cite{dix,Schafer} in the very particular case 
where the paravector component of an arbitrary multivector in $\cl_{0,7}$ is taken into account.

The results in \cite{eu1} are generalized and more possibilities are considered, when the so called directional non-associative 
products are explicitly taken into account in the light of the former formalism \cite{eu1}. Also, more non-equivalent non-associative
products are introduced in order to encompass the exterior and Clifford bundles on $S^7$, in a more general context 
than the cases presented in \cite{eu1}.

Here we try to get the nature of the non-associative structures that can be defined on the exterior and Clifford bundles on $S^7$, and it is verified that all the additional unexpected properties concerning 
the non-associative structure in the Clifford bundle on $S^7$ can not be probed when only the underlying 
structure of the tangent bundle on $S^7$ is considered, like the results in \cite{beng,dix,ced,roo}.
The formalism here presented and studied probes additional properties that cannot be realized only in the tangent bundle on $S^7$.
Naturally, all the well known results concerning the deformed formalism on the tangent bundle on $S^7$ are re-obtained,
when we only consider the particular case where the paravector subspace $\Lambda^0(\RR^{0,7})\oplus\Lambda^1(\RR^{0,7})\simeq \RR\oplus\RR^{0,7}$,
associated to the standard octonionic product, is taken into account. To emphasize, this formalism encompass the most general underlying subspace $\Lambda(\RR^{0,7})$ associated to the 
Clifford algebra constructed on the tangent space at an arbitrary point $X\in S^7$.
We want here to stress that the exterior algebra $\Lambda(\RR^7)$ is a structure that does not depend on a matric structure, but our notation $\Lambda(\RR^{0,7})$ is opted in order to emphasize the underlying vector space $\RR^7$ endowed with a metric diag($-1,-1,-1,-1,-1,-1,-1$). In this space the octonions are naturally described.

Concerning the $X$-product and its equivalent matrix representation  in the Appendix, it was shown in \cite{dix,roo} that 
it introduces the Hopf fibration $S^3\ldots S^7\to S^4$, and we can also search for some extended correspondence between
the products non-associative products, that define a paralellizabe torsion on $S^7$, and some kind of generalized geometric structure that can be led to the Hopf fibration
$S^3\ldots S^7\to S^4$, in the very particular case when $u\in \Lambda^0(\RR^{0,7})\oplus\Lambda^1(\RR^{0,7})\simeq \RR\oplus\RR^{0,7}$. 
In this specific case the product
$\ee_a \circ_u \ee_b$ is made identical to the $X$-product between $\ee_a$ and $\ee_b$.

We can still ask which properties hold, when we consider the exterior and Clifford bundle instead of the tangent bundle on $S^7$ only. The formalism presented in \cite{dix,mart,ced, brink} shows that the octonionic product can be deformed, in order to encompass the parallelizable torsion on $S^7$ \cite{4a,roo}. The $X$-product as presented is exactly twice the torsion components, and we prove that, instead of considering the underlying vector space $\la^0(\RR^{0,7})\op\la^1(\RR^{0,7})$ associated with octonions algebra, it is possible to consider the whole Clifford algebra at an arbitrary point on $S^7$, with the underlying vector space associated with $\cl_{0,7}$.
Possible ramifications of this formalism into its applications may provide manageable models that extends, e.g., \cite{eu1,dix,ced}.

This paper is organized as follows: Section II reviews  some mathematical tools and techniques related to the octonionic algebra in the Clifford algebra arena, and Section III concentrates on the fundamental properties already introduced in \cite{eu1}, and also additional properties on these topics are provided. The new definitions reveal a wealth of unexpected results and the subtle difference arising in the generalization of the $u$-product and the directional $u$-product. These products are introduced with the purpose to get more general non-associative structures on $S^7$. In Section IV we summarize some properties in \cite{eu1} and present some examples elucidating the motivation around the formalism. In Section V,
new classes of non-associative products are introduced in the Clifford bundle on $S^7$, together with the directional non-associative products 
 and some new examples concerning counter-examples on the Moufang identities, that do not hold in our extended formalism. In Section VI, following \cite{roo}, the scalar product between octonions is defined, but now also multivectors fields in
the exterior and Clifford bundles on $S^7$ can be led to their octonionic similar in our formalism. For the sake of
completeness, the matrix representation introduced in \cite{dix} is extended with respect to the $\bu$-product defined in \cite{eu1}. In Section VII, four Propositions are presented and demonstrated. together with two respective extensions, introducing new octonionic involutions induced
by and arbitrary multivector $u\in \cl_{0,7}$. 
 The developments here can settle some open questions addressed in, e.g., \cite{mart,ced}. In Appendices  A-F  the respective demonstrations of Propositions 1-4 are provided. In Appendix G the construction of the tangent bundle on $S^7$ is considered and in Appendix H the matrix representation associated to the $\bu$-product and also to the whole
Clifford algebra $\cl_{0,7}$ basis, acting on an arbitrary point at $S^7$, is presented following previous considerations in \cite{dix}.

\section{Preliminaries}

Let $V$ be a finite $n$-dimensional real vector space and $V^*$ denotes its dual. 
We consider the tensor algebra $\bigoplus_{i=0}^\infty T^i(V)$ from which we
restrict our attention to the space $\Lambda(V) = \bigoplus_{k=0}^n\Lambda^k(V)$ of multivectors over $V$. $\Lambda^k(V)$
denotes the space of the antisymmetric
 $k$-tensors, isomorphic to the  $k$-forms vector space.  Given $\psi\in\Lambda(V)$, $\tilde\psi$ denotes the \emph{reversion}, 
 an algebra antiautomorphism
 given by $\tilde{\psi} = (-1)^{[k/2]}\psi$ ([$k$] denotes the integer part of $k$). $\hat\psi$ denotes 
the \emph{main automorphism or graded involution},  given by 
$\hat{\psi} = (-1)^k \psi$. The \emph{conjugation} is defined as the reversion followed by the main automorphism.
  If $V$ is endowed with a non-degenerate, symmetric, bilinear map $g: V^*\times V^* \rightarrow \RR$, it is 
possible to extend $g$ to $\la(V)$. Given $\psi=\uu^1\w\cdots\w \uu^k$ and $\phi=\vv^1\w\cdots\w \vv^l$, for $\uu^i, \vv^j\in V^*$, one defines $g(\psi,\phi)
 = \det(g(\uu^i,\vv^j))$ if $k=l$ and $g(\psi,\phi)=0$ if $k\neq l$. The projection of a multivector $\psi= \psi_0 + \psi_1 + \cdots + \psi_n$,
 $\psi_k \in \la^k(V)$, on its $p$-vector part is given by $\langle\psi\rangle_p$ = $\psi_p$. 
 Given $\psi,\phi,\xi\in\Lambda(V)$, the  {\it left contraction} is defined implicitly by 
$g(\psi\lrcorner\phi,\xi)=g(\phi,\tilde\psi\w\xi)$. 
 For $a \in \RR$, it follows that 
 ${\bf v} \lrcorner a = 0$. Given $\vv\in V$, the Leibniz rule 
${\bf v}\lrcorner (\psi \w \phi) = ({\bf v} \lrcorner \psi) \w \phi + \hat\psi \w ({\bf v} \lrcorner \phi)$ holds. The 
 {\it right contraction} is analogously defined 
$g(\psi\llcorner\phi,\xi)=g(\phi,\psi\w\tilde\xi)$
 and its associated Leibniz rule $
(\psi \w \phi) \llcorner {\bf v} = \psi \w (\phi \llcorner {\bf v}) + (\psi \llcorner {\bf v}) \w \hat\phi
$ holds. Both contractions are related by 
${\bf v} \lrcorner \psi = -\hat\psi \llcorner {\bf v}$.
The Clifford product between $\ww\in V$ and $\psi\in\la(V)$ is given by $\ww\psi = \ww\w \psi + \ww\lrcorner \psi$.
 The Grassmann algebra $(\la(V),g)$ 
endowed with the Clifford  product is denoted by $\cl(V,g)$ or $\cl_{p,q}$, the Clifford algebra associated with $V\simeq \RR^{p,q},\; p + q = n$.

\section{Octonions}
The octonionic algebra $\mathbb{O}$ can be defined as the paravector space $\RR\op\RR^{0,7}$ \cite{bay2} endowed with the product $\circ \colon (\RR \op \RR^{0, 7}) \times (\RR \op \RR^{0, 7}) \to \RR \op \RR^{0, 7}$, the so called octonionic standard product. The identity $\ee_0=1$ and an orthonormal basis $\left\{\ee_a\right\}^{7}_{a=1}$, in the underlying paravector space $\RR \op \RR^{0, 7}\hookrightarrow \cl_{0, 7}$ associated with $\mathbb{O}$, generate the octonion algebra \cite{baez,ree,8a}. The octonionic product can be constructed using the Clifford algebra $\cl_{0, 7}$ as 
\beq \label{201}A\circ B=\left\langle AB(1-\psi)\right\rangle_{0\op 1}, \quad A,B \in \RR \op \RR^{0, 7},\eeq
where $\psi=\ee_1\ee_2\ee_6+\ee_2\ee_3\ee_7+\ee_3\ee_4\ee_1+\ee_4\ee_5\ee_2+\ee_5\ee_6\ee_3+\ee_6\ee_7\ee_4+\ee_7\ee_1\ee_5\in\la^3(\RR^{0, 7})\hookrightarrow \cl_{0, 7}$, and the juxtaposition denotes the Clifford product \cite{loun}. 
The contrivance of introducing the octonionic product from the Clifford product in this context is to present hereon our formalism using Clifford algebras and the subsequent generalizations to the whole exterior and Clifford bundles on $S^7$. Indeed, as $\mathbb{O}$ is isomorphic to $\RR \op \RR^{0, 7}$ as a vector space, the octonionic product takes two arbitrary elements of the paravector space $\RR \op \RR^{0, 7}$ --- which is itself endowed with the octonionic product --- resulting in another element of the paravector space. But looking at the octonions in the Clifford algebra arena it is possible to go beyond the paravector space and explore the whole  exterior algebra underlying the Clifford algebra, which is one possibility we use to generalize the $X$- and $XY$-products, extending also the results in \cite{eu1}.

The exterior algebra $\la(\RR^{7})$ is denoted by $\la(\RR^{0,7})$, as in \cite{eu1}, to emphasize the underlying octonionic formalism character. It is well known that the exterior algebra is constructed on a vector space, without mention to any metric structure, and the notation $\la(\RR^{0,7})$ seems \emph{a priori} out of context, but we want to emphasize the fact that the underlying vector space is related to $\RR^7$ endowed with the metric $g$ of signature $(0,7)$.

In a close analogy, the octonionic product can be also expressed in terms of the Clifford algebra on the Euclidean space $\RR^{8, 0}$ according to \cite{loun}, in terms of a basis $\left\{e_1, \ldots, e_8\right\}$ of $\RR^{8, 0}$. The octonionic product is given by
\beq \label{202}A\circ B=\left\langle A e_8 B (1+\phi)(1-e_{12\ldots 8})\right\rangle_1, \quad \text{ $A,B \in \RR^{8, 0}$,}\eeq 
where $\phi=e_1e_2e_6+e_2e_3e_7+e_3e_4e_1+e_4e_5e_2+e_5e_6e_3+e_6e_7e_4+e_7e_1e_5 \in \la^3(\RR^{8, 0})$ and $\frac{1}{8}(1+\phi)\frac{1}{2}(1-e_{12\ldots 8})$ is an idempotent. Both the approaches are equivalent: bivectors in $\cl_{8, 0}$ correspond to the paravectors of $\cl_{0,7}$ when the isomorphism $e_\sigma \mapsto e_\sigma e_8 := \ee_\sigma, \; \scriptstyle{\sigma = 1,2,\ldots,7}$ is considered and $e_8e_8 = 1 = \ee_0$ denotes the octonionic unit in $\RR\hko \RR\op\RR^{0,7}$. In fact, 
$\ee_\sigma^2 = (e_\sigma e_8)^2 = -e_\sigma e_\sigma e_8e_8 = -1$. The octonion unit in $\RR\op\RR^{0,7}$ corresponds to the vector $e_8 \in \RR^8$. 
More details can be seen, e.g., in \cite{loun}.

The usual rules between basis elements under the octonionic product are verified when both the Eqs.(\ref{201}) and (\ref{202}) are used, and the definition in Eq.(\ref{201}) is regarded hereupon, in this case $\RR^7$ is considered instead of the usual $\RR^8$ underlying vector space, concerning the definition in Eq.(\ref{202}). In particular in the context of Eq.(\ref{201}) the octonion multiplication table is constructed by
\beq \label{203}\ee_a\circ \ee_b=\epsilon^{c}_{ab}\ee_c-\delta_{ab} \quad (a,b,c=1,\ldots,7),\eeq
where $\epsilon^{c}_{ab}=1$ for the cyclic permutations
$(abc)=(126),(237),(341),(452),(563),(674),(715).$ 
Explicitly, the multiplication is given by Table \ref{table1}, wherein all the relations can be expressed as $\ee_a\circ \ee_{a+1}=\ee_{a+5 \mod 7}$.
\begin{table}
\caption{\label{table1}The octonionic product between units in the $\mathbb{O}_{+5}$ convention.}
\begin{center}
\begin{tabular}{llllllll}
\br
$1$ \,\;&\;\, $\ee_1$ \,\;&\;\, $\ee_2$ \,\;&\;\, $\ee_3$ \,\;&\;\, $\ee_4$ \,\;&\;\, $\ee_5$ \,\;&\;\, $\ee_6$ \,\;&\;\, $\ee_7$ \\
\hline\hline
$\ee_1$ \,\;&\;\, $-1$ \,\;&\;\, $\ee_6$ \,\;&\;\, $\ee_4$ \,\;&\;\, $-\ee_3$ \,\;&\;\, $\ee_7$ \,\;&\;\, $-\ee_2$ \,\;&\;\, $-\ee_5$\\ 
\hline
$\ee_2$ \,\;&\;\, $-\ee_6$ \,\;&\;\, $-1$ \,\;&\;\, $\ee_7$ \,\;&\;\, $\ee_5$ \,\;&\;\, $-\ee_4$ \,\;&\;\, $\ee_1$ \,\;&\;\, $-\ee_3$\\
\hline
$\ee_3$ \,\;&\;\, $-\ee_4$ \,\;&\;\, $-\ee_7$ \,\;&\;\, $-1$ \,\;&\;\, $\ee_1$ \,\;&\;\, $\ee_6$ \,\;&\;\, $-\ee_5$ \,\;&\;\, $\ee_2$\\
\hline
$\ee_4$ \,\;&\;\, $\ee_3$ \,\;&\;\, $-\ee_5$ \,\;&\;\, $-\ee_1$ \,\;&\;\, $-1$ \,\;&\;\, $\ee_2$ \,\;&\;\, $\ee_7$ \,\;&\;\, $-\ee_6$\\
\hline
$\ee_5$ \,\;&\;\, $-\ee_7$ \,\;&\;\, $\ee_4$ \,\;&\;\, $-\ee_6$ \,\;&\;\, $-\ee_2$ \,\;&\;\, $-1$ \,\;&\;\, $\ee_3$ \,\;&\;\, $\ee_1$\\
\hline
$\ee_6$ \,\;&\;\, $\ee_2$ \,\;&\;\, $-\ee_1$ \,\;&\;\, $\ee_5$ \,\;&\;\, $-\ee_7$ \,\;&\;\, $-\ee_3$ \,\;&\;\, $-1$ \,\;&\;\, $\ee_4$\\
\hline
$\ee_7$ \,\;&\;\, $\ee_5$ \,\;&\;\, $\ee_3$ \,\;&\;\, $-\ee_2$ \,\;&\;\, $\ee_6$ \,\;&\;\, $-\ee_1$ \,\;&\;\, $-\ee_4$ \,\;&\;\, $-1$\\
\br
\end{tabular}
\end{center}
\end{table}
\noi 
Some useful identities follow from  Eq.(\ref{203}):
\begin{equation}
\epsilon_{abc}\epsilon_{dcf}+\epsilon_{dbc}\epsilon_{acf}=\delta_{ab}\delta_{df}+\delta_{af}\delta_{db}-2\delta_{ad}\delta_{bf},\nonumber
\end{equation}
and when  an analogous of the Jacobi formula  is computed, it reads
\beq
[\ee_{i},[\ee_{j},\ee_{k}]]+[\ee_{k},[\ee_{i},\ee_{j}]]+[\ee_{j},[\ee_{k},\ee_{i}]] 
=4(\epsilon_{jkm}\epsilon_{imn}+\epsilon_{ijm}\epsilon_{kmn}+\epsilon_{kim}\epsilon_{jmn})\ee_{n} = 3\epsilon_{ijkl}\ee_{l}\eeq\noi where $\epsilon_{ijkl}=-\epsilon_{mij}\epsilon_{mkl}-\delta_{il}\delta_{jk}+\delta_{ik}\delta_{jl}$ \cite{Guna2}. 
 Since the underlying vector space of $\mathbb{O}$ can be considered as being $\RR \op \RR^{0, 7}\hookrightarrow \cl_{0, 7}$, the Clifford conjugation of $X=X^0+X^a\ee_a \in \mathbb{O}$ is given by $\bar{X}=X^0-X^a\ee_a$, where $X^0$ and $X^a$ are real coefficients. The Einstein's summation convention --- asserting that when two equal indices appear, it indicates a sum over this indices, i.e., $X^a\ee_a$ denotes $X^1\ee_1+X^2\ee_2+\cdots +X^7\ee_7=\sum^{7}_{a=1}X^a\ee_a$ --- is used hereon. The underlying structure of the vector space is unable to assert whether the $\mathbb{O}$-conjugation is equivalent to the graded involution of the tensorial algebra, extended to the exterior and Clifford algebras, since the octonionic conjugation $\bar{X}$ can be written either as $\widehat{X}$ or $\bar{X}$, in terms of the Clifford algebra morphisms. But it is well known that the octonionic conjugation $\bar{X}$ is involutive and an anti-automorphism, what  immediately excludes the graded involution.

\section{The $\bu$-product and the $\odot$-product on $S^7$} 
 Given $X,Y \in \RR \op \RR^{0, 7}$ fixed but arbitrary such that $X\bar{X}=\bar{X}X=1=\bar{Y}Y=Y\bar{Y}$ ($X,Y \in S^7$), the $X$-product  is defined by \cite{ced, mart, dix}
\beq \label{205}A\circ_X B:=(A\circ X)\circ(\bar{X}\circ B).\eeq
For a particular value when
 $X=B\in S^7$, the usual octonionic product is obtained $(A\cv B)\cv(\bar{B}\cv B)=A\cv B$. The expressions below are shown in, e.g., \cite{dix, ced}
\beq\label{3p14}(A\circ X)\circ(\bar{X}\circ B)=X\circ ((\bar{X}\circ A)\circ B)=(A\circ(B\circ X))\circ \bar{X}.\eeq
\noi The $XY$-product is defined as
\beq \label{206}A\circ_{X,Y} B:=(A\circ X)\circ(\bar{Y}\circ B),\eeq
\noi and in particular the $(1,X)$-product is given by 
\beq \label{207}A\circ_{1,X} B:=A\circ(\bar{X}\circ B),\eeq
where $X$ is the unit of the $(1,X)$-product above, since $A\circ_{1,X} X=X\circ_{1,X} A=A$ \cite{dix,ced}.

A non-associative product called the $u$-product was introduced in \cite{eu1} as a natural generalization for the $X$-product. We briefly review the respective products obtained in \cite{eu1} for completeness, and subsequently  present a more general class of non-associative products on $S^7$. 
For homogeneous multivectors $u=u_1\ldots u_k\in\la^k(\RR^{0, 7})\hookrightarrow\cl_{0, 7}$, where $\left\{u_p\right\}^{k}_{p=1}\subset\RR^{0, 7}$  and $A\in\RR\op\RR^{0, 7}$, the products $\bu_\llcorner$ and $\bu_\lrcorner$ are  defined by \cite{eu1}
\beq \bu_\llcorner \colon (\RR \op \RR^{0, 7})\times \la^k(\RR^{0, 7})&\to& \RR \op \RR^{0, 7}\n
(A, u)&\mapsto& A\bu_\llcorner u=((\cdots((A\circ u_1)\circ u_2)\circ \cdots)\circ u_{k-1})\circ u_k \label{209}\\
\bu_\lrcorner \colon \la^k(\RR^{0, 7})\times (\RR \op \RR^{0, 7})&\to& \RR \op \RR^{0, 7}\n
(u, A)&\mapsto& u\bu_\lrcorner A=u_1\circ (u_2\cv(\cdots \circ (u_{k-1}\circ (u_k \circ A))\cdots ))\label{210}\eeq
 The symbol $\bu$ uniquely denotes both the products $\bu_\llcorner$ and $\bu_\lrcorner$, in Eqs.(\ref{209}) and (\ref{210}),
since for this product it is implicit where the octonion enters in.
The products (\ref{209}, \ref{210}) are extended to the whole $\Lambda(\RR^{0,7})$ by linearity.
\medbreak
{\bf Remark 1}: The expression $X\bar{X}=\bar{X}X=1$ defines $S^7$, for $X\in\mathbb{O}$.
The product in (\ref{205}) $A\circ_X B=(A\circ X)\circ(\bar{X}\circ B)$ was motivated 
in the sense that it can be written as $A\circ_X B=(A\circ X)\circ({X}^{-1}\circ B)$, since 
for all $X\in S^7$ it follows that $X^{-1} = \bar{X}$. 
 Now, for a multivector $u\in\cl_{0,7}$, we generalized this product in \cite{eu1}, and we must emphasize that
 in order that $(A \bu_\llcorner u)\circ(u^{-1} \bu_\lrcorner B)$ be well defined, there must exist an inverse $u^{-1}$ associated to $u\in\cl_{0,7}$. As $u\bar{u}$
has not only a scalar component in general --- indeed $u\bar{u}\neq\bar{u}u$ in general --- the existence of an invertible element $u\in\cl_{0,7}$ is a necessity to define a generalization of the $X$-product
in order that the term $(A \bu_\llcorner u)\circ(u^{-1} \bu_\lrcorner B)$ is proportional to $(A \bu_\llcorner u)\circ(\bar{u} \bu_\lrcorner B)$.
  
The main aim of this work is to consider how the generalizations of $X\bar{X}=\bar{X}X=1$ defining $S^7$ in the Clifford algebra context for an arbitrary element $u\in\cl_{0,7}$ affect the subsequent deformations 
on the octonionic algebra. In general $u\bar{u}$ and $\bar{u} u$ are not scalars, and the only case we can guarantee that $u^{-1} = \pm \bar{u}$  --- or equivalently $\bar{u}u  = \pm 1$ is regarded ---  in general is to consider the homogeneous and simple elements $u\in\la^k(\RR^{0,7})$, wherein it can be surely asserted that $u^{-1}=\pm u$.
%After, other most general elements of $\la(\RR^{0,7})$ satisfying $u^{-1} = \pm u$ should be considered, but they are rare in $\cl_{0,7}$.
\medbreak
Given an  element $u \in \la(\RR^{0, 7})$, the $u$-product is defined as
\beq \cv_u \colon (\RR \op \RR^{0, 7})\times (\RR \op \RR^{0, 7}) &\to& \RR \op \RR^{0, 7} \n 
(A, B)&\mapsto& A\cv_u B:=(A\bu_\llcorner u)\cv (\bar{u}\bu_\lrcorner B). \label{213}\eeq
The authors in \cite{eu1} ask whether the relations $A\cv_u B:=(A\bu_\llcorner u)\cv (\bar{u}\bu_\lrcorner B)=(A\cv(B\bu_\llcorner u))\bu_\llcorner \bar{u}=u\bu_\lrcorner((\bar{u}\bu_\lrcorner A)\cv B)$
hold, in a context where any similar generalization related to Eq.(\ref{205}) can be constructed in the non-associative  formalism 
induced by the $u$-product, where $u$ can also more generally denote a form field on the exterior bundle on $S^7$.
\medbreak
{\bf Example 1}: Taking $u=\ee_1-\ee_2\ee_3\in\cl_{0,7}$, $A=A^2\ee_2+A^4\ee_4\in \mathbb{O}$, and $B=B^1\ee_1+B^5\ee_5\in\mathbb{O}$, it follows that
\beq (A\bu_\llcorner u)\cv(\bar{u}\bu_\lrcorner B)&=&\left[(A^2\ee_2+A^4\ee_4)\bu_\llcorner(\ee_1-\ee_2\ee_3)\right]\cv\left[\left(\overline{\ee_1-\ee_2\ee_3}\right)\bu_\lrcorner(B^1\ee_1+B^5\ee_5)\right]\n
&=&(-A^2\ee_6+A^2\ee_3+A^4\ee_3-A^4\ee_6)\cv(B^1-B^5\ee_7+B^1\ee_5-B^5\ee_1)\n
&=&2\left(A^2B^5\ee_4+A^2B^1\ee_3+A^4B^1\ee_3+A^4B^5\ee_4\right)\nonumber\eeq
while
\beq (A\cv(B\bu_\llcorner u))\bu_\llcorner \bar{u}&=&\left[(A^2\ee_2+A^4\ee_4)\cv\left[(B^1\ee_1+B^5\ee_5)\bu_\llcorner(\ee_1-\ee_2\ee_3)\right]\right]\bu_\llcorner (\overline{\ee_1-\ee_2\ee_3})\n
&=&\left[(A^2\ee_2+A^4\ee_4)\cv(-B^1-B^1\ee_5-B^5\ee_7+B^5\ee_1)\right]\bu_\llcorner(-\ee_1+\ee_3\ee_2)\n
&=&2\left(-A^2B^1\ee_6-A^2B^1\ee_3-A^4B^5\ee_2+A^4B^5\ee_4\right).\nonumber\eeq
\medbreak
An open question  about the validity of the expression $(A\circ X)\circ(\bar{X}\circ B)=(A\circ(B\circ X))\circ \bar{X}$ for a more general setting  concerns how the use of the $\bu$-product instead of the standard octonionic product affects Eq.(\ref{3p14}), and it was presented in \cite{eu1} in a particular context. Specifically, it has been argued whether the introduction of the $u$-product could allow us to generalize such expression immediately, in order to consider all the exterior algebra constructed on the tangent space on an arbitrary point on $S^7$. Instead of $X \in \RR \op \RR^{0, 7}$,  an analogous expression for $u \in \cl_{0,7}$ that expresses the non-associative algebraic structure related to  the exterior bundle on $S^7$ can be obtained.

In general Eqs.(\ref{3p14}) are not generalizable in terms of a naive immediate substitution, as it can be verified that
\beq (A\bu_\llcorner u)\cv(\bar{u}\bu_\lrcorner B)\neq(A\cv(B\bu_\llcorner u))\bu_\llcorner \bar{u}.\label{contra-exemplo}\eeq
\noi in the Example 1 above. When $u$ is a paravector --- an element of $\RR \op \RR^{0, 7}$ --- it is clear that the $u$-product is equivalent to the $X$-product.

In a close analogy to the $XY$-product and the $(1, X)$-product, respectively defined by Eqs.(\ref{206}) and (\ref{207}), it is also possible to define another product,  the $(1, u)$-product, as
\beq \cv_{1, u}\colon (\RR \op \RR^{0, 7})\times (\RR \op \RR^{0, 7}) &\to& \RR \op \RR^{0, 7} \n
(A, B)&\mapsto& A\cv_{1, u} B:=(A\cv 1)\cv (\bar{u}\bu_\lrcorner B) = A\cv (\bar{u}\bu_\lrcorner B). \label{215}\eeq
Finally, Eq.(\ref{206}) can be generalized for $u, v \in \cl_{0, 7}$ fixed, as follows:
\beq \cv_{u, v}\colon (\RR \op \RR^{0, 7})\times (\RR \op \RR^{0, 7}) &\to& \RR \op \RR^{0, 7} \n
(A, B)&\mapsto& A\cv_{u, v} B:=(A\bu_\llcorner u)\cv (\bar{v}\bu_\lrcorner B). \label{216}\eeq

Given $u=u_1\ldots u_k$ and $v=v_1\ldots v_l\in \cl_{0, 7}$, the non-associative product between Clifford algebra elements was defined in \cite{eu1} as
\beq \odot_\llcorner \colon \cl_{0, 7} \times \cl_{0, 7} &\to& \RR \op \RR^{0, 7}\n
(u, v)&\mapsto& u\odot_\llcorner v:=u_1 \cv (u_2 \cv(\cdots \cv(u_{k-1}\cv(u_k \bu_\llcorner v))\cdots)).\label{219}\\
 \odot_\lrcorner \colon \cl_{0, 7} \times \cl_{0, 7} &\to& \RR\op \RR^{0, 7}\n
(u, v)&\mapsto& u\odot_\lrcorner v:=((\cdots\cv((u\bu_\lrcorner v_1)\cv v_2)\cv\cdots)\cv v_{k-1})\cv v_k.\label{221}\eeq
The symbol $\odot$ denotes both the products $\odot_\llcorner$ and $\odot_\lrcorner$. It is easy to see that, when elements of $\cl_{0, 7}$ are restricted to the paravector space $\RR \op \RR^{0, 7}$, then $A\bu B\equiv A\cv B$ and $A \odot B \equiv A \cv B$, where $A,B \in \RR \op \RR^{0, 7}$.
\medbreak
{\bf Example 2}: Let us calculate the product $(2\ee_1\ee_2-7\ee_5\ee_6)\odot_\llcorner\ee_3\ee_4$:
\beq (2\ee_1\ee_2-7\ee_5\ee_6)\odot_\llcorner\ee_3\ee_4&=&2\ee_1\ee_2\odot_\llcorner\ee_3\ee_4-7\ee_5\ee_6\odot_\llcorner\ee_3\ee_4\n
&=&2\ee_1\cv(\ee_7\cv\ee_4)-7\ee_5\cv(\ee_5\cv\ee_4)=2\ee_1\cv\ee_6-7\ee_5\cv(-\ee_2)\n
&=&-2\ee_2+7\ee_4,\label{220}\eeq
while
\beq (2\ee_1\ee_2-7\ee_5\ee_6)\odot_\lrcorner\ee_3\ee_4&=&2\ee_1\ee_2\odot_\lrcorner\ee_3\ee_4-7\ee_5\ee_6\odot_\lrcorner\ee_3\ee_4\n
&=&2(\ee_1\cv\ee_7)\cv\ee_4-7(\ee_5\cv\ee_5)\cv\ee_4=2(-\ee_5)\cv\ee_4+7\ee_4\n
&=&+2\ee_2+7\ee_4,\label{eq2}\eeq
so Eqs.(\ref{220},\ref{eq2}) can not be mapped into another through automorphisms or antiautomorphisms of $\cl_{0,7}$.
After we shall see that it is possible to induce an involution on a deformed octonionic algebra, induced by an arbitrary Clifford element, that can make some extension of the Moufang identities possible. 
\medbreak
In what follows it is implicit that $u \in \cl_{0, 7}$ is not a scalar, since in this case would be nothing new to prove, since $A=1\cv A=1\bu A=1\odot A,\quad \forall A\in \mathbb{O}$.

\section{Generalized non-associative products on $S^7$} 
All the possible products obtained from the combinations between $\cv, \bu_\llcorner, \bu_\lrcorner, \odot_\llcorner$, and $\odot_\lrcorner$ are listed below. Some examples are given to illustrate the different results obtained if only one of those products are interchanged. Moreover, the generalizations for the $u$-, $(1,u)$-, and $(u,v)$- products are provided by means of the directional octonionic products.

The definitions above allow us to see that the $(1, u)$-product can be generalized to encompass and include multivectors of $\cl_{0, 7}$ in the first or second entry, so the $(1, u)$-product  for Clifford multivectors in the first entry is
given by
\beq\cv_{1,u}\colon \cl_{0,7}\times\mathbb{O} &\to& \mathbb{O}\n
(\psi,A)&\mapsto& \psi\circ_{1,u} A:= (\psi\bu_\lrcorner 1)\cv(\bar{u}\bu_\lrcorner A),\eeq
which is the immediate generalization for the $(1,u)$-product defined in Eq.(\ref{215}), since $A\cv 1=A,$ $\forall A\in\mathbb{O}$ and $\psi\bu_\lrcorner 1\neq\psi,$ $\forall \psi\in \cl_{0,7}\setminus(\RR\op\RR^{0,7})$. Therefore, the following  product can be also defined
\beq\blacktriangledown_{1,u}\colon \cl_{0,7}\times\mathbb{O} &\to& \mathbb{O}\n
(\psi,A)&\mapsto& \psi\blacktriangledown_{1,u} A:= \psi\bu_\lrcorner(\bar{u}\bu_\lrcorner A).\eeq\noi Such  a product is  a generalization that is not analogous to the $(1,u)$-product defined in Eq.(\ref{215}), but it is exactly the immediate generalization of the standard octonionic formalism, given by Eq.(\ref{207}).

For a Clifford multivector in the right, two non-equivalent possibilities can be introduced by defining them in terms of the $\odot_\lrcorner$ or the $\odot_\llcorner$-product as well:
\beq \cv^{\llcorner}_{1,u}\colon \mathbb{O} \times \cl_{0,7} &\to& \mathbb{O}\n
(A,\psi)&\mapsto& A\cv^{\llcorner}_{1,u} \psi:= (A\cv 1)\cv(\bar{u}\odot_\llcorner \psi) = A\cv(\bar{u}\odot_\llcorner \psi),\label{t11}\eeq
while for the Clifford multivector in the right and the $\odot_\lrcorner$-product in the left it reads
\beq \cv^{\lrcorner}_{1,u}\colon \mathbb{O} \times \cl_{0,7} &\to& \mathbb{O}\n
(A,\psi)&\mapsto& A\cv^{\lrcorner}_{1,u} \psi:= (A\cv 1)\cv(\bar{u}\odot_\lrcorner \psi) = A\cv(\bar{u}\odot_\lrcorner \psi)\label{t22}.\eeq
The last extension of the $(1, u)$-product for given $\psi, \phi \in \cl_{0,7}$, fixed but arbitrary, is defined by
\beq \cv^{\llcorner}_{1,u}\colon\cl_{0,7}\times\cl_{0,7} &\to& \mathbb{O}\n
(\psi,\phi)&\mapsto& \psi\cv^{\llcorner}_{1,u} \phi:= (\psi\bu_\lrcorner 1)\cv(\bar{u}\odot_\llcorner \phi),\\ 
\cv^{\lrcorner}_{1,u}\colon\cl_{0,7}\times\cl_{0,7} &\to& \mathbb{O}\n
(\psi,\phi)&\mapsto& \psi\cv^{\lrcorner}_{1,u} \phi:= (\psi\bu_\lrcorner 1)\cv(\bar{u}\odot_\lrcorner \phi).\eeq 
And, for the non-similar generalization it follows
\beq \blacktriangledown^{\llcorner}_{1,u}\colon\cl_{0,7}\times\cl_{0,7} &\to& \mathbb{O}\n
(\psi,\phi)&\mapsto& \psi\blacktriangledown^{\llcorner}_{1,u} \phi:= \psi\bu_\lrcorner(\bar{u}\odot_\llcorner \phi),\\ 
\blacktriangledown^{\lrcorner}_{1,u}\colon\cl_{0,7}\times\cl_{0,7} &\to& \mathbb{O}\n
(\psi,\phi)&\mapsto& \psi\blacktriangledown^{\lrcorner}_{1,u} \phi:=\psi\bu_\lrcorner(\bar{u}\odot_\lrcorner \phi).\eeq 
\medbreak
{\bf Remark 2}: Note that, $A\cv^{\llcorner}_{1,u} \psi=A\blacktriangledown^{\llcorner}_{1,u} \psi$ and $A\cv^{\lrcorner}_{1,u} \psi=A\blacktriangledown^{\lrcorner}_{1,u} \psi$, while $\psi\cv^{\llcorner}_{1,u} \phi\neq\psi\blacktriangledown^{\llcorner}_{1,u} \phi$, and $\psi\cv^{\lrcorner}_{1,u} \phi\neq\psi\blacktriangledown^{\lrcorner}_{1,u} \phi$, i.e., where an octonion appear in the first entry the equality between the $\cv_{1,u}$- and $\blacktriangledown_{1,u}$-products is satisfied. Indeed, 
\beq A\cv^{\llcorner}_{1,u} \psi=(A\cv 1)\cv(u\odot_\llcorner \psi)=A\cv(u\odot_\llcorner\psi)=A\blacktriangledown^{\llcorner}_{1,u}\psi\n
\psi\cv^{\llcorner}_{1,u}\phi=(\psi\bu_\lrcorner 1)\cv(\bar{u}\odot_{\llcorner}\phi)\neq\psi\bu_\lrcorner(\bar{u}\odot_{\llcorner}\phi)=\psi\blacktriangledown^{\llcorner}_{1,u}\phi\nonumber\eeq
since $\psi\bu_\lrcorner 1$ is in general an octonion and $\psi$ is a Clifford multivector, the result of $A\cv^{\lrcorner}_{1,u}\psi=A\blacktriangledown^{\lrcorner}_{1,u}\psi$ and $\psi\cv^{\lrcorner}_{1,u}\phi\neq\psi\blacktriangledown^{\lrcorner}_{1,u}\phi$ follows analogously.
\medbreak
In order to generalize the $u$-product for a Clifford multivector, an octonion is replaced by a Clifford multivector in one of the entries. Firstly a Clifford multivector is placed in the first entry and using the $\odot$-product in the right and in the left, and subsequently for a Clifford multivector in the second entry:
\beq 
\cv^{\llcorner \bu}_u\colon \cl_{0,7}\times\mathbb{O}&\to&\mathbb{O}\n
(\psi,A)&\mapsto& \psi\cv^{\llcorner \bu}_u A:= (\psi\odot_\llcorner u)\cv(\bar{u}\bu_\lrcorner A),\\ 
\cv^{\lrcorner \bu}_u\colon \cl_{0,7}\times\mathbb{O}&\to&\mathbb{O}\n
(\psi,A)&\mapsto& \psi\cv^{\lrcorner \bu}_u A:= (\psi\odot_\lrcorner u)\cv(\bar{u}\bu_\lrcorner A),\\
\cv^{\bu \llcorner}_u\colon \mathbb{O}\times\cl_{0,7}&\to&\mathbb{O}\n
(A,\psi)&\mapsto& A\cv^{\bu \llcorner}_u \psi:= (A\bu_\llcorner u)\cv(\bar{u}\odot_\llcorner \psi),\\ 
\cv^{\bu \lrcorner}_u\colon \mathbb{O}\times\cl_{0,7}&\to&\mathbb{O}\n
(A,\psi)&\mapsto& A\cv^{\bu \lrcorner}_u \psi:= (A\bu_\llcorner u)\cv(\bar{u}\odot_\lrcorner \psi).\eeq 
Now an arbitrary multivector $\psi \in \cl_{0,7}$ is taken into account in both entries, and a choice needs to be made for the  $\odot$-product direction, introducing the directional non-associative products. For Eq.(\ref{ex3}) in Example 3 there are explicit computations illustrating how to perform the products $\circ^{\llcorner \llcorner}_{u}$ and $\circ^{\llcorner \lrcorner}_{u}$.
\beq
\cv^{\llcorner \llcorner}_u\colon \cl_{0,7}\times\cl_{0,7}&\to&\mathbb{O}\n
(\psi,\phi)&\mapsto& \psi\cv^{\llcorner \llcorner}_u \phi:= (\psi\odot_\llcorner u)\cv(\bar{u}\odot_\llcorner \phi),\label{ex3}\\
\cv^{\llcorner \lrcorner}_u\colon \cl_{0,7}\times\cl_{0,7}&\to&\mathbb{O}\n
(\psi,\phi)&\mapsto& \psi\cv^{\llcorner \lrcorner}_u \phi:= (\psi\odot_\llcorner u)\cv(\bar{u}\odot_\lrcorner \phi),\\
\cv^{\lrcorner \llcorner}_u\colon \cl_{0,7}\times\cl_{0,7}&\to&\mathbb{O}\n
(\psi,\phi)&\mapsto& \psi\cv^{\lrcorner \llcorner}_u \phi:= (\psi\odot_\lrcorner u)\cv(\bar{u}\odot_\llcorner \phi),\\
\cv^{\lrcorner \lrcorner}_u\colon \cl_{0,7}\times\cl_{0,7}&\to&\mathbb{O}\n
(\psi,\phi)&\mapsto& \psi\cv^{\lrcorner \lrcorner}_u \phi:= (\psi\odot_\lrcorner u)\cv(\bar{u}\odot_\lrcorner \phi).\eeq
The $(u, v)$-product is now generalized \cite{eu1}, hence the product is considered with entries in $\mathbb{O}$ and $\cl_{0,7}$ where in both the cases the $\odot$-product is performed in two directions.
\beq
\cv^{\llcorner \bu}_{u,v}\colon\cl_{0,7}\times\mathbb{O} &\to& \mathbb{O}\n
(\psi,A)&\mapsto& \psi\cv^{\llcorner \bu}_{u,v} A:= (\psi\odot_\llcorner u)\cv(\bar{v}\bu_\lrcorner A),\\
\cv^{\lrcorner \bu}_{u,v}\colon\cl_{0,7}\times\mathbb{O} &\to& \mathbb{O}\n
(\psi,A)&\mapsto& \psi\cv^{\lrcorner \bu}_{u,v} A:= (\psi\odot_\lrcorner u)\cv(\bar{v}\bu_\lrcorner A),\\
\cv^{\bu \llcorner}_{u,v}\colon \mathbb{O}\times\cl_{0,7}&\to&\mathbb{O}\n
(A,\psi)&\mapsto& A\cv^{\bu \llcorner}_{u,v} \psi:= (A\bu_\llcorner u)\cv(\bar{v}\odot_\llcorner \psi),\\ 
\cv^{\bu \lrcorner}_{u,v}\colon \mathbb{O}\times\cl_{0,7}&\to&\mathbb{O}\n
(A,\psi)&\mapsto& A\cv^{\bu \lrcorner}_{u,v} \psi:= (A\bu_\llcorner u)\cv(\bar{v}\odot_\lrcorner \psi).\eeq
Finally, the $(u, v)$-product can be generalized for Clifford multivectors in both entries, as in \cite{eu1}, but now explicitly considering
both the directions related to the $\odot$-product:
\beq
\cv^{\llcorner \llcorner}_{u,v}\colon \cl_{0,7}\times\cl_{0,7}&\to&\mathbb{O}\n
(\psi,\phi)&\mapsto& \psi\cv^{\llcorner \llcorner}_{u,v} \phi:= (\psi\odot_\llcorner u)\cv(\bar{v}\odot_\llcorner \phi),\\
\cv^{\llcorner \lrcorner}_{u,v}\colon \cl_{0,7}\times\cl_{0,7}&\to&\mathbb{O}\n
(\psi,\phi)&\mapsto& \psi\cv^{\llcorner \lrcorner}_{u,v} \phi:= (\psi\odot_\llcorner u)\cv(\bar{v}\odot_\lrcorner \phi),\\
\cv^{\lrcorner \llcorner}_{u,v}\colon \cl_{0,7}\times\cl_{0,7}&\to&\mathbb{O}\n
(\psi,\phi)&\mapsto& \psi\cv^{\lrcorner \llcorner}_{u,v} \phi:= (\psi\odot_\lrcorner u)\cv(\bar{v}\odot_\llcorner \phi),\\
\cv^{\lrcorner \lrcorner}_{u,v}\colon \cl_{0,7}\times\cl_{0,7}&\to&\mathbb{O}\n
(\psi,\phi)&\mapsto& \psi\cv^{\lrcorner \lrcorner}_{u,v} \phi:= (\psi\odot_\lrcorner u)\cv(\bar{v}\odot_\lrcorner \phi).\eeq
In addition, the non-associative products between Clifford algebra arbitrary elements in \cite{eu1} are now completely constructed in the present context.  
\medbreak
{\bf Example 3}: Let us calculate $\psi\circ^{\llcorner \llcorner}_{u} \phi$ and $\psi\circ^{\llcorner \lrcorner}_{u} \phi$ for $u=\ee_6\ee_7\ee_1\ee_3\ee_2+2\ee_4\ee_5$, $\bar{u}=-\ee_2\ee_3\ee_1\ee_7\ee_6-2\ee_5\ee_4$, $\psi=\ee_7\ee_3$, and $\phi=\ee_5\ee_4$:
\beq (\psi \odot_\llcorner u)&\cv&(\bar{u}\odot_\llcorner \phi)\n
&=&\left[\ee_7\ee_3\odot_\llcorner(\ee_6\ee_7\ee_1\ee_3\ee_2+2\ee_4\ee_5)\right]\cv\left[(-\ee_2\ee_3\ee_1\ee_7\ee_6-2\ee_5\ee_4)\odot_\llcorner\ee_5\ee_4\right]\n
&=&\left[\ee_7\cv(\ee_3 \bu_\llcorner \ee_6\ee_7\ee_1\ee_3\ee_2+2\ee_4\ee_5)+2\ee_7\cv(\ee_3 \bu_\llcorner \ee_4\ee_5)\right]\cv\n
&&\left[-\ee_2\cv(\ee_3\cv(\ee_1\cv(\ee_7\cv(\ee_6\bu_\llcorner \ee_5\ee_4))))-2\ee_5\cv(\ee_4\bu_\llcorner\ee_5\ee_4)\right]\n
&=&\left[\ee_7\cv((((-\ee_5\cv\ee_7)\cv\ee_1)\cv\ee_3)\cv\ee_2)+2\ee_7\cv(\ee_1\cv\ee_5)\right]\cv\n
&&\left[-\ee_2\cv(\ee_3\cv(\ee_1\cv(\ee_7\cv(-\ee_3\cv \ee_4))))-2\ee_5\cv(\ee_2\cv\ee_4)\right]\n
&=&\left[\ee_7\cv(((\ee_1\cv\ee_1)\cv\ee_3)\cv\ee_2)+2\ee_7\cv\ee_7)\right]\cv\left[-\ee_2\cv(\ee_3\cv(\ee_1\cv(\ee_7\cv(-\ee_1))))-2\ee_5\cv\ee_5)\right]\n
&=&\left[\ee_7\cv(-\ee_3\cv\ee_2)-2\right]\cv\left[-\ee_2\cv(\ee_3\cv(\ee_1\cv(\ee_5)))+2\right]\n
&=&\left[\ee_7\cv(-\ee_7)-2\right]\cv\left[-\ee_2\cv(\ee_3\cv(-\ee_7))+2\right]=\left[1-2\right]\cv\left[-\ee_2\cv(-\ee_2)+2\right]\n
&=&\left[-1\right]\cv\left[-1+2\right]=\left[-1\right]\cv\left[1\right]=-1.\eeq
On the another hand,
\beq (\psi \odot_\llcorner u)\cv(\bar{u}\odot_\lrcorner \phi)&=&\left[\ee_7\ee_3\odot_\llcorner(\ee_6\ee_7\ee_1\ee_3\ee_2+2\ee_4\ee_5)\right]\cv\left[(-\ee_2\ee_3\ee_1\ee_7\ee_6-2\ee_5\ee_4)\odot_\lrcorner\ee_5\ee_4\right]\n
&=&\left[-1\right]\cv\left[(-\ee_2\ee_3\ee_1\ee_7\ee_6\bu_\lrcorner\ee_5)\cv\ee_4-2(\ee_5\ee_4\bu_\lrcorner\ee_5)\cv\ee_4\right]\n
&=&\left[-1\right]\cv\left[(-\ee_2\cv(\ee_3\cv(\ee_1\cv(\ee_7\cv(-\ee_3)))))\cv\ee_4-2(\ee_5\cv\ee_2)\cv\ee_4\right]\n
&=&\left[-1\right]\cv\left[(-\ee_2\cv(\ee_3\cv(\ee_1\cv\ee_2)))\cv\ee_4-2\ee_4\cv\ee_4\right]\n
&=&\left[-1\right]\cv\left[(-\ee_2\cv(\ee_3\cv\ee_6))\cv\ee_4+2\right]=\left[-1\right]\cv\left[(-\ee_2\cv(-\ee_5))\cv\ee_4+2\right]\n
&=&\left[-1\right]\cv\left[-\ee_4\cv\ee_4+2\right]=\left[-1\right]\cv\left[1+2\right]=(-1)\cv(3)=-3.\eeq
\medbreak
\noi This example shows the importance of the direction in the $\odot$-product, because in this case just by changing one direction different results 
are achieved for the same initial data. 
\medbreak
{\bf Remark 3}: The Moufang identities \cite{23, dix} for the octonionic algebra are listed below 
for all $ A,B,C\in\mathbb{O}$:
\beq
(A\cv B\cv A)\cv C = A\cv(B\cv(A\cv C)),\label{mou1}\\
C\cv(A\cv B\cv A) = ((C\cv A)\cv B)\cv A,\label{mou3}\\
(A\cv B)\cv(C\cv A) = A\cv (B\cv C)\cv A.\label{mou0}\eeq\noi 
\medbreak\noi
In the case of the products $\bu_\llcorner\colon\mathbb{O}\times\la^k(\RR^{0,7}) \to \mathbb{O}$ and  $\bu_\lrcorner\colon \la^k(\RR^{0,7})\times \mathbb{O} \to \mathbb{O}$,
the Moufang identities cannot be generalized only using conjugation and graded involution. 
Two counter-examples are presented below.
\medbreak
{\bf Example 4}: One of the Moufang identities for octonions is expressed as 
\bege
(A\cv B\cv A)\cv C = A\cv(B \cv(A\cv C)),\quad A,B,C\in\mathbb{O}.\enge\noi  
Suppose that an immediate generalization could be accomplished, by naively replacing the standard octonion product by the $\bu$-product, as
\bege\label{101}
(u\bu_\lrcorner B\bu_\llcorner u)\bu C = u\bu_\lrcorner(B \bu(u\bu_\lrcorner C)),\quad u\in\cl_{0,7},
\enge\noi or even as $(u\bu B\bu u)\bu C = \widehat{u}\bu(B \bu(u\bu C))$,  $(u\bu B\bu u)\bu C = \bar{u}\bu(B \bu(u\bu C))$, 
or the product above with any composition of graded involution and/or the Clifford conjugation over $u$. 
For an easy understanding of the expressions, Eq.(\ref{101}) is written down as
\bege\label{relga}
(u\bu_\lrcorner B\bu_\llcorner u)\cv C = u\bu_\lrcorner(B \cv(u\bu_\lrcorner C)),\quad u\in\cl_{0,7},
\enge\noi since the $\bu$-product between octonions is identical to the $\cv$-product.
Take $u = \ee_2\ee_7\ee_4$, $B = \ee_1$ and $C = \ee_3$. First, 
\beq (\ee_2\ee_7\ee_4\bu_\lrcorner \ee_1 \bu_\llcorner \ee_2\ee_7\ee_4)\cv \ee_3 &=&((\ee_2\cv(\ee_7\cv\ee_3))\bu_\llcorner \ee_2\ee_7\ee_4)\cv\ee_3=((\ee_2\cv(-\ee_2))\bu_\llcorner \ee_2\ee_7\ee_4)\cv \ee_3\n
&=&((\ee_2\cv\ee_7)\cv\ee_4)\cv\ee_3=(-\ee_3\cv\ee_4)\cv\ee_3 =-\ee_1\cv\ee_3\n
&=&-\ee_4,\eeq\noi while
\beq \ee_2\ee_7\ee_4\bu_\lrcorner (\ee_1\cv (\ee_2\ee_7\ee_4\bu_\lrcorner \ee_3))&=&\ee_2\ee_7\ee_4\bu_\lrcorner(\ee_1\cv(\ee_2\cv(\ee_7\cv(-\ee_1))))\n&=&\ee_2\ee_7\ee_4\bu_\lrcorner(\ee_1\cv(\ee_2\cv(-\ee_5)))\n
=\ee_2\ee_7\ee_4\bu_\lrcorner(\ee_1\cv \ee_4)\n&=&\ee_2\ee_7\ee_4\bu_\lrcorner(-\ee_3) =\ee_2\cv(\ee_7\cv \ee_1) =\ee_2\cv \ee_5 =-\ee_4.\eeq\noi 
On the another hand, by taking
$u = \ee_3\ee_5\ee_7$, $B = \ee_2$ and $C = \ee_1$, it follows that: 
\beq (\ee_3\ee_5\ee_7\bu_\lrcorner \ee_2\bu_\llcorner \ee_3\ee_5\ee_7)\cv \ee_1&=&((\ee_3\cv(\ee_5\cv\ee_3))\bu_\llcorner\ee_3\ee_5\ee_7)\cv\ee_1\n
&=&(\ee_5\bu_\llcorner\ee_3\ee_5\ee_7)\cv\ee_1=(\ee_3\cv\ee_7)\cv\ee_1 =\ee_2\cv\ee_1\n 
&=&-\ee_6,\eeq\noi while 
\beq \ee_3\ee_5\ee_7\bu_\lrcorner (\ee_2\cv (\ee_3\ee_5\ee_7\bu_\lrcorner\ee_1))&=&\ee_3\ee_5\ee_7\bu_\lrcorner(\ee_2\cv(\ee_3\cv(\ee_5\cv\ee_5)))=\ee_3\ee_5\ee_7\bu_\lrcorner(-\ee_7)\n
&=& \ee_3\cv \ee_5 \ee_6.\eeq\noi 
It can be seen that for distinct elements representing $u\in\la^3(\RR^{0,7})$, 
both the relations $(u\bu_\lrcorner B\bu_\llcorner u)\cv C = u\bu_\lrcorner(B \cv(u\bu_\lrcorner C))$ and $(u\bu_\lrcorner B\bu_\llcorner u)\cv C = -u\bu_\lrcorner(B \cv(u\bu_\lrcorner C))$ are obtained.
These last two relations cannot be mutually satisfied for elements with the same degree in $\cl_{0,7}$, the same for the product given by Eq.(\ref{101}) with any composition of the graded involution and/or the Clifford conjugation on $u$. 
Using the same counter-example above it can be shown that the other Moufang identities can not be generalized using Clifford conjugation and graded involution only.
\medbreak
{\bf Example 5}: The Moufang identities could not be generalized for the products $\odot_\llcorner$ and $\odot_\lrcorner$ defined in Eqs.(\ref{219},\ref{221}) respectively. Indeed, computing the product $(u\odot_\llcorner B\odot_\llcorner u)\odot_\llcorner C$ and $u\odot_\llcorner(B\odot_\llcorner(u\odot_\llcorner C))$ for $u=\ee_1\ee_3\ee_7$, $B=\ee_2\ee_5\ee_6$ and $C=\ee_3\ee_5\ee_7$ it follows that:
\beq
(\ee_1\ee_3\ee_7\odot_\llcorner\ee_2\ee_5\ee_6\odot_\llcorner\ee_1\ee_3\ee_7)\odot_\llcorner\ee_3\ee_5\ee_7&=&(\ee_1\cv(\ee_3\cv(\ee_7\bu_\llcorner\ee_2\ee_5\ee_6))\odot_\llcorner\ee_1\ee_3\ee_7)\odot_\llcorner\ee_3\ee_5\ee_7\n
&=&(\ee_1\cv(\ee_3\cv((\ee_3\cv\ee_5)\cv\ee_6))\odot_\llcorner\ee_1\ee_3\ee_7)\odot_\llcorner\ee_3\ee_5\ee_7\n
&=&(\ee_1\cv(\ee_3\cv(\ee_6\cv\ee_6))\odot_\llcorner\ee_1\ee_3\ee_7)\odot_\llcorner\ee_3\ee_5\ee_7\n
&=&(\ee_1\cv(-\ee_3)\odot_\llcorner\ee_1\ee_3\ee_7)\odot_\llcorner\ee_3\ee_5\ee_7\n
&=&(-\ee_4\bu_\llcorner\ee_1\ee_3\ee_7)\odot_\llcorner\ee_3\ee_5\ee_7\n&=&((-\ee_3\cv\ee_3)\cv\ee_7)\odot_\llcorner\ee_3\ee_5\ee_7\n
&=&\ee_7\odot_\llcorner\ee_3\ee_5\ee_7=\ee_7\bu_\llcorner\ee_3\ee_5\ee_7\n&=&(-\ee_2\cv\ee_5)\cv\ee_7=\ee_4\cv\ee_7\n
&=&-\ee_6\nonumber\eeq
On the another hand we have:
\beq
\ee_1\ee_3\ee_7\odot_\llcorner(\ee_2\ee_5\ee_6\odot_\llcorner(\ee_1\ee_3\ee_7\odot_\llcorner\ee_3\ee_5\ee_7))&=&\ee_1\ee_3\ee_7\odot_\llcorner(\ee_2\ee_5\ee_6\odot_\llcorner(\ee_1\cv(\ee_3\cv(\ee_7\bu_\llcorner\ee_3\ee_5\ee_7))))\n
&=&\ee_1\ee_3\ee_7\odot_\llcorner(\ee_2\ee_5\ee_6\odot_\llcorner(\ee_1\cv(\ee_3\cv((-\ee_2\cv\ee_5)\cv\ee_7))))\n
&=&\ee_1\ee_3\ee_7\odot_\llcorner(\ee_2\ee_5\ee_6\odot_\llcorner(\ee_1\cv(\ee_3\cv(\ee_4\cv\ee_7))))\n
&=&\ee_1\ee_3\ee_7\odot_\llcorner(\ee_2\ee_5\ee_6\odot_\llcorner(\ee_1\cv(\ee_3\cv(-\ee_6))))\n
&=&\ee_1\ee_3\ee_7\odot_\llcorner(\ee_2\ee_5\ee_6\odot_\llcorner(\ee_1\cv\ee_5))\n
&=&\ee_1\ee_3\ee_7\odot_\llcorner(\ee_2\cv(\ee_5\cv(\ee_6\cv\ee_7)))\n
&=&\ee_1\ee_3\ee_7\odot_\llcorner(\ee_2\cv(-\ee_2))=\ee_1\ee_3\ee_7\odot_\llcorner 1\n&=&\ee_1\ee_3\ee_7\bu_\lrcorner 1\n
&=&\ee_1\cv(\ee_3\cv(\ee_7\cv 1))=\ee_1\cv(\ee_3\cv\ee_7)=\ee_1\cv\ee_2\n
&=&\ee_6\nonumber\eeq
\medbreak 
Analogously, another counter-example with the product $\odot_\lrcorner$ can be presented to show that the Moufang identities given by Eqs.(\ref{mou1}, \ref{mou3}, \ref{mou0}) are not generalized in this context.
 
\section{$\mathbb{O}$-Scalar Product and $S^7$ Tangent Bundle Basis}
For $A, B \in \mathbb{O}$ the scalar product between two octonions is defined as \cite{roo,dix,loun,baez}:
\beq \left\langle A, B\right\rangle=\frac{1}{2}(\bar{A}\cv B+\bar{B}\cv A)=\frac{1}{2}(A\cv\bar{B}+B\cv\bar{A}).\nonumber\eeq
Considering an arbitrary unit octonion $X=X^0+X^a\ee_a$,  its squared norm is given by
$\|X\|^2 = \left\langle X, X\right\rangle=(X^0)^2+(X^1)^2+(X^2)^2+(X^3)^2+(X^4)^2+(X^5)^2+(X^6)^2+(X^7)^2=1$,
and the elements $\{\ee_a\cv X\}$ constitute  the frame bundle on $S^7$, obtained by the left multiplication of $X\in S^7$ by an octonion basis $\{\ee_a\}$. They satisfy the relations
\beq\left\langle \ee_a\cv X, \ee_b\cv X\right\rangle&=&\delta_{ab},\qquad\qquad\left\langle \ee_a\cv X, X\right\rangle = 0,\label{103}\\
\left\langle \ee_a\ee_b\bu_\lrcorner X, X\right\rangle&=&0.\label{104}\eeq
Concerning the octonions $\ee_a\cv X$, Eqs.(\ref{103}) tell us that these vectors are orthogonal to each other and more, these vectors lie in the  tangent space at the arbitrary point $X\in S^7$, where they form an orthonormal basis. In the Appendix \ref{matrix rep} the $\{e_a\cv X\}$ representations are explicitly constructed. For more details see \cite{roo, dix}.

Note that the product $\ee_a\ee_b\bu_\lrcorner X$ also presents an associated matrix representation --- in the \ref{matrix rep} all are constructed, and some of them were first listed in \cite{dix} ---  it is possible to prove that $\forall a,b=1,\ldots,7$, $a \neq b$,
\beq\left\langle \ee_a\ee_b\bu_\lrcorner X, X\right\rangle = \left\langle \ee_a\cv (\ee_b\cv X), X\right\rangle =\left\langle \ee_b\cv X, \bar{\ee_a} \cv X\right\rangle = - \left\langle \ee_b\cv X, \ee_a\cv X \right\rangle = 0,\eeq\noi 
where Eq.(\ref{103}) is used in the last one, i.e., $\ee_a\cv X$ is a basis for the tangent space in an arbitrary point on $S^7$. It is clear to see 
such property, in particular when the properties in the Appendix \ref{proof of} are taken into account.
 
Eqs.(\ref{103}) and (\ref{104}) can be also easily demonstrated by computing all the cases for all $a,b=1,\ldots,7$, $a \neq b$, 
which shows that 
Eq.(\ref{103}) relates a set of elements orthogonal to $X\in S^7$ for $a\neq b$, and then a basis for the tangent space is obtained. 

In order to illustrate, in \ref{proof of} particular cases of Eqs.(\ref{103}) are explicitly demonstrated.
Also in general for $a\neq b\neq c$, 
\beq\left\langle \ee_a\ee_b\ee_c\bu_\lrcorner X, X\right\rangle\neq 0.\label{1114}\eeq
Indeed, concerning the element $\ee_1\ee_2\ee_3$, it reads
\beq \left\langle \ee_1\ee_2\ee_3\bu_\lrcorner X, X\right\rangle &=& 1/2[(\overline{\ee_1\ee_2\ee_3\bu_\lrcorner X})\cv X+\bar{X}\cv (\ee_1\ee_2\ee_3\bu_\lrcorner X)]\n
&=& 1/2[(+X^0\ee_5+X^1\ee_7-X^2\ee_4+X^3\ee_6-X^4\ee_2-X^5+X^6\ee_3+X^7\ee_1)\cv \n
& &(+X^0+X^1\ee_1+X^2\ee_2+X^3\ee_3+X^4\ee_4+X^5\ee_5+X^6\ee_6+X^7\ee_7)+\n
& &(+X^0-X^1\ee_1-X^2\ee_2-X^3\ee_3-X^4\ee_4-X^5\ee_5-X^6\ee_6-X^7\ee_7)\cv\n
& &(-X^0\ee_5-X^1\ee_7+X^2\ee_4-X^3\ee_6+X^4\ee_2-X^5-X^6\ee_3-X^7\ee_1)]\n
&=&2(-X^0X^5-X^1X^7+X^2X^4-X^3X^6).\nonumber\eeq

\section{Properties of the $\bu$-product}
\label{add}
Some additional properties concerning the $\bu$-product are asserted and proved for $u\in\cl_{0,7}$ homogeneous and simple, the only cases we can guarantee that there exists an inverse $u^{-1}$, in the light of the points presented in Remark 1.
Such properties are helpful in case we want to obtain the most general expression that generalizes Eq.(\ref{3p14}) 
and emulates it when $u\in\cl_{0,7}$ is considered instead of $X\in\RR\oplus\RR^{0,7}$.

\medbreak
{\bf Proposition 1}: \emph{Given $\ee_a\in \mathbb{O}$ and $u=\ee_{i_1}\ee_{i_2}\ldots \ee_{i_k}\in\la^k(\RR^{0,7})$, $k=1,\ldots,6$ with either $\ee_a\notin\{\ee_{i_1}, \ee_{i_2},\ldots, \ee_{i_k}\}$ or $\{\ee_a, \ee_{i_{j_1}}, \ee_{i_{j_2}}\}$ not a $\HH$-triple, for $\{i_{j_1}, i_{j_2}\}=\{i_1,\ldots,i_6\}$, then
\beq(\ee_a\bu_\llcorner u)\cv \overline{(1\bu_\llcorner \tilde{u})} = (-1)^{(|u|-1)(|u|-2)/2}\ee_a\eeq}
\medbreak
{\bf Proposition 2}: \emph{Given $\ee_a\in \mathbb{O}$ and $u=\ee_{i_1}\ee_{i_2}\ldots \ee_{i_k}\in\la^k(\RR^{0,7})$, $k=1,\ldots,6$  then 
\beq(\bar{u}\bu_\lrcorner \ee_a)\circ(1\bu_\llcorner \tilde{u})=(-1)^{|u|(|u|+1)/2}\ee_a\eeq}
\medbreak
{\bf Observation 1}:
Propositions 1 and 2 when $\ee_a=\ee_0=1$ can be asserted as 
\medbreak {\bf Proposition 1$^\prime$}: \emph{Given $u=\ee_{i_1}\ee_{i_2}\ldots \ee_{i_k}\in\la^k(\RR^{0,7})$, $k=1,\ldots,6$ then
\beq(1\bu_\llcorner u)\cv \overline{(1\bu_\llcorner \tilde{u})} = (-1)^{|u|(r_{|u|} - r_{|u|-1}) + (8-|u|)(7-|u|)/2}
\eeq
\noi where $r_j$ are the Radon-Hurwitz numbers defined by the following table 
\begin{table}
\caption{\label{table2}Radon-Hurwitz numbers.}
\begin{center}
\begin{tabular}{lllllllll}
\br
$j$&0&1&2&3&4&5&6&7\\
$r_j$&0&1&2&2&3&3&3&3\\
\br
\end{tabular}
\end{center}
\end{table}
\noi and the recurrence relation $r_{j+8} = r_j + 4$.}
\medbreak
{\bf Proposition 2$^\prime$}: \emph{Given $u=\ee_{i_1}\ee_{i_2}\ldots \ee_{i_k}\in\la^k(\RR^{0,7})$, $k=1,\ldots,6$  then 
\beq(\bar{u}\bu_\lrcorner 1)\circ(1\bu_\llcorner \tilde{u})=(-1)^{|u|(r_{|u|} - r_{|u|-1}) + (8-|u|)(7-|u|)/2}\eeq}
\medbreak
{\bf Proposition 3}: \emph{Given $\ee_a\in \mathbb{O}$ and $u=\ee_{i_1}\ee_{i_2}\ldots \ee_{i_k}\in\la^k(\RR^{0,7})$, $k=1,\ldots,6$ with either $\ee_a\notin\{\ee_{i_1}, \ee_{i_2},\ldots, \ee_{i_k}\}$ or $\{\ee_a, \ee_{i_{j_1}}, \ee_{i_{j_2}}\}$ not a $\HH$-triple, for $\{i_{j_1}, i_{j_2}\}=\{i_1,\ldots,i_6\}$, then
\beq\ee_a\bu_\llcorner u=(-1)^{|u|+1}\ee_a\cv(1\bu_\llcorner u)\eeq}
\medbreak
{\bf Proposition 4}: \emph{Given $\psi\in\cl_{0,7}$ and $u\in \la^k(\RR^{0,7})$, $k=1,\ldots,6$ then 
\beq\psi\odot_\llcorner u=(-1)^{|u|+1}\psi\bu_\lrcorner(1\bu_\llcorner u)\eeq\noi 
In all the cases above $|u|$ denotes the degree of $u$: if $u\in\la^k(\RR^{0,7})$, then $|u|$ = $k$.} 
\medbreak
Note that the Proposition 4 is the generalization of the Proposition 3. The proofs for all the Propositions are given by making computations for all cases that can be completely checked in Appendices A, B, C, D, E, and F, for the Propositions 1, $1^\prime$, 2, $2^\prime$, 3, and 4, respectively.
\medbreak
 {\bf Observation 2}: 
Proposition 1 elicits an $u$-induced involution on $\mathbb{O}$. By denoting it by 
\beq
\dashv_u (\ee_a):=(\ee_a\bu_\llcorner u)\cv \overline{(1\bu_\llcorner \tilde{u})}\label{dashv}\eeq\noi 
it is immediate to see that $\dashv_u \dashv_u (\ee_a) = (-1)^{(|u|-1)(|u|-2)}\ee_a = \ee_a$.

Also, Proposition 2 provides another $u$-induced involution on $\mathbb{O}$. By denoting it by 
\beq
\vdash_u (\ee_a):= (\bar{u}\bu_\lrcorner \ee_a)\circ(1\bu_\llcorner \tilde{u})\label{vdash}\eeq\noi 
it is immediate to see that $\vdash_u \vdash_u (\ee_a) = (-1)^{|u|(|u|+1)}\ee_a = \ee_a$.

Those two involutions bring some new light on the generalization of Eq.(\ref{3p14}) in the context of the $\bu$-product.
\section{Generalized Non-Associative Products}
In this Section a more general class of non-associative products is introduced, presenting similarity to the $u$-product between Clifford multivectors, but constructed without an additional element. The difference is that the product is computed without the insertion of any additional element $u$, but now a vector $u_j$ is taken out the Clifford multivector itself, instead. This $\mathbb{O}$-valued map is called the non-associative shear of a Clifford multivector, and it cogently describes how to split
an arbitrary multivector in a non-associative manner into an octonion.

Given $\psi = u_1u_2\ldots u_k\in\Lambda^k(\RR^{0,7})$ and $\phi = v_1v_2\ldots v_j\in\Lambda^j(\RR^{0,7})$, where $u_{i_j}\in \RR^{0,7}$ and  $v_{i_k}\in\RR^{0,7}$, we define the $\rhd$- and $\lhd$-product that merge the $u$- and the $\bu$-product, as: 
\beq 
\rhd: \cl_{0,7}\times\cl_{0,7} &\to& \mathbb{O}\n
(\psi,\phi)&\mapsto& \psi \rhd \phi:= ((u_1u_2\ldots u_{k-1})\bu_\lrcorner u_k)\cv (\bar{u_k}\bu_\llcorner (v_1v_2\ldots v_j))\\ 
\lhd: \cl_{0,7}\times\cl_{0,7} &\to& \mathbb{O}\n
(\psi,\phi)&\mapsto& \psi \lhd \phi:= ((u_1u_2\ldots u_{k})\bu_\lrcorner v_1)\cv (\bar{v_1}\bu_\llcorner (v_2v_3\ldots v_j))\eeq 
\medbreak
{\bf Example 6}: Let $\psi=\ee_1\ee_2\ee_3$ and $\phi=\ee_4\ee_5$, and calculate the above products for this case. On the one hand, 
\beq \ee_{1}\ee_2\ee_3 \rhd \ee_4\ee_5 &=& (\ee_1\ee_2\bu_\lrcorner\ee_3)\cv(\bar{\ee_3}\bu_\llcorner\ee_4\ee_5)=(\ee_1\ee_2\bu_\lrcorner\ee_3)\cv(-\ee_3\bu_\llcorner\ee_4\ee_5)\n
&=&(\ee_1\cv\ee_7)\cv(-\ee_1\cv\ee_5)=-\ee_5\cv(-\ee_7)=\ee_1.\nonumber\eeq
In the other hand,
\beq \ee_{1}\ee_2\ee_3 \lhd \ee_4\ee_5 &=& (\ee_1\ee_2\ee_3\bu_\lrcorner\ee_4)\cv(\bar{\ee_4}\cv\ee_5)=(\ee_1\ee_2\ee_3\bu_\lrcorner\ee_4)\cv(-\ee_4\cv\ee_5)\n
&=&(\ee_1\cv(\ee_2\cv\ee_1))\cv(-\ee_2)=(\ee_1\cv(-\ee_6))\cv(-\ee_2)=\ee_2\cv(-\ee_2)=1.\nonumber\eeq
\medbreak\noi Those products can be expressed using the $\cv^{\lrcorner\llcorner}_{u}$ product as 
\beq
\psi \rhd \phi = (\psi u_k^{-1}) \cv^{\lrcorner\llcorner}_{u_k} \phi,\qquad\quad \psi \lhd \phi = \psi \cv^{\lrcorner\llcorner}_{v_1} (v_1^{-1}\phi).
\eeq

\section{Concluding remarks and outlook}

As Hopf fibrations can be accomplished by the use of the $X$-product \cite{dix}, a natural task should be to ascertain about the 
geometric meaning --- in the context of Hopf fibrations --- of particular products of the type $\ee_a \cv_u \ee_b$.
The formalism presented here may make --- for instance --- the Hopf fibration $S^3\ldots S^7\rightarrow S^4$ and the parallelizable torsion on $S^7$ to arise from the immediate deformation of the octonionic product.
The parameter $A\circ_X B = (A\circ X)\circ(\bar{X}\circ B) = \bar{X}\circ((X\circ A)\circ B)$ is twice the parallelizing torsion \cite{24,roo} which components are given by $T_{ijk}(X)=[(\bar{\ee_i}\circ\bar{X})\circ(X\circ {\ee_j})]\circ \ee_k$, which is exactly the $\bar{X}$-product between 
$\bar{\ee_i}$ and ${\ee_j}$, 
and the $S^7$ algebra can be written as $[\delta_i,\delta_j]=2T_{ijk}(X)\delta_k$, where $\delta_AX = X\cv A$, and the variation $\delta$ is indeed the parallelizing covariant derivative \cite{ced,dix,beng,brink,roo}. 
The question whether the analogous of the above structure of the form
$(A\cv(B\bu_\llcorner u))\bu_\llcorner \bar{u}$, for $u\in\cl_{0,7}$, $A, B\in \mathbb{O}$
also may allow for non-trivial central extensions \cite{schwinger} remains open.
Also, it is well know the classical version of the $S^7$ Kac-Moody algebra, and  the question
concerning whether a generalization of the $X$-product as accomplished along this paper could 
ascertain the validity of immediate generalizations of some expressions 
%$\wickcontract{{\mathfrak J}_A(z)}{{\mathfrak J}_B(\zeta)}={\frac{1}{z-\zeta}}{\mathfrak J}_{[A, B]_X}$ 
in \cite{ced}. 
We can ask what is the meaning
of the $u$-commutator $[A,B]_u = A\cv_u B - B\cv_u A$ instead of the usual term $[A,B]_X = A\cv_X B - B\cv_X A$.

The innate difficulties in computing non-associative products are circumvented, when these products can be incorporated 
in the multivector Clifford algebra structure. This is a strong property reminiscent of the assumption
that there is defined an extension of the octonionic product in order to encompass also non-associative products between octonions and Clifford multivectors, and between Clifford multivectors themselves. By means of the $\odot$-product, all the arbitrary number of octonionic subsequent products
are regarded as the $\odot$-product involving the Clifford multivector associated with the subsequent octonionic product as defined in Eqs.(\ref{209},\ref{210},\ref{219}). The arbitrary number of octonionic products can be encoded in an unique product --- the $\odot$-product ---
and their associated Clifford multivector structure.

When we deal with the homogeneous simple multivectors in $\cl_{0,7}$, the non-associative structure related to the subsequent $\mathbb{O}$-products
among the $\mathbb{O}$-units can be regarded in the anti-commutative structure in the underlying exterior algebra $\Lambda(\RR^{0,7})\hko \cl_{0,7}$.
It is not a quite straightforward task to consider the reversed non-associative products. For instance, given $\alpha_0 = \frac{1}{4}(-3+\ee_4\ee_7\ee_6+\ee_5\ee_1\ee_7+\ee_6\ee_2\ee_1+\ee_7\ee_3\ee_2+\ee_1\ee_4\ee_3+\ee_2\ee_5\ee_4+\ee_3\ee_6\ee_5)$ \cite{dix}, 
it is possible to show that 
\beq -(\ee_a\cv \ee_b)\cv X=-(\alpha_0 \bu_\lrcorner(X\cv{\ee_b}))\cv \ee_a + (\alpha_0 \bu_\lrcorner(X\cv\bar{\ee_a}))\cv \bar{\ee_b} -(X\cv {\ee_b})\cv \ee_a,\;\; \forall X\in\mathbb{O}.
\eeq\noi The left actions introduced, in e.g.\cite{dix}, can be completely described in the $\bu$-product formalism, since it explicitly 
shows the manifest exterior algebra character of the subsequent left action. We completely formalize the framework introduced in \cite{dix,ced}
in a robust platform provided by the Clifford bundle on $S^7$.
The possibility of performing non-associative products between arbitrary multivectors of $\cl_{0,7}$ naturally arises in our formalism that completes 
\cite{eu1}, and it also generalizes the formalism introduced in \cite{dix}, concerning the original $X$-product.

The authors in \cite{eu1} introduced the octonionic algebra in the Clifford algebra arena and dealt with non-associative products, firstly introduced by Dixon \cite{dix} with the $X$- and $XY$-products between octonions and also in \cite{eu1}  the $u$-, $\bu$-,
and $\odot$-products not only but also with an octonion and a Clifford multivector, and between  Clifford multivectors as well. In this paper  those products are generalized with respect to the direction, where more possibilities to make the product appear and a full list of them is presented. Moreover, four important Propositions with respect to the $\bu$-product are shown, and a generalization for the non-associative products is provided, using both the $u$- and $\bu$-product. Such propositions may be germane for the extension of Eq.(\ref{3p14}) in order to encompass an arbitrary $u\in\cl_{0,7}$ instead of $X\in\RR\oplus\RR^{0,7}$. 
Although the Example 1 asserts that such a naive substitution of $X\in\mathbb{O}$ by $u\in\cl_{0,7}$ does not hold in general, we conjecture that the most general expression holding in this case must be of the form 
\[(A\bu_\llcorner u)\cv (\bar{u}\bu_\lrcorner B) = [\star_{u}^1({A})\cv ({\star_{u}^2}({B})\bu_\llcorner u)]\cv \overline{(1\bullet_\lrcorner \tilde{u})},\]
where $\star_{u}^1$ and $\star_{u}^2$  are  $u$-induced involutions on $\mathbb{O}$, distinct from the $\mathbb{O}$-conjugation. In addition, 
$f(u)$ is some $\mathbb{O}$-valued function involving $u\in \cl_{0,7}$ and eventually some of the generalized non-associative products
defined heretofore.

Finally, the non-associative shear was introduced as a map that takes into account the splitting
of a multivector into an octonion, subsequently choosing its components to perform non-associative products
among themselves, following a variety of possibilities as illustrated the tables througuout the text.

We emphasize that conformal field theory and the Kac-Moody algebra, e.g. in \cite{ced} are not concerned and are very far beyond the scope
of the present paper, since the main aim here was to introduce and investigate a more general class  of non-associative products on $S^7$.
Also, the framework here introduced is a promising tool for considering its interplay with some applications, as 
introduced in \cite{loun,pi5SU3,Wiedersehen, eld, edld1,gromollmeyer,shes,joshi,lassig}.

One more comment is worthwhile. As triality stabilizes the Lie algebra $\mathfrak{g}_2$  --- the derivations of the octonions algebra --- pointwise, the formalism presented 
in this paper can bring some new light on some deformations related to the exceptional group $G_2$. Some topological consequences can be addressed as, for example the cubic roots of the unit which describe two latitudes of the sphere $S^6$. 
In addition, the $\bu$-product introduces deformations in $\mk{g}_2$, and an open and fundamental question remains,  regarding the construction of the derivation algebra of the octonions, but now the octonionic products are the extended non-associative $u$-, $(1, u)$-, and $(u, v)$-products.
For the $X$-product particular case, such a question was completely posed and solved in \cite{dix}, for the deformations induced by the paravectors in $\Lambda^{0}(\RR^{0,7})\oplus \Lambda^{1}(\RR^{0,7})$, but the general case still claims for a solution. In addition, 
the (Leech) lattices leaving the $X$-product equal to the original octonionic product, and their interrelations to the exceptional Lie groups can be investigated using this formalism.
Finally, the formalism presented here can reveal and bring some new light on the $S^7$ Hopf bundle of Yang-Mills fields over 
compactified $\RR^4$. As the exotic versions are no longer principal SU(2) bundles, but rather associated bundles with group SO(4), the formalism
of the $\bu$- and $\odot$-products introduced may provide some understanding of the differential geometric 
restrictions inherent to exotic structures, and how to possibly extended it to a more general formalism.
The formalism introduced here originates more general non-associative structures on $S^7$, which arise from 
the products already introduced in \cite{eu1}.  Further applications may concern some aspects of \cite{top,24,pi5SU3,Wiedersehen,gromollmeyer,exoticGR,11a,9a, 10,okubo,smolin,12}, which are beyond the scope of this paper.

\ack
R. da Rocha is grateful to CNPq 472903/2008-0 and 304862/2009-6 for financial support, and to  Funda\c c\~ao de Amparo à Pesquisa do Estado de S\~ao Paulo (FAPESP) 2011/08710-1 that supported financially my attendance to QTS7.

\appendix
\section{Proposition 1}
In the next six Appendices the Propositions aforementioned in Section \ref{add}, and  their respective demonstrations, are provided in details. Special attention for the case where $u\in\la^7(\RR^{0,7})$ is the volume element, which is not taken into account, since the elements not having common factors with $\ee_a$ must be taken, and in this case $u\in\la^7(\RR^{0,7})$ is led back  to the case where $u\in\la^6(\RR^{0,7})$. The volume element must have an $\ee_a$ term that obviously commute with the $\ee_a$ in the Proposition statement.
\medbreak
{\bf Proposition 1}: \emph{Given $\ee_a\in \mathbb{O}$ and $u=\ee_{i_1}\ee_{i_2}\ldots \ee_{i_k}\in\la^k(\RR^{0,7})$, $k=1,\ldots,6$ with either $\ee_a\notin\{\ee_{i_1}, \ee_{i_2},\ldots, \ee_{i_k}\}$ or $\{\ee_a, \ee_{i_{j_1}}, \ee_{i_{j_2}}\}$ not a $\HH$-triple, for $\{i_{j_1}, i_{j_2}\}=\{i_1,\ldots,i_6\}$, then
\beq(\ee_a\bu_\llcorner u)\cv \overline{(1\bu_\llcorner \tilde{u})} = (-1)^{(|u|-1)(|u|-2)/2}\ee_a\eeq}
\medbreak
{\bf Proof}: Hereon the acronym MI in parenthesis denotes  the Moufang identity given by Eq.(\ref{mou0}).
\begin{enumerate}
\item[0)] For $u \in \la^0(\RR^{0,7})$:
\beq (\ee_a\bu_\llcorner u)\cv\overline{(1\bu_\llcorner \tilde{u})}&=&(\ee_a\bu_\llcorner \ee_0)\cv \overline{(1\bu_\llcorner \tilde{\ee_0})}=(\ee_a\cv \ee_0)\cv \overline{(1\cv \ee_0)}=(\ee_a\cv \ee_0)\cv \ee_0=\ee_a\nonumber\eeq
\item[1)] For $u \in \la^1(\RR^{0,7})$:
\begin{enumerate}
\item[(a)] $a\neq b$:
\beq(\ee_a\bu_\llcorner u)\cv\overline{(1\bu_\llcorner \tilde{u})}&=&(\ee_a\bu_\llcorner \ee_b)\cv \overline{(1\bu_\llcorner \tilde{\ee_b})}=(\ee_a\cv \ee_b)\cv \overline{(1\cv \ee_b)}\nonumber\\&=&(\ee_b\cv \ee_a)\cv \ee_b=\ee_a\nonumber\eeq
\item[(b)] $a=b$:
\beq(\ee_a\bu_\llcorner u)\cv\overline{(1\bu_\llcorner \tilde{u})}&=&(\ee_a\bu_\llcorner \ee_b)\cv \overline{(1\bu_\llcorner \tilde{\ee_b})}=(\ee_a\bu_\llcorner \ee_a)\cv \overline{(1\bu_\llcorner \tilde{\ee_a})}\n&=&(\ee_a\cv \ee_a)\cv \overline{(1\cv \ee_a)}\n
&=&-(\ee_a\cv \ee_a)\cv \ee_a=\ee_a \nonumber\eeq
\end{enumerate}
\item[2)] For $u \in \la^2(\RR^{0,7})$:
\begin{enumerate}
\item[(a)] $a\notin \{b, c\}$, and ($abc$) is not a $\HH$-triple:
\beq(\ee_a\bu_\llcorner u)\cv\overline{(1\bu_\llcorner \tilde{u})}&=&(\ee_a\bu_\llcorner \ee_{bc})\cv \overline{(1\bu_\llcorner \widetilde{\ee_{bc}})}=(\ee_a\bu_\llcorner \ee_{bc})\cv \overline{(1\bu_\llcorner (-\ee_{bc}))}\n &=&(\ee_a\bu_\llcorner \ee_{bc})\cv (1\bu_\llcorner\ee_{bc}) = ((\ee_a\cv \ee_b)\cv\ee_c)\cv (\ee_b\cv\ee_c)\n &=&-(\ee_c\cv(\ee_a\cv \ee_b))\cv (\ee_b\cv\ee_c)\overset{MI}{=}\n &&-(\ee_a\cv \ee_b)\cv \ee_b {=}\ee_a\nonumber\eeq
\item[(b)] $a=b$, or $a=c$, or ($abc$) is a $\HH$-triple. Without loss of generality,  consider $a=b$:
\beq(\ee_a\bu_\llcorner u)\cv\overline{(1\bu_\llcorner \tilde{u})}&=&(\ee_a\bu_\llcorner \ee_{bc})\cv \overline{(1\bu_\llcorner \widetilde{\ee_{bc}})}=(\ee_a\bu_\llcorner \ee_{ac})\cv \overline{(1\bu_\llcorner \widetilde{\ee_{ac}})}\n &=&(\ee_a\bu_\llcorner \ee_{ac})\cv \overline{(1\bu_\llcorner (-\ee_{ac}))} =(\ee_a\bu_\llcorner \ee_{ac})\cv (1\bu_\llcorner\ee_{ac})\n &=&((\ee_a\cv \ee_a)\cv\ee_c)\cv (\ee_a\cv\ee_c)\n&=&-(\ee_c\cv(\ee_a\cv \ee_a))\cv (\ee_a\cv\ee_c)\n
&\overset{MI}{=}&-(\ee_a\cv \ee_a)\cv \ee_a = \ee_a\nonumber\eeq
\end{enumerate}
\item[3)] For $u \in \la^3(\RR^{0,7})$:
\begin{enumerate}
\item[(a)] $a\notin \{b, c\}$, and ($abc$) is not a $\HH$-triple:
\beq(\ee_a\bu_\llcorner u)\cv\overline{(1\bu_\llcorner \tilde{u})}&=&(\ee_a\bu_\llcorner \ee_{bcd})\cv \overline{(1\bu_\llcorner \widetilde{\ee_{bcd}})}\n&=&(\ee_a\bu_\llcorner \ee_{bcd})\cv \overline{(1\bu_\llcorner (-\ee_{bcd}))}\n
&=&(\ee_a\bu_\llcorner \ee_{bcd})\cv (1\bu_\llcorner \ee_{bcd})\n&=&(((\ee_a\cv \ee_b)\cv \ee_c)\cv \ee_d)\cv((\ee_b \cv \ee_c)\cv \ee_d)\n
&=&-(\ee_d\cv((\ee_a\cv \ee_b)\cv \ee_c))\cv ((\ee_b \cv \ee_c)\cv \ee_d)\n&\overset{MI}{=}&-((\ee_a\cv \ee_b)\cv \ee_c)\cv (\ee_b \cv \ee_c) {=}-\ee_a\nonumber\eeq
\item[(b)] $a=b$ or $a=c$ or ($abc$) is a $\HH$-triple. Without loss of generality,  consider $a=b$:
\beq(\ee_a\bu_\llcorner u)\cv\overline{(1\bu_\llcorner \tilde{u})}&=&(\ee_a\bu_\llcorner \ee_{bcd})\cv \overline{(1\bu_\llcorner \widetilde{\ee_{bcd}})}\n&=&(\ee_a\bu_\llcorner \ee_{acd})\cv \overline{(1\bu_\llcorner \widetilde{\ee_{acd}})}\n&=&(\ee_a\bu_\llcorner \ee_{acd})\cv \overline{(1\bu_\llcorner (-\ee_{acd}))}\n
&=&(\ee_a\bu_\llcorner \ee_{acd})\cv (1\bu_\llcorner \ee_{acd})\n&=&(((\ee_a\cv \ee_a)\cv \ee_c)\cv \ee_d)\cv((\ee_a \cv \ee_c)\cv \ee_d)\n
&=&-(\ee_d((\ee_a\cv \ee_a)\cv \ee_c))\cv ((\ee_a \cv \ee_c)\cv \ee_d)\n&\overset{MI}{=}&-((\ee_a\cv \ee_a)\cv \ee_c)\cv (\ee_a \cv \ee_c)\overset{(2)(b)}{=}-\ee_a\nonumber\eeq
\end{enumerate}
\item[4)] For $u \in \la^4(\RR^{0,7})$:
\begin{enumerate}
\item[a)] $a\notin \{b, c\}$, and ($abc$) is not a $\HH$-triple:
\beq(\ee_a\bu_\llcorner u)\cv\overline{(1\bu_\llcorner \tilde{u})}&=&(\ee_a\bu_\llcorner \ee_{bcdf})\cv \overline{(1\bu_\llcorner \widetilde{\ee_{bcdf}})}\n&=&(\ee_a\bu_\llcorner \ee_{bcdf})\cv \overline{(1\bu_\llcorner \ee_{bcdf})}\n
&=&-((((\ee_a\cv \ee_b)\cv \ee_c)\cv \ee_d)\cv \ee_f)\cv (((\ee_b\cv \ee_c)\cv \ee_d)\cv \ee_f)\n
&=&(\ee_f\cv(((\ee_a\cv \ee_b)\cv \ee_c)\cv \ee_d))\cv (((\ee_b\cv \ee_c)\cv \ee_d)\cv \ee_f)\n
&\overset{MI}{=}&(((\ee_a\cv \ee_b)\cv \ee_c)\cv \ee_d)\cv ((\ee_b\cv \ee_c)\cv \ee_d)\overset{(3)(a)}{=}-\ee_a\n
\nonumber\eeq
\item[(b)] $a=b$ or $a=c$ or ($abc$) is a $\HH$-triple. Without loss of generality,  consider $a=b$:
\beq\hspace*{-0.2cm}(\ee_a\bu_\llcorner u)\cv\overline{(1\bu_\llcorner \tilde{u})}&=&(\ee_a\bu_\llcorner \ee_{bcdf})\cv \overline{(1\bu_\llcorner \widetilde{\ee_{bcdf}})}\n&=&(\ee_a\bu_\llcorner \ee_{acdf})\cv \overline{(1\bu_\llcorner \widetilde{\ee_{acdf}})}\n
&=&-((((\ee_a\cv \ee_a)\cv \ee_c)\cv \ee_d)\cv \ee_f)\cv (((\ee_a\cv \ee_c)\cv \ee_d)\cv \ee_f)\n
&=&(\ee_f\cv(((\ee_a\cv \ee_a)\cv \ee_c)\cv \ee_d))\cv (((\ee_a\cv \ee_c)\cv \ee_d)\cv \ee_f)\n
&\overset{MI}{=}&(((\ee_a\cv \ee_a)\cv \ee_c)\cv \ee_d)\cv ((\ee_a\cv \ee_c)\cv \ee_d)\overset{(3)(b)}{=}-\ee_a\n
\nonumber\eeq
\end{enumerate}
\item[5)] For $u \in \la^5(\RR^{0,7})$:
\begin{enumerate}
\item[(a)] $a\notin \{b, c\}$, and ($abc$) is not a $\HH$-triple:
\beq\hspace*{-0.2cm}(\ee_a\bu_\llcorner u)\cv\overline{(1\bu_\llcorner \tilde{u})}&=&(\ee_a\bu_\llcorner \ee_{bcdfg})\cv \overline{(1\bu_\llcorner \widetilde{\ee_{bcdfg}})}\n&=&(\ee_a\bu_\llcorner \ee_{bcdfg})\cv \overline{(1\bu_\llcorner \ee_{bcdfg})}\n
&=&-(\ee_a\bu_\llcorner \ee_{bcdfg})\cv(1\bu_\llcorner \ee_{bcdfg})\n
&=&-(((((\ee_a\cv \ee_b)\cv \ee_c)\cv \ee_d)\cv \ee_f)\cv \ee_g)\n&&\cv ((((\ee_b\cv \ee_c)\cv \ee_d)\cv \ee_f)\cv \ee_g)\n
&=&(\ee_g\cv((((\ee_a\cv \ee_b)\cv \ee_c)\cv \ee_d)\cv \ee_f))\n&&\cv ((((\ee_b\cv \ee_c)\cv \ee_d)\cv \ee_f)\cv \ee_g)\n
&\overset{MI}{=}&((((\ee_a\cv \ee_b)\cv \ee_c)\cv \ee_d)\cv \ee_f)\cv (((\ee_b\cv \ee_c)\cv \ee_d)\cv \ee_f)\n&\overset{(4)(a)}{=}&\ee_a
\nonumber\eeq
\item[(b)] $a=b$ or $a=c$ or ($abc$) is a $\HH$-triple. Without loss of generality,  consider $a=b$:
\beq(\ee_a\bu_\llcorner u)\cv\overline{(1\bu_\llcorner \tilde{u})}&=&(\ee_a\bu_\llcorner \ee_{bcdfg})\cv \overline{(1\bu_\llcorner \widetilde{\ee_{bcdfg}})}\n&=&(\ee_a\bu_\llcorner \ee_{acdfg})\cv \overline{(1\bu_\llcorner \widetilde{\ee_{acdfg}})}\n
&=&-(\ee_a\bu_\llcorner \ee_{acdfg})\cv(1\bu_\llcorner \ee_{acdfg})\n
&=&-(((((\ee_a\cv \ee_a)\cv \ee_c)\cv \ee_d)\cv \ee_f)\cv \ee_g)\n&&\cv ((((\ee_a\cv \ee_c)\cv \ee_d)\cv \ee_f)\cv \ee_g)\n
&=&(\ee_g\cv((((\ee_a\cv \ee_a)\cv \ee_c)\cv \ee_d)\cv \ee_f))\n&&\cv ((((\ee_a\cv \ee_c)\cv \ee_d)\cv \ee_f)\cv \ee_g)\n
&\overset{MI}{=}&((((\ee_a\cv \ee_a)\cv \ee_c)\cv \ee_d)\cv \ee_f)\n&&\cv (((\ee_a\cv \ee_c)\cv \ee_d)\cv \ee_f)\overset{(4)(b)}{=}\ee_a
\nonumber\eeq
\end{enumerate}
\item[6)] For $u \in \la^6(\RR^{0,7})$:
\begin{enumerate}
\item[(a)] $a\notin \{b, c\}$, and ($abc$) is not a $\HH$-triple:
\beq(\ee_a\bu_\llcorner u)\cv\overline{(1\bu_\llcorner \tilde{u})}&=&(\ee_a\bu_\llcorner \ee_{bcdfgh})\cv \overline{(1\bu_\llcorner \widetilde{\ee_{bcdfgh}})}\n&&=(\ee_a\bu_\llcorner \ee_{bcdfgh})\cv (1\bu_\llcorner \ee_{bcdfgh})\n
&\overset{MI}{=}&((((\ee_a\cv \ee_b)\cv \ee_c)\cv \ee_d)\cv \ee_f)\n&&\cv (((\ee_b\cv \ee_c)\cv \ee_d)\cv \ee_f)\overset{(5)(a)}{=}\ee_a
\nonumber\eeq
\item[(b)] $a=b$ or $a=c$ or ($abc$) is a $\HH$-triple. Without loss of generality,  consider $a=b$:
\beq(\ee_a\bu_\llcorner u)\cv\overline{(1\bu_\llcorner \tilde{u})}&=&(\ee_a\bu_\llcorner \ee_{bcdfgh})\cv \overline{(1\bu_\llcorner \widetilde{\ee_{bcdfgh}})}\n&=&(\ee_a\bu_\llcorner \ee_{acdfgh})\cv \overline{(1\bu_\llcorner \widetilde{\ee_{acdfgh}})}\n
&=&(\ee_a\bu_\llcorner \ee_{acdfgh})\cv \overline{(1\bu_\llcorner (-\ee_{acdfgh}))}\n&=&(\ee_a\bu_\llcorner \ee_{acdfgh})\cv (1\bu_\llcorner \ee_{acdfgh})\n
&\overset{MI}{=}&-(((((\ee_a\cv \ee_a)\cv \ee_c)\cv \ee_d)\cv \ee_f)\cv \ee_g)\n&&\cv ((((\ee_a\cv \ee_c)\cv \ee_d)\cv \ee_f)\cv \ee_g)\n&\overset{(5)(b)}{=}&\ee_a\nonumber\eeq
\end{enumerate}\end{enumerate}
%Thus, in general
%\beq(\ee_a\bu_\llcorner u)\cv \overline{(1\bu_\llcorner \tilde{u})} = \pm\ee_a\nonumber\eeq
\section{Proposition 1$^\prime$}
{\bf Proposition 1$^\prime$}: \emph{Given $u=\ee_{i_1}\ee_{i_2}\ldots \ee_{i_k}\in\la^k(\RR^{0,7})$, $k=1,\ldots,6$ then
\beq(1\bu_\llcorner u)\cv \overline{(1\bu_\llcorner \tilde{u})} = (-1)^{|u|(r_{|u|} - r_{|u|-1}) + (8-|u|)(7-|u|)/2}
\eeq
\noi where $r_j$ are the Radon-Hurwitz numbers defined by the following table 
\begin{table}
\caption{\label{table3}Radon-Hurwitz numbers.}
\begin{center}
\begin{tabular}{lllllllll}
\br
$j$&0&1&2&3&4&5&6&7\\
\hline
$r_j$&0&1&2&2&3&3&3&3\\
\br
\end{tabular}
\end{center}
\end{table}
\noi and the recurrence relation $r_{j+8} = r_j + 4$.}\medbreak
{\bf Proof}: Hereon the acronym MI in parenthesis denotes  the Moufang identity given by Eq.(\ref{mou0}).
\begin{enumerate}
\item[0)] For $u \in \la^0(\RR^{0,7})$:
\beq (1\bu_\llcorner u)\cv\overline{(1\bu_\llcorner \tilde{u})}&=&(1\bu_\llcorner \ee_0)\cv \overline{(1\bu_\llcorner \tilde{\ee_0})}=\ee_0\cv \ee_0=1\nonumber\eeq
\item[1)] For $u \in \la^1(\RR^{0,7})$:
\beq(1\bu_\llcorner u)\cv\overline{(1\bu_\llcorner \tilde{u})}&=&(1\bu_\llcorner \ee_a)\cv \overline{(1\bu_\llcorner\tilde{\ee_a})}=-\ee_a\cv \ee_a=1\nonumber\eeq
\item[2)] For $u \in \la^2(\RR^{0,7})$:
\beq(1\bu_\llcorner u)\cv\overline{(1\bu_\llcorner \tilde{u})}&=&(1\bu_\llcorner \ee_a\ee_b)\cv \overline{(1\bu_\llcorner\widetilde{\ee_a\ee_b})}\n&=&(\ee_a\cv\ee_b)\cv\overline{(1\bu_\llcorner(-\ee_a\ee_b))}=(\ee_a\cv\ee_b)\cv(\ee_a\cv\ee_b)\n
&=&-(\ee_b\cv\ee_a)\cv(\ee_a\cv\ee_b)\overset{MI}{=}\ee_b\cv\ee_b=-1\nonumber\eeq
\item[3)] For $u \in \la^3(\RR^{0,7})$:
\begin{enumerate}
\item[(a)] $(abc)$ is not a $\HH$-triple:
\beq(\ee_a\bu_\llcorner u)\cv\overline{(1\bu_\llcorner \tilde{u})}&=&(1\bu_\llcorner \ee_a\ee_b\ee_c)\cv \overline{(1\bu_\llcorner \widetilde{\ee_a\ee_b\ee_c})}\n&=&((\ee_a\cv\ee_b)\cv\ee_c)\cv\overline{(1\bu_\llcorner (-\ee_a\ee_b\ee_c))}\n
&=&((\ee_a\cv\ee_b)\cv\ee_c)\cv((\ee_a\cv\ee_b)\cv\ee_c)\n&=&-(\ee_c\cv(\ee_a\cv\ee_b))\cv((\ee_a\cv\ee_b)\cv\ee_c)\n
&\overset{MI}{=}&-(\ee_a\cv\ee_b)\cv(\ee_a\cv\ee_b)\overset{(2)}{=}1\nonumber\eeq
\item[(b)] $(abc)$ is a $\HH$-triple:
\beq(\ee_a\bu_\llcorner u)\cv\overline{(1\bu_\llcorner \tilde{u})}&=&(1\bu_\llcorner \ee_a\ee_b\ee_c)\cv \overline{(1\bu_\llcorner \widetilde{\ee_a\ee_b\ee_c})}\n&=&((\ee_a\cv\ee_b)\cv\ee_c)\cv\overline{(1\bu_\llcorner (-\ee_a\ee_b\ee_c))}\n
&=&((\ee_a\cv\ee_b)\cv\ee_c)\cv((\ee_a\cv\ee_b)\cv\ee_c)\n&=&(\ee_c\cv(\ee_a\cv\ee_b))\cv((\ee_a\cv\ee_b)\cv\ee_c)\n
&\overset{MI}{=}&(\ee_a\cv\ee_b)\cv(\ee_a\cv\ee_b)\overset{(2)}{=}-1\nonumber\eeq
\end{enumerate}
\item[4)] For $u \in \la^4(\RR^{0,7})$:
\begin{enumerate}
\item[(a)] $(abc)$ is not a $\HH$-triple:
\beq(1\bu_\llcorner u)\cv\overline{(1\bu_\llcorner \tilde{u})}&=&(1\bu_\llcorner \ee_a\ee_b\ee_c\ee_d)\cv \overline{(1\bu_\llcorner \widetilde{\ee_a\ee_b\ee_c\ee_d})}\n&=&(((\ee_a\cv\ee_b)\cv\ee_c)\cv\ee_d)\cv \overline{(1\bu_\llcorner \ee_a\ee_b\ee_c\ee_d)}\n
&=&-(((\ee_a\cv\ee_b)\cv\ee_c)\cv\ee_d)\cv(((\ee_a\cv\ee_b)\cv\ee_c)\cv\ee_d)\n
&=&(\ee_d\cv((\ee_a\cv\ee_b)\cv\ee_c))\cv(((\ee_a\cv\ee_b)\cv\ee_c)\cv\ee_d)\n
&\overset{MI}{=}&((\ee_a\cv\ee_b)\cv\ee_c)\cv((\ee_a\cv\ee_b)\cv\ee_c)\overset{(3)(a)}{=}1\nonumber\eeq
\item[(b)] $(abc)$ is a $\HH$-triple:
\beq(1\bu_\llcorner u)\cv\overline{(1\bu_\llcorner \tilde{u})}&=&(1\bu_\llcorner \ee_a\ee_b\ee_c\ee_d)\cv \overline{(1\bu_\llcorner \widetilde{\ee_a\ee_b\ee_c\ee_d})}\n&=&(((\ee_a\cv\ee_b)\cv\ee_c)\cv\ee_d)\cv \overline{(1\bu_\llcorner \ee_a\ee_b\ee_c\ee_d)}\n
&=&-(((\ee_a\cv\ee_b)\cv\ee_c)\cv\ee_d)\cv(((\ee_a\cv\ee_b)\cv\ee_c)\cv\ee_d)\n
&=&(\ee_d\cv((\ee_a\cv\ee_b)\cv\ee_c))\cv(((\ee_a\cv\ee_b)\cv\ee_c)\cv\ee_d)\n
&\overset{MI}{=}&((\ee_a\cv\ee_b)\cv\ee_c)\cv((\ee_a\cv\ee_b)\cv\ee_c)\overset{(3)(b)}{=}-1\nonumber\eeq
\end{enumerate}
\item[5)] For $u \in \la^5(\RR^{0,7})$:
\begin{enumerate}
\item[(a)] $(abc)$ is not a $\HH$-triple:
\beq(1\bu_\llcorner u)\cv\overline{(1\bu_\llcorner \tilde{u})}&=&(1\bu_\llcorner \ee_a\ee_b\ee_c\ee_d\ee_f)\cv \overline{(1\bu_\llcorner \widetilde{\ee_a\ee_b\ee_c\ee_d\ee_f})}\n
&=&((((\ee_a\cv\ee_b)\cv\ee_c)\cv\ee_d)\cv\ee_f)\n&&\cv \overline{(1\bu_\llcorner \ee_a\ee_b\ee_c\ee_d\ee_f)}\n
&=&-((((\ee_a\cv\ee_b)\cv\ee_c)\cv\ee_d)\cv\ee_f)\cv((((\ee_a\cv\ee_b)\cv\ee_c)\cv\ee_d)\cv\ee_f)\n
&=&(\ee_f\cv(((\ee_a\cv\ee_b)\cv\ee_c)\cv\ee_d))\cv((((\ee_a\cv\ee_b)\cv\ee_c)\cv\ee_d)\cv\ee_f)\n
&\overset{MI}{=}&(((\ee_a\cv\ee_b)\cv\ee_c)\cv\ee_d)\cv(((\ee_a\cv\ee_b)\cv\ee_c)\cv\ee_d)\overset{(4)(a)}{=}-1\nonumber\eeq
\item[(b)] $(abc)$ is a $\HH$-triple:
\beq(1\bu_\llcorner u)\cv\overline{(1\bu_\llcorner \tilde{u})}&=&(1\bu_\llcorner \ee_a\ee_b\ee_c\ee_d\ee_f)\cv \overline{(1\bu_\llcorner \widetilde{\ee_a\ee_b\ee_c\ee_d\ee_f})}\n
&=&((((\ee_a\cv\ee_b)\cv\ee_c)\cv\ee_d)\cv\ee_f)\cv \overline{(1\bu_\llcorner \ee_a\ee_b\ee_c\ee_d\ee_f)}\n
&=&-((((\ee_a\cv\ee_b)\cv\ee_c)\cv\ee_d)\cv\ee_f)\n&&\cv((((\ee_a\cv\ee_b)\cv\ee_c)\cv\ee_d)\cv\ee_f)\n
&=&(\ee_f\cv(((\ee_a\cv\ee_b)\cv\ee_c)\cv\ee_d))\cv((((\ee_a\cv\ee_b)\cv\ee_c)\cv\ee_d)\cv\ee_f)\n
&\overset{MI}{=}&(((\ee_a\cv\ee_b)\cv\ee_c)\cv\ee_d)\cv(((\ee_a\cv\ee_b)\cv\ee_c)\cv\ee_d)\overset{(4)(b)}{=}1\nonumber\eeq
\end{enumerate}
\item[6)] For $u \in \la^6(\RR^{0,7})$:
\begin{enumerate}
\item[(a)] $(abc)$ is not a $\HH$-triple:
\beq(1\bu_\llcorner u)\cv\overline{(1\bu_\llcorner \tilde{u})}&=&(1\bu_\llcorner \ee_a\ee_b\ee_c\ee_d\ee_f\ee_g)\cv \overline{(1\bu_\llcorner \widetilde{\ee_a\ee_b\ee_c\ee_d\ee_f\ee_g})}\n
&=&-(((((\ee_a\cv\ee_b)\cv\ee_c)\cv\ee_d)\cv\ee_f)\cv\ee_g)\cv \overline{(1\bu_\llcorner \ee_a\ee_b\ee_c\ee_d\ee_f\ee_g)}\n&\overset{MI}{=}&-((((\ee_a\cv\ee_b)\cv\ee_c)\cv\ee_d)\cv\ee_f)\n&&\cv((((\ee_a\cv\ee_b)\cv\ee_c)\cv\ee_d)\cv\ee_f)\n&\overset{(5)(a)}{=}&-1\nonumber\eeq
\item[(b)] $(abc)$ is a $\HH$-triple:
\beq(1\bu_\llcorner u)\cv\overline{(1\bu_\llcorner \tilde{u})}&=&(1\bu_\llcorner \ee_a\ee_b\ee_c\ee_d\ee_f\ee_g)\cv \overline{(1\bu_\llcorner \widetilde{\ee_a\ee_b\ee_c\ee_d\ee_f\ee_g})}\n
&=&-(((((\ee_a\cv\ee_b)\cv\ee_c)\cv\ee_d)\cv\ee_f)\cv\ee_g)\cv \overline{(1\bu_\llcorner \ee_a\ee_b\ee_c\ee_d\ee_f\ee_g)}\n
&\overset{MI}{=}&-((((\ee_a\cv\ee_b)\cv\ee_c)\cv\ee_d)\cv\ee_f)\cv((((\ee_a\cv\ee_b)\cv\ee_c)\cv\ee_d)\cv\ee_f)\n&\overset{(5)(b)}{=}&1\nonumber\eeq
\end{enumerate}
\item[7)] For $u \in \la^7(\RR^{0,7})$:
\begin{enumerate}
\item[(a)] $(abc)$ is not a $\HH$-triple:
\beq(1\bu_\llcorner u)\cv\overline{(1\bu_\llcorner \tilde{u})}&&\overset{(6)(a)}{=}1\nonumber\eeq
\item[(b)] $(abc)$ is a $\HH$-triple:
\beq(1\bu_\llcorner u)\cv\overline{(1\bu_\llcorner \tilde{u})}&&\overset{(6)(b)}{=}-1\nonumber\eeq
\end{enumerate}
\end{enumerate}
%Thus, in general
%\beq(1\bu_\llcorner u)\cv \overline{(1\bu_\llcorner \tilde{u})} = \pm 1.\nonumber\eeq
\section{Proposition 2}
{\bf Proposition 2}: \emph{Given $\ee_a\in \mathbb{O}$ and $u=\ee_{i_1}\ee_{i_2}\ldots \ee_{i_k}\in\la^k(\RR^{0,7})$, $k=1,\ldots,6$  then 
\beq(\bar{u}\bu_\lrcorner \ee_a)\circ(1\bu_\llcorner \tilde{u})=(-1)^{|u|(|u|+1)/2}\ee_a\eeq}
\medbreak
{\bf Proof}: $(\bar{u}\bu_\lrcorner \ee_a)\circ(1\bu_\llcorner \tilde{u})$ is computed explicitly for all the homogeneous multivectors that do not have common factors with $\ee_a$. 
\begin{enumerate}
\item[0)] For $u \in \la^0(\RR^{0,7})$:
\beq(\bar{u}\bu_\lrcorner \ee_a)\cv(1\bu_\llcorner \tilde{u})&=&(\bar{\ee_0}\bu_\lrcorner \ee_a)\cv(1\bu_\llcorner \tilde{\ee_0})=(\ee_0\cv \ee_a)\cv(1\cv \ee_0)\n&=&(\ee_0\cv \ee_a)\cv \ee_0=\ee_a\nonumber\eeq
\item[1)] For $u \in \la^1(\RR^{0,7})$:
\beq(\bar{u}\bu_\lrcorner \ee_a)\cv(1\bu_\llcorner \tilde{u})&=&(\bar{\ee_b}\bu_\lrcorner \ee_a)\cv(1\bu_\llcorner \tilde{\ee_b})=(-\ee_b\cv \ee_a)\cv(1\cv \ee_b)\n&=&-(\ee_b\cv \ee_a)\cv \ee_b=-\ee_a\nonumber\eeq
\item[2)] For $u \in \la^2(\RR^{0,7})$:
\beq(\bar{u}\bu_\lrcorner \ee_a)\cv(1\bu_\llcorner \tilde{u})&=&(\overline{\ee_{bc}}\bu_\lrcorner \ee_a)\cv(1\bu_\llcorner \widetilde{\ee_{bc}})=(\ee_{cb}\bu_\lrcorner \ee_a)\cv(1\bu_\llcorner (-\ee_{bc}))\n
&=&-(\ee_c\cv(\ee_b\cv \ee_a))\cv(\ee_b\cv \ee_c)\overset{MI}{=}-(\ee_b\cv \ee_a)\cv \ee_b=-\ee_a\nonumber\eeq
\item[3)] For $u \in \la^3(\RR^{0,7})$:
\beq(\bar{u}\bu_\lrcorner \ee_a)\cv(1\bu_\llcorner \tilde{u})&=&(\overline{\ee_{bcd}}\bu_\lrcorner \ee_a)\cv(1\bu_\llcorner \widetilde{\ee_{bcd}})=(-\ee_{dcb}\bu_\lrcorner \ee_a)\cv(1\bu_\llcorner (-\ee_{bcd}))\n
&=&(\ee_{dcb}\bu_\lrcorner\ee_a)\cv(1\bu_\llcorner\ee_{bcd})\n&=&(\ee_d\cv(\ee_c\cv(\ee_b\cv \ee_a)))\cv((\ee_b\cv \ee_c)\cv \ee_d)\n
&\overset{MI}{=}&(\ee_c\cv(\ee_b\cv \ee_a))\cv(\ee_b\cv\ee_c)\overset{(2)}{=}\ee_a\nonumber\eeq
\item[4)] For $u \in \la^4(\RR^{0,7})$:
\beq(\bar{u}\bu_\lrcorner \ee_a)\cv(1\bu_\llcorner \tilde{u})&=&(\overline{\ee_{bcdf}}\bu_\lrcorner \ee_a)\cv(1\bu_\llcorner \widetilde{\ee_{bcdf}})=(\ee_{fdcb}\bu_\lrcorner \ee_a)\cv(1\bu_\llcorner \ee_{bcdf})\n
&=&(\ee_f\cv(\ee_d \cv(\ee_c\cv (\ee_b \cv \ee_a))))\cv(((\ee_b\cv \ee_c)\cv \ee_d)\cv \ee_f)\n
&\overset{MI}{=}&(\ee_d \cv(\ee_c\cv (\ee_b \cv \ee_a)))\cv((\ee_b\cv \ee_c)\cv \ee_d)\overset{(3)}{=}\ee_a\nonumber\eeq
\item[5)] For $u \in \la^5(\RR^{0,7})$:
\beq(\bar{u}\bu_\lrcorner \ee_a)\cv(1\bu_\llcorner \tilde{u})&=&(\overline{\ee_{bcdfg}}\bu_\lrcorner \ee_a)\cv(1\bu_\llcorner \widetilde{\ee_{bcdfg}})=(-\ee_{gfdcb}\bu_\lrcorner \ee_a)\cv(1\bu_\llcorner \ee_{bcdfg})\n
&=&-\ee_a\nonumber\eeq
\item[6)] For $u \in \la^6(\RR^{0,7})$:
\beq(\bar{u}\bu_\lrcorner \ee_a)\cv(1\bu_\llcorner \tilde{u})&=&(\overline{\ee_{bcdfgh}}\bu_\lrcorner \ee_a)\cv(1\bu_\llcorner \widetilde{\ee_{bcdfgh}})=(\ee_{hgfdcb}\bu_\lrcorner \ee_a)\cv(1\bu_\llcorner (-\ee_{bcdfgh}))\n
&\overset{MI}{=}&-(\ee_g\cv(\ee_f\cv(\ee_d\cv(\ee_c\cv(\ee_b\cv \ee_a)))))\cv((((\ee_b\cv \ee_c)\cv \ee_d)\cv \ee_f)\cv \ee_g)\n&\overset{(5)}{=}&-\ee_a\nonumber\eeq
\end{enumerate}
All the cases above can be evinced as
\beq(\bar{u}\bu_\lrcorner \ee_a)\circ(1\bu_\llcorner \tilde{u})=\pm\ee_a\nonumber\eeq
\section{Proposition 2$^\prime$}
{\bf Proposition 2$^\prime$}: \emph{Given $u=\ee_{i_1}\ee_{i_2}\ldots \ee_{i_k}\in\la^k(\RR^{0,7})$, $k=1,\ldots,6$  then 
\beq(\bar{u}\bu_\lrcorner 1)\circ(1\bu_\llcorner \tilde{u})=(-1)^{|u|(r_{|u|} - r_{|u|-1}) + (8-|u|)(7-|u|)/2}\eeq}
\medbreak
{\bf Proof}:
\begin{enumerate}
\item[0)] For $u \in \la^0(\RR^{0,7})$:
\beq(\bar{u}\bu_\lrcorner 1)\cv(1\bu_\llcorner \tilde{u})&=&(\bar{\ee_0}\bu_\lrcorner 1)\cv(1\bu_\llcorner \tilde{\ee_0})=\ee_0\cv\ee_0=1
\nonumber\eeq
\item[1)] For $u \in \la^1(\RR^{0,7})$:
\beq(\bar{u}\bu_\lrcorner 1)\cv(1\bu_\llcorner \tilde{u})&=&(\bar{\ee_a}\bu_\lrcorner 1)\cv(1\bu_\llcorner \tilde{\ee_a})=(-\ee_a\cv 1)\cv(1\cv \ee_a)\n&=&-\ee_a\cv \ee_a=1\nonumber\eeq
\item[2)] For $u \in \la^2(\RR^{0,7})$:
\beq(\bar{u}\bu_\lrcorner 1)\cv(1\bu_\llcorner \tilde{u})&=&(\overline{\ee_a\ee_b}\bu_\lrcorner 1)\cv(1\bu_\llcorner \widetilde{\ee_a\ee_b})\n&=&(\ee_b\ee_a\bu_\lrcorner 1)\cv(1\bu_\llcorner (-\ee_a\ee_b))\n
&=&-(\ee_b\cv \ee_a)\cv(\ee_a\cv \ee_b)\overset{MI}{=}\ee_b\cv\ee_b=-1\nonumber\eeq
\item[3)] For $u \in \la^3(\RR^{0,7})$:
\beq(\bar{u}\bu_\lrcorner 1)\cv(1\bu_\llcorner \tilde{u})&=&(\overline{\ee_a\ee_b\ee_c}\bu_\lrcorner 1)\cv(1\bu_\llcorner \widetilde{\ee_a\ee_b\ee_c})\n&=&(-\ee_c\ee_b\ee_a\bu_\lrcorner 1)\cv(1\bu_\llcorner (-\ee_a\ee_b\ee_c))\n
&=&(\ee_c\cv(\ee_b\cv\ee_a))\cv((\ee_a\cv\ee_b)\cv\ee_c)\n&\overset{MI}{=}&(\ee_b\cv\ee_a)\cv(\ee_a\cv\ee_b)\overset{(2)}{=}1\nonumber\eeq
\item[4)] For $u \in \la^4(\RR^{0,7})$:
\beq(\bar{u}\bu_\lrcorner 1)\cv(1\bu_\llcorner \tilde{u})&=&(\overline{\ee_a\ee_b\ee_c\ee_d}\bu_\lrcorner 1)\cv(1\bu_\llcorner \widetilde{\ee_a\ee_b\ee_c\ee_d})\n&=&(\ee_d\ee_c\ee_b\ee_a\bu_\lrcorner 1)\cv(1\bu_\llcorner \ee_a\ee_b\ee_c\ee_d)\n
&=&(\ee_d \cv(\ee_c\cv (\ee_b \cv \ee_a)))\cv(((\ee_a\cv \ee_b)\cv \ee_c)\cv \ee_d)\n &\overset{MI}{=}&(\ee_c\cv (\ee_b \cv \ee_a))\cv((\ee_a\cv \ee_b)\cv \ee_c)\overset{(3)}{=}1\nonumber\eeq
\item[5)] For $u \in \la^5(\RR^{0,7})$:
\beq(\bar{u}\bu_\lrcorner 1)\cv(1\bu_\llcorner \tilde{u})&=&(\overline{\ee_a\ee_b\ee_c\ee_d\ee_f}\bu_\lrcorner 1)\cv(1\bu_\llcorner \widetilde{\ee_a\ee_b\ee_c\ee_d\ee_f})\n&=&(-\ee_f\ee_d\ee_c\ee_b\ee_a\bu_\lrcorner 1)\cv(1\bu_\llcorner\ee_a\ee_b\ee_c\ee_d\ee_f)\n
&=&-(\ee_f\cv(\ee_d\cv(\ee_c\cv(\ee_b\cv\ee_a))))\cv((((\ee_a\cv\ee_b)\cv\ee_c)\cv\ee_d)\cv\ee_f)\n
&\overset{MI}{=}&-(\ee_d\cv(\ee_c\cv(\ee_b\cv\ee_a)))\cv(((\ee_a\cv\ee_b)\cv\ee_c)\cv\ee_d)\n&\overset{(4)}{=}&-1\nonumber\eeq
\item[6)] For $u \in \la^6(\RR^{0,7})$:
\beq(\bar{u}\bu_\lrcorner 1)\cv(1\bu_\llcorner \tilde{u})&=&(\overline{\ee_a\ee_b\ee_c\ee_d\ee_f\ee_g}\bu_\lrcorner 1)\cv(1\bu_\llcorner \widetilde{\ee_a\ee_b\ee_c\ee_d\ee_f\ee_g})\n
&=&(\ee_g\ee_f\ee_d\ee_c\ee_b\ee_a\bu_\lrcorner 1)\cv(1\bu_\llcorner (-\ee_a\ee_b\ee_c\ee_d\ee_f\ee_g))\n
&=&-(\ee_g\cv(\ee_f\cv(\ee_d\cv(\ee_c\cv(\ee_b\cv \ee_a)))))\n&&\cv(((((\ee_a\cv \ee_b)\cv \ee_c)\cv \ee_d)\cv \ee_f)\cv \ee_g)\n
&\overset{MI}{=}&-(\ee_f\cv(\ee_d\cv(\ee_c\cv(\ee_b\cv \ee_a))))\cv((((\ee_a\cv \ee_b)\cv \ee_c)\cv \ee_d)\cv \ee_f)\n&\overset{(5)}{=}&-1\nonumber\eeq
\item[7)] For $u \in \la^7(\RR^{0,7})$:
\beq(\bar{u}\bu_\lrcorner 1)\cv(1\bu_\llcorner \tilde{u})&=&(\overline{\ee_a\ee_b\ee_c\ee_d\ee_f\ee_g\ee_h}\bu_\lrcorner 1)\cv(1\bu_\llcorner \widetilde{\ee_a\ee_b\ee_c\ee_d\ee_f\ee_g\ee_h})\n
&=&(-\ee_h\ee_g\ee_f\ee_d\ee_c\ee_b\ee_a\bu_\lrcorner 1)\cv(1\bu_\llcorner (-\ee_a\ee_b\ee_c\ee_d\ee_f\ee_g\ee_h))\n
&\overset{MI}{=}&(\ee_g\cv(\ee_f\cv(\ee_d\cv(\ee_c\cv(\ee_b\cv \ee_a)))))\n&&\cv(((((\ee_a\cv \ee_b)\cv \ee_c)\cv \ee_d)\cv \ee_f)\cv\ee_g)\n&\overset{(6)}{=}&1\nonumber\eeq
\end{enumerate}
%All the cases above can be evinced as
%\beq(\bar{u}\bu_\lrcorner 1)\circ(1\bu_\llcorner \tilde{u})=\pm 1\nonumber\eeq
\section{Proposition 3}
\medbreak
{\bf Proposition 3}: \emph{Given $\ee_a\in \mathbb{O}$ and $u=\ee_{i_1}\ee_{i_2}\ldots \ee_{i_k}\in\la^k(\RR^{0,7})$, $k=1,\ldots,6$ with either $\ee_a\notin\{\ee_{i_1}, \ee_{i_2},\ldots, \ee_{i_k}\}$ or $\{\ee_a, \ee_{i_{j_1}}, \ee_{i_{j_2}}\}$ not a $\HH$-triple, for $\{i_{j_1}, i_{j_2}\}=\{i_1,\ldots,i_6\}$, then
\beq\ee_a\bu_\llcorner u=(-1)^{|u|+1}\ee_a\cv(1\bu_\llcorner u)\nonumber\eeq}
\medbreak
{\bf Proof}: \begin{enumerate}
\item[0)] When $u = \ee_0\in\la^0(\RR^{0,7})$ it follows that:
\beq \ee_a\bu_\llcorner\ee_0&=&\ee_a\cv\ee_0=\ee_a\cv(1\cv\ee_0)=\ee_a\cv(1\bu_\llcorner u)\nonumber\eeq
\item[1)] When $u = \ee_b\in\la^1(\RR^{0,7})$ it follows that:
\begin{enumerate}
\item[(a)] $a\neq b$: 
\beq 
\ee_a\bu_\llcorner u &=& \ee_a\bu_\llcorner\ee_b = \ee_a\cv (1 \cv \ee_b)= \ee_a\cv (1\bu_\llcorner u)\nonumber\eeq 
\item[(b)] $a= b$: 
\beq 
\ee_a\bu_\llcorner u &=& \ee_a\bu_\llcorner\ee_b = \ee_a\cv\ee_a= \ee_a\cv (1 \cv \ee_a)=\ee_a\cv (1\bu_\llcorner u)\nonumber\eeq \end{enumerate}
\item[2)] When $u = \ee_{bc}\in\la^2(\RR^{0,7})$ it follows that:
\begin{enumerate}
\item[(a)] $a\notin \{b, c\}$, and ($abc$) is not a $\HH$-triple:  
\beq\ee_a\bu_\llcorner u &=& \ee_a\bu_\llcorner \ee_{bc} = (\ee_a\cv \ee_b)\cv \ee_c\n &=& -\ee_a\cv (\ee_b\cv \ee_c)=-\ee_a\cv (1\bu_\llcorner \ee_{bc})=-\ee_a\cv (1\bu_\llcorner u)\nonumber\eeq 
\item[(b)] $a\in \{b, c\}$ or ($abc$) is a $\HH$-triple. Without loss of generality consider $a=b$: 
\beq\ee_a\bu_\llcorner u &=& \ee_a\bu_\llcorner\ee_{bc}=\ee_a\bu_\llcorner \ee_{ac} = (\ee_a\cv \ee_a)\cv \ee_c\n &=& \ee_a\cv (\ee_a\cv \ee_c)=\ee_a\cv (1\bu_\llcorner \ee_{ac})\n&=&\ee_a\cv (1\bu_\llcorner u)\label{252}\eeq\end{enumerate}
\item[3)] When $u = \ee_{bcd}\in\la^3(\RR^{0,7})$, follows that:
\begin{enumerate}
\item[(a)] $a\notin \{b, c, d\}$, and ($ijk$) is not a $\HH$-triple where $i,j,k \in \{a,b,c,d\}$: 
\beq 
\ee_a\bu_\llcorner u &=& \ee_a\bu_\llcorner \ee_{bcd} = ((\ee_a\cv \ee_b)\cv \ee_c)\cv \ee_d =  -(\ee_a\cv (\ee_b\cv \ee_c))\cv \ee_d\n 
&=&\ee_a\cv ((\ee_b\cv \ee_c)\cv \ee_d) = \ee_a\cv (1\bu_\llcorner\ee_{bcd})= \ee_a\cv (1\bu_\llcorner u)\nonumber\eeq
\item[(b)] $a\in \{b, c, d\}$ or ($abc$) is a $\HH$-triple. Without loss of generality consider $a=b$: 
\beq 
\ee_a\bu_\llcorner u &=& \ee_a\bu_\llcorner \ee_{bcd} = \ee_a\bu_\llcorner\ee_{acd}=((\ee_a\cv \ee_a)\cv \ee_c)\cv \ee_d \n&=& (\ee_a\cv (\ee_a\cv \ee_c))\cv \ee_d=-\ee_a\cv ((\ee_a\cv \ee_c)\cv \ee_d) = -\ee_a\cv (1\bu_\llcorner\ee_{acd})\n
&=&-\ee_a\cv (1\bu_\llcorner u)\nonumber\eeq
\item[(c)] ($ijk$) is a $\HH$-triple where $i,j,k\in \{a,b, c, d\}$ and ($ijk)\neq(abc$). Let us suppose that ($abd$) is a $\HH$-triple: 
\beq 
\ee_a\bu_\llcorner u &=& \ee_a\bu_\llcorner \ee_{bcd} = ((\ee_a\cv \ee_b)\cv \ee_c)\cv \ee_d = -(\ee_a\cv (\ee_b\cv \ee_c))\cv \ee_d \n 
&=&\ee_a\cv ((\ee_b\cv \ee_c)\cv \ee_d)=\ee_a(1\bu_\llcorner\ee_{bcd})=\ee_a\cv (1\bu_\llcorner u)\nonumber\eeq
\end{enumerate}
\item[4)]  When $u = \ee_{bcdf}\in\la^4(\RR^{0,7})$, follows that:
\begin{enumerate}
\item[(a)] $a\notin \{b, c, d, f\}$, and ($ijk$) is not a $\HH$-triple where $i,j,k \in \{a,b,c,d,f\}$: 
\beq 
\ee_a\bu_\llcorner u &=& \ee_a\bu_\llcorner \ee_{bcdf} = (((\ee_a\cv \ee_b)\cv \ee_c)\cv \ee_d)\cv \ee_f = -((\ee_a\cv (\ee_b\cv \ee_c))\cv \ee_d)\cv \ee_f \n &=& 
(\ee_a\cv ((\ee_b\cv \ee_c)\cv \ee_d))\cv \ee_f = -\ee_a\cv (((\ee_b\cv \ee_c)\cv \ee_d)\cv \ee_f)\n&=& -\ee_a\cv (1\bu_\llcorner\ee_{bcdf})\n
&=&-\ee_a \cv (1\bu_\llcorner u)
\nonumber\eeq
\item[(b)] $a\in \{b, c, d, f\}$ or ($abc$) is a $\HH$-triple. Without loss of generality consider $a=b$: 
\beq 
\ee_a\bu_\llcorner u &=& \ee_a\bu_\llcorner \ee_{bcdf}=\ee_a\bu_\llcorner\ee_{acdf} \n&=& (((\ee_a\cv \ee_a)\cv \ee_c)\cv \ee_d)\cv \ee_f = ((\ee_a\cv (\ee_a\cv \ee_c))\cv \ee_d)\cv \ee_f \n 
&=&-(\ee_a\cv ((\ee_a\cv \ee_c)\cv \ee_d))\cv \ee_f\n& =& \ee_a\cv (((\ee_a\cv \ee_c)\cv \ee_d)\cv \ee_f)=\ee_a\cv(1\bu_\llcorner\ee_{acdf})\n
&=& \ee_a \cv (1\bu_\llcorner u)\label{4b}\eeq
\item[(c)] ($ijk$) is a $\HH$-triple where $i,j,k\in \{a,b, c, d, f\}$ and ($ijk)\neq(abc$). Let us suppose that ($abd$) is a $\HH$-triple:
\beq 
\ee_a\bu_\llcorner u &=& \ee_a\bu_\llcorner \ee_{bcdf}\n &=& (((\ee_a\cv \ee_b)\cv \ee_c)\cv \ee_d)\cv \ee_f = -((\ee_a\cv (\ee_b\cv \ee_c))\cv \ee_d)\cv \ee_f \n 
&=& (\ee_a\cv ((\ee_b\cv \ee_c)\cv \ee_d))\cv \ee_f = -\ee_a\cv (((\ee_b\cv \ee_c)\cv \ee_d)\cv \ee_f)\n&=&-\ee_a\cv(1\bu_\llcorner\ee_{bcdf})\n
&=&-\ee_a \cv (1\bu_\llcorner u)\nonumber\eeq\end{enumerate}
\item[5)]  When $u = \ee_{bcdfg}\in\la^5(\RR^{0,7})$, follows that:
\begin{enumerate}
\item[(a)] $a\notin \{b, c, d, f, g\}$, and ($ijk$) is not a $\HH$-triple where $i,j,k \in \{a,b,c,d,f,g\}$:
\beq 
\ee_a\bu_\llcorner u &=& \ee_a\bu_\llcorner \ee_{bcdfg} = ((((\ee_a\cv \ee_b)\cv \ee_c)\cv \ee_d)\cv \ee_f)\cv \ee_g\n &=& -(((\ee_a\cv (\ee_b\cv \ee_c))\cv \ee_d)\cv \ee_f)\cv \ee_g \n 
&=&((\ee_a\cv ((\ee_b\cv \ee_c)\cv \ee_d))\cv \ee_f)\cv \ee_g\n& =& -(\ee_a\cv (((\ee_b\cv \ee_c)\cv \ee_d)\cv \ee_f))\cv \ee_g\n
&=& \ee_a\cv ((((\ee_b\cv \ee_c)\cv \ee_d)\cv \ee_f)\cv \ee_g)\n &=& \ee_a\cv(1\bu_\llcorner\ee_{bcdfg})=\ee_a \cv (1\bu_\llcorner u)\nonumber\eeq
\item[(b)] $a\in \{b, c, d, f, g\}$ or ($abc$) is a $\HH$-triple. Without loss of generality,  consider $a=b$: 
\beq 
\ee_a\bu_\llcorner u &=& \ee_a\bu_\llcorner \ee_{bcdfgh}=\ee_a\bu_\llcorner\ee_{acdfgh} = ((((\ee_a\cv \ee_a)\cv \ee_c)\cv \ee_d)\cv \ee_f)\cv \ee_g\n
&=&(((\ee_a\cv (\ee_a\cv \ee_c))\cv \ee_d)\cv \ee_f)\cv \ee_g\n&=&-((\ee_a\cv ((\ee_a\cv \ee_c)\cv \ee_d))\cv \ee_f)\cv \ee_g\n 
&=&(\ee_a\cv (((\ee_a\cv \ee_c)\cv \ee_d)\cv \ee_f))\cv \ee_g\n&=&-\ee_a\cv ((((\ee_a\cv \ee_c)\cv \ee_d)\cv \ee_f)\cv \ee_g)\n
&=&-\ee_a\cv(1\bu_\llcorner\ee_{acdfgh})=-\ee_a \cv (1\bu_\llcorner u)\nonumber\eeq
\item[(c)] ($ijk$) is a $\HH$-triple where $i,j,k\in \{a,b, c, d, f, g\}$ and ($ijk)\neq(abc$). Let us suppose that ($abd$) is a $\HH$-triple:
\beq 
\ee_a\bu_\llcorner u &=& \ee_a\bu_\llcorner \ee_{bcdfg} = ((((\ee_a\cv \ee_b)\cv \ee_c)\cv \ee_d)\cv \ee_f)\cv \ee_g\n &=& -(((\ee_a\cv (\ee_b\cv \ee_c))\cv \ee_d)\cv \ee_f)\cv \ee_g \n 
&=& ((\ee_a\cv ((\ee_b\cv \ee_c)\cv \ee_d))\cv \ee_f)\cv \ee_g\n& =& -(\ee_a\cv (((\ee_b\cv \ee_c)\cv \ee_d)\cv \ee_f))\cv \ee_g\n
&=& \ee_a\cv ((((\ee_b\cv \ee_c)\cv \ee_d)\cv \ee_f)\cv \ee_g)\n& =& \ee_a\cv(1\bu_\llcorner\ee_{bcdfg})=\ee_a \cv (1\bu_\llcorner u)\nonumber\eeq
\item[(d)] ($ijk$) and $(lmn$) are $\HH$-triples where $i,j,k,l,m,n \in \{a,b,c,d,f,g\}$, $\{i,j,k\}\neq\{l,m,n\}$ and ($ijk)\neq(abc$). Let us suppose that ($abd$) and ($cfg$) are $\HH$-triples: 
\beq 
\ee_a\bu_\llcorner u &=& \ee_a\bu_\llcorner \ee_{bcdfg} = ((((\ee_a\cv \ee_b)\cv \ee_c)\cv \ee_d)\cv \ee_f)\cv \ee_g\n &=& -(((\ee_a\cv (\ee_b\cv \ee_c))\cv \ee_d)\cv \ee_f)\cv \ee_g \n 
&=&((\ee_a\cv ((\ee_b\cv \ee_c)\cv \ee_d))\cv \ee_f)\cv \ee_g = -(\ee_a\cv (((\ee_b\cv \ee_c)\cv \ee_d)\cv \ee_f))\cv \ee_g\n
&=& \ee_a\cv ((((\ee_b\cv \ee_c)\cv \ee_d)\cv \ee_f)\cv \ee_g) = \ee_a\cv(1\bu_\llcorner\ee_{bcdfg})=\ee_a \cv (1\bu_\llcorner u)\nonumber\eeq
\end{enumerate}
\item[6)]  When $u = \ee_{bcdfgh}\in\la^6(\RR^{0,7})$, follows that:
\begin{enumerate}
\item[(a)] $a\notin \{b, c, d, f, g, h\}$, and ($ijk$) and $(lmn$) are not $\HH$-triples where $i,j,k,l,m,n \in \{a,b,c,d,f,g,h\}$ and $\{i,j,k\}\neq\{l,m,n\}$:
\beq 
\ee_a\bu_\llcorner u &=& \ee_a\bu_\llcorner \ee_{bcdfgh} = (((((\ee_a\cv \ee_b)\cv \ee_c)\cv \ee_d)\cv \ee_f)\cv \ee_g)\cv \ee_h \n
&=&-((((\ee_a\cv (\ee_b\cv \ee_c))\cv \ee_d)\cv \ee_f)\cv \ee_g)\cv \ee_h\n&=&(((\ee_a\cv ((\ee_b\cv \ee_c)\cv \ee_d))\cv \ee_f)\cv \ee_g)\cv \ee_h\n 
&=&-((\ee_a\cv (((\ee_b\cv \ee_c)\cv \ee_d)\cv \ee_f))\cv \ee_g)\cv \ee_h\n&=&(\ee_a\cv ((((\ee_b\cv \ee_c)\cv \ee_d)\cv \ee_f)\cv \ee_g))\cv \ee_h\n
&=&-\ee_a\cv (((((\ee_b\cv \ee_c)\cv \ee_d)\cv \ee_f)\cv \ee_g)\cv \ee_h)\n&=&-\ee_a\cv(1\bu_\llcorner\ee_{bcdfgh})=-\ee_a \cv (1\bu_\llcorner u)\nonumber\eeq
\item[(b)] $a\in \{b, c, d, f, g, h\}$ or ($abc$) is a $\HH$-triple. Without loss of generality,  consider $a=b$: 
\beq 
\ee_a\bu_\llcorner u &=& \ee_a\bu_\llcorner \ee_{bcdfgh}=\ee_a\bu_\llcorner\ee_{acdfgh}\n& =& (((((\ee_a\cv \ee_a)\cv \ee_c)\cv \ee_d)\cv \ee_f)\cv \ee_g)\cv \ee_h \n
&=&((((\ee_a\cv (\ee_a\cv \ee_c))\cv \ee_d)\cv \ee_f)\cv \ee_g)\cv \ee_h\n&=&-(((\ee_a\cv ((\ee_a\cv \ee_c)\cv \ee_d))\cv \ee_f)\cv \ee_g)\cv \ee_h\n 
&=&((\ee_a\cv (((\ee_a\cv \ee_c)\cv \ee_d)\cv \ee_f))\cv \ee_g)\cv \ee_h\n&=&-(\ee_a\cv ((((\ee_a\cv \ee_c)\cv \ee_d)\cv \ee_f)\cv \ee_g))\cv \ee_h\n
&=&\ee_a\cv (((((\ee_a\cv \ee_c)\cv \ee_d)\cv \ee_f)\cv \ee_g)\cv \ee_h)\n&=&\ee_a\cv(1\bu_\llcorner\ee_{acdfgh})=\ee_a \cv (1\bu_\llcorner u)\nonumber\eeq
\item[(c)] ($ijk$) is a $\HH$-triple where $i,j,k\in \{a,b, c, d, f, g, h\}$ and ($ijk)\neq(abc$). Let us suppose that ($abd$) is a $\HH$-triple:
\beq 
\ee_a\bu_\llcorner u &=& \ee_a\bu_\llcorner \ee_{bcdfgh} = (((((\ee_a\cv \ee_b)\cv \ee_c)\cv \ee_d)\cv \ee_f)\cv \ee_g)\cv \ee_h \n
&=&-((((\ee_a\cv (\ee_b\cv \ee_c))\cv \ee_d)\cv \ee_f)\cv \ee_g)\cv \ee_h\n&=&(((\ee_a\cv ((\ee_b\cv \ee_c)\cv \ee_d))\cv \ee_f)\cv \ee_g)\cv \ee_h\n 
&=&-((\ee_a\cv (((\ee_b\cv \ee_c)\cv \ee_d)\cv \ee_f))\cv \ee_g)\cv \ee_h\n&=&(\ee_a\cv ((((\ee_b\cv \ee_c)\cv \ee_d)\cv \ee_f)\cv \ee_g))\cv \ee_h\n
&=&-\ee_a\cv (((((\ee_b\cv \ee_c)\cv \ee_d)\cv \ee_f)\cv \ee_g)\cv \ee_h)\n&=&-\ee_a\cv(1\bu_\llcorner\ee_{bcdfgh})=-\ee_a \cv (1\bu_\llcorner u)\nonumber\eeq
\item[(d)] ($ijk$) and $(lmn$) are $\HH$-triples where $i,j,k,l,m,n \in \{a,b,c,d,f,g,h\}$, $\{i,j,k\}\neq\{l,m,n\}$ and ($ijk)\neq(abc$). Let us suppose that ($abd$) and ($cfg$) are $\HH$-triples:
\beq 
\ee_a\bu_\llcorner u &=& \ee_a\bu_\llcorner \ee_{bcdfgh} = (((((\ee_a\cv \ee_b)\cv \ee_c)\cv \ee_d)\cv \ee_f)\cv \ee_g)\cv \ee_h \n
&=&((((\ee_a\cv (\ee_b\cv \ee_c))\cv \ee_d)\cv \ee_f)\cv \ee_g)\cv \ee_h\n&=&-(((\ee_a\cv ((\ee_b\cv \ee_c)\cv \ee_d))\cv \ee_f)\cv \ee_g)\cv \ee_h\n 
&=&((\ee_a\cv (((\ee_b\cv \ee_c)\cv \ee_d)\cv \ee_f))\cv \ee_g)\cv \ee_h\n&=&-(\ee_a\cv ((((\ee_b\cv \ee_c)\cv \ee_d)\cv \ee_f)\cv \ee_g))\cv \ee_h\n
&=&\ee_a\cv (((((\ee_b\cv \ee_c)\cv \ee_d)\cv \ee_f)\cv \ee_g)\cv \ee_h)\n&=&\ee_a\cv(1\bu_\llcorner\ee_{bcdfgh})=\ee_a \cv (1\bu_\llcorner u)\nonumber\eeq
\end{enumerate}
\end{enumerate}
%Thus, in general
%\beq\ee_a\bu_\llcorner u=\pm\ee_a\cv(1\bu_\llcorner u)\nonumber\eeq
On the hypothesis above $u\in\la(\RR^{0,7})$ should not be homogeneous. For instance, by taking Eqs.(\ref{252}) and (\ref{4b}). When $u = \ee_b\ee_c + \ee_b\ee_c\ee_d\in\la^2(\RR^{0,7})\op \la^3(\RR^{0,7})$ is defined, the Propositions shown in cases 2(b) and 4(b) above imply that $\ee_a\bu_\llcorner u =\ee_a\cv (1\bu_\llcorner u)$, and therefore in many cases the refereed Proposition holds for elements that are not homogeneous. Taking similar examples, it can shown that there exists some cases where $u\in \Lambda^k(\RR^{0,7})\op \Lambda^{k+1}(\RR^{0,7})$ and it satisfies the Proposition 3.

\section{Proposition 4}
Here the result of the Proposition 3 for $\ee_a\in\mathbb{O}$ and $u\in\cl_{0,7}$ is considered, $\ee_a\bu_\llcorner u=\pm\ee_a\cv(1\bu_\llcorner u)$. Hence the Proposition 4 is a generalization of the Proposition 3. 
\medbreak
{\bf Proposition 4}: \emph{Given $\psi\in\cl_{0,7}$ and $u\in \la^k(\RR^{0,7})$, $k=1,\ldots,6$ then 
\beq\psi\odot_\llcorner u=(-1)^{|u|+1}\psi\bu_\lrcorner(1\bu_\llcorner u)\eeq\noi 
In all the cases above $|u|$ denotes the degree of $u$: if $u\in\la^k(\RR^{0,7})$, then $|u|$ = $k$.}
\medbreak
{\bf Proof}: It must be evaluated the computation for all the cases, but considering Proposition 3 the work is reduced.
\begin{enumerate}
\item[0)] For $\psi=\ee_0 \in \la^0(\RR^{0,7})$:
\beq \ee_0\odot_\llcorner u&=&\ee_0\bu_\llcorner u=\ee_0\cv(1\bu_\llcorner u)=\psi\bu_\lrcorner(1\bu_\llcorner u)\nonumber\eeq
\item[1)] For $\psi=\ee_a \in \la^1(\RR^{0,7})$:
\beq \ee_a\odot_\llcorner u&=&\ee_a\bu_\llcorner u=\pm\ee_a\cv(1\bu_\llcorner u)=\pm\psi\bu_\lrcorner(1\bu_\llcorner u)\nonumber\eeq
\item[2)] For $\psi=\ee_{ab} \in \la^2(\RR^{0,7})$:
\beq \ee_{ab}\odot_\llcorner u&=&\ee_a\cv(\ee_b\bu_\llcorner u)=\pm\ee_a\cv(\ee_b\cv(1\bu_\llcorner u))=\pm\ee_{ab}\bu_\lrcorner(1\bu_\llcorner u)=\pm\psi\bu_\lrcorner(1\bu_\llcorner u)\nonumber\eeq
\item[3)] For $\psi=\ee_{abc} \in \la^3(\RR^{0,7})$:
\beq \ee_{abc}\odot_\llcorner u&=&\ee_a\cv(\ee_b\cv(\ee_c\bu_\llcorner u))\n&=&\pm\ee_a\cv(\ee_b\cv(\ee_c\cv(1\bu_\llcorner u)))=\pm\ee_{abc}\bu_\lrcorner(1\bu_\llcorner u)=\pm\psi\bu_\lrcorner(1\bu_\llcorner u)\nonumber\eeq
\item[4)] For $\psi=\ee_{abcd} \in \la^4(\RR^{0,7})$:
\beq \ee_{abcd}\odot_\llcorner u&=&\ee_a\cv(\ee_b\cv(\ee_c\cv(\ee_d\bu_\llcorner u)))\n&=&\pm\ee_a\cv(\ee_b\cv(\ee_c\cv(\ee_d\cv(1\bu_\llcorner u)))=\pm\ee_{abcd}\bu_\lrcorner(1\bu_\llcorner u)\n
&=&\pm\psi\bu_\lrcorner(1\bu_\llcorner u)\nonumber\eeq
\item[5)] For $\psi=\ee_{abcdf} \in \la^5(\RR^{0,7})$:
\beq \ee_{abcdf}\odot_\llcorner u&=&\ee_a\cv(\ee_b\cv(\ee_c\cv(\ee_d\cv(\ee_f\bu_\llcorner u))))\n&=&\pm\ee_a\cv(\ee_b\cv(\ee_c\cv(\ee_d\cv(\ee_f\cv(1\bu_\llcorner u)))=\pm\ee_{abcdf}\bu_\lrcorner(1\bu_\llcorner u)\n
&=&\pm\psi\bu_\lrcorner(1\bu_\llcorner u)\nonumber\eeq
\item[6)] For $\psi=\ee_{abcdfg} \in \la^6(\RR^{0,7})$:
\beq \ee_{abcdfg}\odot_\llcorner u&=&\ee_a\cv(\ee_b\cv(\ee_c\cv(\ee_d\cv(\ee_f\cv(\ee_g\bu_\llcorner u))))\n&=&\pm\ee_a\cv(\ee_b\cv(\ee_c\cv(\ee_d\cv(\ee_f\cv(1\bu_\llcorner u)))))\n
&=&\pm\ee_{abcdfg}\bu_\lrcorner(1\bu_\llcorner u)=\pm\psi\bu_\lrcorner(1\bu_\llcorner u)\nonumber\eeq
\end{enumerate}
Note that the indices of $\psi$ and $u$ must be different, otherwise the signal changes. 
%Thus, in general
%\beq\psi\odot_\llcorner u=\pm\psi\bu_\lrcorner(1\bu_\llcorner u)\nonumber\eeq
%\noi and the sign respects the statement.
\section{The frame $\{\ee_a\cv X\}$}
\label{proof of}
As $X\in\mathbb{O}$ does not take any privileged unit, without loss of generality, by setting $\ee_a=\ee_1$, once the process is similar for all $\ee_a$, $a=1,\ldots,7$, it shall be checked that $\left\langle \ee_a\cv X, X\right\rangle=0$. 
\beq \left\langle \ee_1\cv X, X\right\rangle &=& 1/2[(\overline{\ee_1\cv X})\cv X+\bar{X}\cv (\ee_1\cv X)]\n
&=& 1/2[(-X^0\ee_1-X^1-X^2\ee_6-X^3\ee_4+X^4\ee_3-X^5\ee_7+X^6\ee_2+X^7\ee_5)\cv \n
& &(X^0+X^1\ee_1+X^2\ee_2+X^3\ee_3+X^4\ee_4+X^5\ee_5+X^6\ee_6+X^7\ee_7)+\n
& &(X^0-X^1\ee_1-X^2\ee_2-X^3\ee_3-X^4\ee_4-X^5\ee_5-X^6\ee_6-X^7\ee_7)\cv\n
& &(X^0\ee_1-X^1+X^2\ee_6+X^3\ee_4-X^4\ee_3+X^5\ee_7-X^6\ee_2-X^7\ee_5)]\n
&=&0.\nonumber\eeq

Now, it will be shown that $\left\langle \ee_a\ee_b\bu_\lrcorner X, X\right\rangle=0$. Again, without loss of generality, choosing $\ee_1\ee_2$ comes:
\beq \left\langle \ee_1\ee_2\bu_\lrcorner X, X\right\rangle &=& 1/2[(\overline{\ee_1\ee_2\bu_\lrcorner X})\cv X+\bar{X}\cv (\ee_1\ee_2\bu_\lrcorner X)]\n
&=& 1/2[(-X^0\ee_6-X^1\ee_2+X^2\ee_1+X^3\ee_5-X^4\ee_7-X^5\ee_3-X^6+X^7\ee_4)\cv \n
& &(X^0+X^1\ee_1+X^2\ee_2+X^3\ee_3+X^4\ee_4+X^5\ee_5+X^6\ee_6+X^7\ee_7)+\n
& &(X^0-X^1\ee_1-X^2\ee_2-X^3\ee_3-X^4\ee_4-X^5\ee_5-X^6\ee_6-X^7\ee_7)\cv\n
& &(X^0\ee_6+X^1\ee_2-X^2\ee_1-X^3\ee_5+X^4\ee_7+X^5\ee_3-X^6-X^7\ee_4)]\n
&=&0.\nonumber\eeq
\section{The $X$-product representations}
\label{matrix rep}
It is a well known assertion (see e.g. \cite{dix}) that the octonionic algebra is a non-associative algebra, not being possible to represent it on a matrix algebra. The adjoint algebras of the left and right actions on octonions itself are associative. In our case, the $\bu$-product on the  right and left is presented in Eqs.(\ref{209},\ref{210}) respectively. 

The main purpose of this Section is to evince the matrix representation for the left actions on the octonions in our formalism\footnote{In fact, Dixon exhibited  the matrix representations for the left and right actions on the octonions by computing ---  in his notation --- $\ee_{La}, \ee_{Lab}, \ee_{Labc}$, and also $\ee_{Ra}, \ee_{Rab}, \ee_{Rabc}$, for octonions provided from both the $\ee_a\ee_{a+1}=\ee_{a+5 \mod 7}$ and $\ee_a\ee_{a+1}=\ee_{a+3 \mod 7}$ rules.} \cite{dix}. The following matrices generate $\mathcal{M}(8,\RR)$ and all $u\bu_\lrcorner X$ representations for $u\in\cl_{0,7}$. Indeed, the Hodge dual in $\cl_{0,7}$ can be expressed as
\beq\star u = \tilde{u}\ee_1\ee_2\ee_3\ee_4\ee_5\ee_6\ee_7,\nonumber\eeq
therefore in order to generate the matrices associated to the set $$\{\ee_a\ee_b\ee_c\ee_d, \ee_a\ee_b\ee_c\ee_d\ee_f, \ee_a\ee_b\ee_c\ee_d\ee_f\ee_g, \ee_1\ee_2\ee_3\ee_4\ee_5\ee_6\ee_7\}$$ that acts on $X\in\mathbb{O}$ by $\bu$-product --- it  must just to be considered respectively  the correspondence to $\{\star 1, \star \ee_a, \star \ee_a\ee_b, \star \ee_a\ee_b\ee_c\}$ in this order. Moreover, the set $\{1, \ee_a, \ee_a\ee_b, \ee_a\ee_b\ee_c\}$ has dimension equal to 64.

 An octonion $X = X^0 + X^a\ee_a$  can be written as $(X^0, X^1, \ldots, X^7)^T$ representing its vector space underlying structure.
Below we present the $\circ$-product by the action of matrices:
\beq\ee_1\circ(X) &=& \ee_1\circ(X^0+X^1\ee_1+X^2\ee_2+X^3\ee_3+X^4\ee_4+X^5\ee_5+X^6\ee_6+X^7\ee_7)\n
&=& X^0\ee_1-X^1+X^2\ee_6+X^3\ee_4-X^4\ee_3+X^5\ee_7-X^6\ee_2-X^7\ee_5\n&=&\left(\begin{smallmatrix}0 & -1 & 0 & 0 & 0 & 0 & 0 & 0\\ 1 & 0 & 0 & 0 & 0 & 0 & 0 & 0\\0 & 0 & 0 & 0 & 0 & 0 & -1 & 0\\0 & 0 & 0 & 0 & -1 & 0 & 0 & 0\\0 & 0 & 0 & 1 & 0 & 0 & 0 & 0\\0 & 0 & 0 & 0 &0 & 0 & 0 & -1\\0 & 0 & 1 & 0 &0 & 0 & 0 & 0\\0 & 0 & 0 & 0 &0 & 1 & 0 & 0\end{smallmatrix}\right)\left(\begin{smallmatrix}X^0\\X^1\\X^2\\X^3\\X^4\\X^5\\X^6\\X^7\end{smallmatrix}\right)\nonumber\eeq\noi Analogously, 
each $\ee_a\cv (\,\cdot\,)$ action corresponds respectively to the following matrices:
\beq\ee_2\circ(\cdot) \mapsto\left(\begin{smallmatrix}0 & 0 & -1 & 0 & 0 & 0 & 0 & 0\\ 0 & 0 & 0 & 0 & 0 & 0 & 1 & 0\\1 & 0 & 0 & 0 & 0 & 0 & 0 & 0\\0 & 0 & 0 & 0 & 0 & 0 & 0 & -1\\0 & 0 & 0 & 0 & 0 & -1 & 0 & 0\\0 & 0 & 0 & 0 &1 & 0 & 0 & 0\\0 & -1 & 0 & 0 &0 & 0 & 0 & 0\\0 & 0 & 0 & 1 &0 & 0 & 0 & 0\end{smallmatrix}\right)\quad\ee_3\circ(\cdot) \mapsto\left(\begin{smallmatrix}0 & 0 & 0 & -1 & 0 & 0 & 0 & 0\\ 0 & 0 & 0 & 0 & 1 & 0 & 0 & 0\\0 & 0 & 0 & 0 & 0 & 0 & 0 & 1\\1 & 0 & 0 & 0 & 0 & 0 & 0 & 0\\0 & -1 & 0 & 0 & 0 & 0 & 0 & 0\\0 & 0 & 0 & 0 &0 & 0 & -1 & 0\\0 & 0 & 0 & 0 &0 & 1 & 0 & 0\\0 & 0 & -1 & 0 &0 & 0 & 0 & 0\end{smallmatrix}\right)\nonumber\eeq
\beq\quad\ee_4\circ(\cdot) \mapsto\left(\begin{smallmatrix}0 & 0 & 0 & 0 & -1 & 0 & 0 & 0\\ 0 & 0 & 0 & -1 & 0 & 0 & 0 & 0\\0 & 0 & 0 & 0 & 0 & 1 & 0 & 0\\0 & 1 & 0 & 0 & 0 & 0 & 0 & 0\\1 & 0 & 0 & 0 & 0 & 0 & 0 & 0\\0 & 0 & -1 & 0 &0 & 0 & 0 & 0\\0 & 0 & 0 & 0 &0 & 0 & 0 & -1\\0 & 0 & 0 & 0 &0 & 0 & 1 & 0\end{smallmatrix}\right)\quad\ee_5\circ(\cdot) \mapsto\left(\begin{smallmatrix}0 & 0 & 0 & 0 & 0 & -1 & 0 & 0\\ 0 & 0 & 0 & 0 & 0 & 0 & 0 & 1\\0 & 0 & 0 & 0 & -1& 0 & 0 & 0\\0 & 0 & 0 & 0 & 0 & 0 & 1 & 0\\0 & 0 & 1 & 0 & 0 & 0 & 0 & 0\\1 & 0 & 0 & 0 &0 & 0 & 0 & 0\\0 & 0 & 0 & -1 &0 & 0 & 0 & 0\\0 & -1 & 0 & 0 &0 & 0 & 0 & 0\end{smallmatrix}\right)\nonumber\eeq\beq\ee_6\circ(\cdot) \mapsto\left(\begin{smallmatrix}0 & 0 & 0 & 0 & 0 & 0 & -1 & 0\\ 0 & 0 & -1 & 0 & 0 & 0 & 0 & 0\\0 & 1 & 0 & 0 & 0& 0 & 0 & 0\\0 & 0 & 0 & 0 & 0 & -1 & 0 & 0\\0 & 0 & 0 & 0 & 0 & 0 & 0 & 1\\0 & 0 & 0 & 1 &0 & 0 & 0 & 0\\1 & 0 & 0 & 0 &0 & 0 & 0 & 0\\0 & 0 & 0 & 0 &-1 & 0 & 0 & 0\end{smallmatrix}\right)\quad\ee_7\circ(\cdot) \mapsto\left(\begin{smallmatrix}0 & 0 & 0 & 0 & 0 & 0 & 0 & -1\\ 0 & 0 & 0 & 0 & 0 & -1 & 0 & 0\\0 & 0 & 0 & -1 & 0& 0 & 0 & 0\\0 & 0 & 1 & 0 & 0 & 0 & 0 & 0\\0 & 0 & 0 & 0 & 0 & 0 & -1 & 0\\0& 1 & 0 & 0 &0 & 0 & 0 & 0\\0 & 0 & 0 & 0 &1 & 0 & 0 & 0\\1 & 0 & 0 & 0 &0 & 0 & 0 & 0\end{smallmatrix}\right)\nonumber\eeq
Also, the $\bu_\lrcorner$-product can be represented by their matrices left actions. Below we present explicitly 
all the $\Lambda(V)$ basis vectors action\beq(\ee_1\ee_2)\bu_\lrcorner(X) &=& \ee_1\circ(\ee_2\circ(X^0+X^1\ee_1+X^2\ee_2+X^3\ee_3+X^4\ee_4+X^5\ee_5+X^6\ee_6+X^7\ee_7))\n
&=& \ee_1\circ(X^0\ee_2-X^1\ee_6-X^2+X^3\ee_7+X^4\ee_5-X^5\ee_4+X^6\ee_1-X^7\ee_3)\n
&=& X^0\ee_6+X^1\ee_2-X^2\ee_1-X^3\ee_5+X^4\ee_7+X^5\ee_3-X^6-X^7\ee_4\n&=&\left(\begin{smallmatrix}0 & 0 & 0 & 0 & 0 & 0 & -1 & 0\\ 0 & 0 & -1 & 0 & 0 & 0 & 0 & 0\\0 & 1 & 0 & 0 & 0 & 0 & 0 & 0\\0 & 0 & 0 & 0 & 0 & 1 & 0 & 0\\0 & 0 & 0 & 0 & 0 & 0 & 0 & -1\\0 & 0 & 0 & -1 &0 & 0 & 0 & 0\\1 & 0 & 0 & 0 &0 & 0 & 0 & 0\\0 & 0 & 0 & 0 &1 & 0 & 0 & 0\end{smallmatrix}\right)\left(\begin{smallmatrix}X^0\\X^1\\X^2\\X^3\\X^4\\X^5\\X^6\\X^7\end{smallmatrix}\right)\nonumber\eeq\noi Analogously, 
each $\ee_a\ee_b\bu_\lrcorner (\,\cdot\,)$ action corresponds respectively to the following matrices:
\beq(\ee_1\ee_3)\bu_\lrcorner(\cdot) \mapsto\left(\begin{smallmatrix}0 & 0 & 0 & 0 & -1& 0 & 0 & 0\\ 0 & 0 & 0 & -1 & 0 & 0 & 0 & 0\\0 &0 & 0 & 0 & 0 & -1 & 0 & 0\\0 & 1 & 0 & 0 & 0 & 0 & 0 & 0\\1 & 0 & 0 & 0 & 0 & 0 & 0 &0\\0 & 0 & 1 & 0 &0 & 0 & 0 & 0\\0 & 0 & 0 & 0 &0 & 0 & 0 & 1\\0 & 0 & 0 & 0 &0 & 0 & -1 & 0\end{smallmatrix}\right)\quad(\ee_1\ee_4)\bu_\lrcorner(\cdot) \mapsto\left(\begin{smallmatrix}0 & 0 & 0 & 1 & 0& 0 & 0 & 0\\ 0 & 0 & 0 & 0 & -1 & 0 & 0 & 0\\0 &0 & 0 & 0 & 0 & 0 & 0 & 1\\-1 & 0 & 0 & 0 & 0 & 0 & 0 & 0\\0 & 1 & 0 & 0 & 0 & 0 & 0 &0\\0 & 0 & 0 & 0 &0 & 0 & -1 & 0\\0 & 0 & 0 & 0 &0 & 1 & 0 & 0\\0 & 0 & -1 & 0 &0 & 0 & 0 & 0\end{smallmatrix}\right)\nonumber\eeq
\beq(\ee_1\ee_6)\bu_\lrcorner(\cdot) \mapsto\left(\begin{smallmatrix}0 & 0 & 1& 0 & 0& 0 & 0 & 0\\ 0 & 0 & 0 & 0 & 0 & 0 &-1 & 0\\-1 &0 & 0 & 0 & 0 & 0 & 0 & 0\\0 & 0 & 0 & 0 & 0 & 0 & 0 & -1\\0 & 0 & 0 & 0 & 0 & -1 & 0&0\\0 & 0 & 0 & 0 &1 & 0 & 0 & 0\\0 & 1 & 0 & 0 &0 & 0 & 0 & 0\\0 & 0 & 0 & 1 &0 & 0 & 0 & 0\end{smallmatrix}\right)\quad(\ee_1\ee_7)\bu_\lrcorner(\cdot) \mapsto\left(\begin{smallmatrix}0 & 0 & 0& 0 & 0& 1 & 0 & 0\\ 0 & 0 & 0 & 0 & 0 & 0 &0 & -1\\0 &0 & 0 & 0 & -1 & 0 & 0 & 0\\0 & 0 & 0 & 0 & 0 & 0 & 1 & 0\\0 & 0 & 1 & 0 & 0 & 0 & 0&0\\-1 & 0 & 0 & 0 &0 & 0 & 0 & 0\\0 & 0 & 0 & -1 &0 & 0 & 0 & 0\\0 & 1 & 0 & 0 &0 & 0 & 0 & 0\end{smallmatrix}\right)\nonumber\eeq
\beq \quad(\ee_1\ee_5)\bu_\lrcorner(\cdot) \mapsto\left(\begin{smallmatrix}0 & 0 & 0 & 0 & 0& 0 & 0 & -1\\ 0 & 0 & 0 & 0 & 0 & -1 & 0 & 0\\0 &0 & 0 & 1 & 0 & 0 & 0 & 0\\0 & 0 & -1 & 0 & 0 & 0 & 0 & 0\\0 & 0 & 0 & 0 & 0 & 0 & 1&0\\0 & 1 & 0 & 0 &0 & 0 & 0 & 0\\0 & 0 & 0 & 0 &-1 & 0 & 0 & 0\\1 & 0 & 0 & 0 &0 & 0 & 0 & 0\end{smallmatrix}\right)\quad(\ee_2\ee_3)\bu_\lrcorner(\cdot) \mapsto\left(\begin{smallmatrix}0 & 0 & 0& 0 & 0& 0 & 0 & -1\\ 0 & 0 & 0 & 0 & 0 & 1 &0 & 0\\0 &0 & 0 & -1 & 0 & 0 & 0 & 0\\0 & 0 & 1 & 0 & 0 & 0 & 0 & 0\\0 & 0 & 0 & 0 & 0 & 0 & 1&0\\0 & -1 & 0 & 0 &0 & 0 & 0 & 0\\0 & 0 & 0 & 0 &-1& 0 & 0 & 0\\1 & 0 & 0 & 0 &0 & 0 & 0 & 0\end{smallmatrix}\right)\nonumber\eeq
\beq(\ee_2\ee_4)\bu_\lrcorner(\cdot) \mapsto\left(\begin{smallmatrix}0 & 0 & 0& 0 & 0& -1 & 0 & 0\\ 0 & 0 & 0 & 0 & 0 & 0 &0 & -1\\0 &0 & 0 & 0 & -1 & 0 & 0 & 0\\0 & 0 & 0 & 0 & 0 & 0 & -1 & 0\\0 & 0 & 1 & 0 & 0 & 0 & 0&0\\1 & 0& 0 & 0 &0 & 0 & 0 & 0\\0 & 0 & 0 & 1 &0& 0 & 0 & 0\\0 & 1 & 0 & 0 &0 & 0 & 0 & 0\end{smallmatrix}\right)\quad(\ee_2\ee_5)\bu_\lrcorner(\cdot) \mapsto\left(\begin{smallmatrix}0 & 0 & 0& 0 & 1& 0 & 0 & 0\\ 0 & 0 & 0 & -1 & 0 & 0 &0 & 0\\0 &0 & 0 & 0 & 0 & -1 & 0 & 0\\0 & 1 & 0 & 0 & 0 & 0 & 0 & 0\\-1 & 0 & 0 & 0 & 0 & 0 & 0&0\\0& 0& 1 & 0 &0 & 0 & 0 & 0\\0 & 0 & 0 & 0 &0& 0 & 0 & -1\\0 & 0 & 0 & 0 &0 & 0 & 1 & 0\end{smallmatrix}\right)\nonumber\eeq
\beq(\ee_2\ee_7)\bu_\lrcorner(\cdot) \mapsto\left(\begin{smallmatrix}0 & 0 & 0& 1 & 0& 0 & 0 & 0\\ 0 & 0 & 0 & 0 & 1 & 0 &0 & 0\\0 &0 & 0 & 0 & 0 & 0 & 0& -1\\-1 & 0 & 0 & 0 & 0 & 0 & 0 & 0\\0 & -1 & 0 & 0 & 0 & 0 & 0&0\\0& 0& 0 & 0 &0 & 0 & -1 & 0\\0 & 0 & 0 & 0 &0& 1 & 0 & 0\\0 & 0 & 1 & 0 &0 & 0 & 0 & 0\end{smallmatrix}\right)\quad(\ee_3\ee_4)\bu_\lrcorner(\cdot) \mapsto\left(\begin{smallmatrix}0 & -1 & 0& 0 & 0& 0 & 0 & 0\\ 1 & 0 & 0 & 0 & 0 & 0 &0 & 0\\0 &0 & 0 & 0 & 0 & 0 & 1& 0\\0 & 0 & 0 & 0 & -1 & 0 & 0 & 0\\0 & 0 & 0 & 1 & 0 & 0 & 0&0\\0& 0& 0 & 0 &0 & 0 & 0 & 1\\0 & 0 & -1 & 0 &0& 0 & 0 & 0\\0 & 0 & 0 & 0 &0 & -1 & 0 & 0\end{smallmatrix}\right)\nonumber\eeq\beq\quad(\ee_2\ee_6)\bu_\lrcorner(\cdot) \mapsto\left(\begin{smallmatrix}0 & -1 & 0& 0 & 0& 0 & 0 & 0\\ 1 & 0 & 0 & 0 & 0 & 0 &0 & 0\\0 &0 & 0 & 0 & 0 & 0 & -1& 0\\0 & 0 & 0 & 0 & 1 & 0 & 0 & 0\\0 & 0 & 0 & -1 & 0 & 0 & 0&0\\0& 0& 0 & 0 &0 & 0 & 0 & 1\\0 & 0 & 1 & 0 &0& 0 & 0 & 0\\0 & 0 & 0 & 0 &0 & -1 & 0 & 0\end{smallmatrix}\right)\quad(\ee_3\ee_5)\bu_\lrcorner(\cdot) \mapsto\left(\begin{smallmatrix}0 & 0 & 0& 0 & 0& 0 & -1 & 0\\ 0 & 0 & 1 & 0 & 0 & 0 &0 & 0\\0 &-1 & 0 & 0 & 0 & 0 & 0& 0\\0 & 0 & 0 & 0 & 0 & -1& 0 & 0\\0 & 0 & 0 & 0 & 0 & 0 & 0&-1\\0& 0& 0 & 1 &0 & 0 & 0 & 0\\1 & 0 & 0 & 0 &0& 0 & 0 & 0\\0 & 0 & 0 & 0 &1 & 0 & 0 & 0\end{smallmatrix}\right)\nonumber\eeq
\beq(\ee_3\ee_6)\bu_\lrcorner(\cdot) \mapsto\left(\begin{smallmatrix}0 & 0 & 0& 0 & 0& 1 & 0 & 0\\ 0 & 0 &0 & 0 & 0 & 0 &0 & 1\\0 &0 & 0 & 0 & -1 & 0 & 0& 0\\0 & 0 & 0 & 0 & 0 & 0& -1 & 0\\0 & 0 & 1 & 0 & 0 & 0 & 0&0\\-1& 0& 0 & 0 &0 & 0 & 0 & 0\\0 & 0 & 0 & 1 &0& 0 & 0 & 0\\0 & -1 & 0 & 0 &0 & 0 & 0 & 0\end{smallmatrix}\right)\quad(\ee_3\ee_7)\bu_\lrcorner(\cdot) \mapsto\left(\begin{smallmatrix}0 & 0 & -1& 0 & 0& 0 & 0 & 0\\ 0 & 0 &0 & 0 & 0 & 0 &-1 & 0\\1 &0 & 0 & 0 & 0 & 0 & 0& 0\\0 & 0 & 0 & 0 & 0 & 0& 0 & -1\\0 & 0 & 0 & 0 & 0 & 1 & 0&0\\0& 0& 0 & 0 &-1 & 0& 0 & 0\\0 & 1 & 0 & 0 &0& 0 & 0 & 0\\0 & 0 & 0 & 1 &0 & 0 & 0 & 0\end{smallmatrix}\right)\quad\nonumber\eeq
\beq(\ee_4\ee_6)\bu_\lrcorner(\cdot) \mapsto\left(\begin{smallmatrix}0 & 0 & 0& 0 & 0& 0 & 0 & -1\\ 0 & 0 &0 & 0 & 0 & 1 &0 & 0\\0 &0 & 0 & 1 & 0 & 0 & 0& 0\\0 & 0 & -1 & 0 & 0 & 0& 0 & 0\\0 & 0 & 0 & 0 & 0 & 0 & -1&0\\0& -1& 0 & 0 &0 & 0& 0 & 0\\0 & 0 & 0 & 0 &1& 0 & 0 & 0\\1 & 0 & 0 & 0 &0 & 0 & 0 & 0\end{smallmatrix}\right)\quad(\ee_4\ee_7)\bu_\lrcorner(\cdot) \mapsto\left(\begin{smallmatrix}0 & 0 & 0& 0 & 0& 0 & 1 & 0\\ 0 & 0 &-1 & 0 & 0 & 0 &0 & 0\\0 &1 & 0 & 0 & 0 & 0 & 0& 0\\0 & 0 & 0 & 0 & 0 & -1& 0 & 0\\0 & 0 & 0 & 0 & 0 & 0 & 0&-1\\0& 0& 0 & 1 &0 & 0& 0 & 0\\-1 & 0 & 0 & 0 &0& 0 & 0 & 0\\0 & 0 & 0 & 0 &1 & 0 & 0 & 0\end{smallmatrix}\right)\nonumber\eeq\beq
(\ee_4\ee_5)\bu_\lrcorner(\cdot) \mapsto\left(\begin{smallmatrix}0 & 0 & -1& 0 & 0& 0 & 0 & 0\\ 0 & 0 &0 & 0 & 0 & 0 &-1 & 0\\1 &0 & 0 & 0 & 0 & 0 & 0& 0\\0 & 0 & 0 & 0 & 0 & 0& 0 & 1\\0 & 0 & 0 & 0 & 0 & -1 & 0&0\\0& 0& 0 & 0 &1 & 0& 0 & 0\\0 & 1 & 0 & 0 &0& 0 & 0 & 0\\0 & 0 & 0 & -1 &0 & 0 & 0 & 0\end{smallmatrix}\right)\quad(\ee_5\ee_6)\bu_\lrcorner(\cdot) \mapsto\left(\begin{smallmatrix}0 & 0 & 0& -1 & 0& 0 & 0 & 0\\ 0 & 0 &0 & 0 & -1 & 0 &0 & 0\\0 &0 & 0 & 0 & 0 & 0 & 0& -1\\1 & 0 & 0 & 0 & 0 & 0& 0 & 0\\0 & 1 & 0 & 0 & 0 & 0 & 0&0\\0& 0& 0 & 0 &0 & 0& -1 & 0\\0 & 0 & 0 & 0 &0& 1 & 0 & 0\\0 & 0 & 1 & 0 &0 & 0 & 0 & 0\end{smallmatrix}\right)\nonumber\eeq
\beq(\ee_5\ee_7)\bu_\lrcorner(\cdot) \mapsto\left(\begin{smallmatrix}0 & -1 & 0& 0 & 0& 0 & 0 & 0\\ 1 & 0 &0 & 0 & 0 & 0 &0 & 0\\0 &0 & 0 & 0 & 0 & 0 & 1& 0\\0 & 0 & 0 & 0 & 1 & 0& 0 & 0\\0 & 0 & 0 & -1 & 0 & 0 & 0&0\\0& 0& 0 & 0 &0 & 0& 0 & -1\\0 & 0 & -1 & 0 &0& 0 & 0 & 0\\0 & 0 & 0 & 0 &0 & 1 & 0 & 0\end{smallmatrix}\right)\quad(\ee_6\ee_7)\bu_\lrcorner(\cdot) \mapsto\left(\begin{smallmatrix}0 & 0 & 0& 0 & -1& 0 & 0 & 0\\ 0 & 0 &0 & 1 & 0 & 0 &0 & 0\\0 &0 & 0 & 0 & 0 & -1& 0& 0\\0 & -1 & 0 & 0 & 0 & 0& 0 & 0\\1 & 0 & 0 & 0 & 0 & 0 & 0&0\\0& 0& 1 & 0 &0 & 0& 0 & 0\\0 & 0 & 0 & 0 &0& 0 & 0 & -1\\0 & 0 & 0 & 0 &0 & 0 & 1 & 0\end{smallmatrix}\right)\nonumber\eeq
\beq(\ee_1\ee_2\ee_3)&\bu_\lrcorner&(X)\n
&=& \ee_1\circ(\ee_2\circ(\ee_3\circ(X^0+X^1\ee_1+X^2\ee_2+X^3\ee_3+X^4\ee_4+X^5\ee_5+X^6\ee_6+X^7\ee_7)))\n
&=& \ee_1\circ(\ee_2\circ(X^0\ee_3-X^1\ee_4-X^2\ee_7-X^3+X^4\ee_1+X^5\ee_6-X^6\ee_5+X^7\ee_2))\n
&=& \ee_1\circ(X^0\ee_7-X^1\ee_5+X^2\ee_3-X^3\ee_2-X^4\ee_6+X^5\ee_1+X^6\ee_4-X^7)\n
&=& -X^0\ee_5-X^1\ee_7+X^2\ee_4-X^3\ee_6+X^4\ee_2-X^5-X^6\ee_3-X^7\ee_1\n&=&\left(\begin{smallmatrix}0 & 0 & 0& 0 & 0& -1 & 0 & 0\\ 0 & 0 &0 & 0 & 0 & 0 &0 & -1\\0 &0 & 0 & 0 & 1 & 0& 0& 0\\0 & 0 & 0 & 0 & 0 & 0& -1 & 0\\0 & 0 & 1 & 0 & 0 & 0 & 0&0\\-1& 0& 0 & 0 &0 & 0& 0 & 0\\0 & 0 & 0 & -1 &0& 0 & 0 & 0\\0 & -1 & 0 & 0 &0 & 0 & 0 & 0\end{smallmatrix}\right)\left(\begin{smallmatrix}X^0\\X^1\\X^2\\X^3\\X^4\\X^5\\X^6\\X^7\end{smallmatrix}\right)\nonumber\eeq\noi Analogously, 
each $\ee_a\ee_b\ee_c\bu_\lrcorner (\,\cdot\,)$ action corresponds respectively to the following matrices:
\beq(\ee_1\ee_2\ee_4)\bu_\lrcorner(\cdot) \mapsto\left(\begin{smallmatrix}0 & 0 & 0& 0 & 0& 0 & 0 & 1\\ 0 & 0 &0 & 0 & 0 & -1 &0 & 0\\0 &0 & 0 & -1 & 0 & 0& 0& 0\\0 & 0 & -1 & 0 & 0 & 0& 0 & 0\\0 & 0 & 0 & 0 & 0 & 0 & -1&0\\0& -1& 0 & 0 &0 & 0& 0 & 0\\0 & 0 & 0 & 0 &-1& 0 & 0 & 0\\1 & 0 & 0 & 0 &0 & 0 & 0 & 0\end{smallmatrix}\right)\quad(\ee_1\ee_2\ee_5)\bu_\lrcorner(\cdot) \mapsto=\left(\begin{smallmatrix}0 & 0 & 0& 1 & 0& 0 & 0 & 0\\ 0 & 0 &0 & 0 & 1 & 0 &0 & 0\\0 &0 & 0 & 0 & 0 & 0& 0& 1\\1 & 0 & 0 & 0 & 0 & 0& 0 & 0\\0 & 1 & 0 & 0 & 0 & 0 & 0&0\\0& 0& 0 & 0 &0 & 0& -1 & 0\\0 & 0 & 0 & 0 &0& -1 & 0 & 0\\0 & 0 & 1 & 0 &0 & 0 & 0 & 0\end{smallmatrix}\right)\nonumber\eeq
\beq(\ee_1\ee_2\ee_6)\bu_\lrcorner(\cdot) \mapsto\left(\begin{smallmatrix}-1 & 0 & 0& 0 & 0& 0 & 0 & 0\\ 0 & -1 &0 & 0 & 0 & 0 &0 & 0\\0 &0 & -1 & 0 & 0 & 0& 0& 0\\0 & 0 & 0 & 1 & 0 & 0& 0 & 0\\0 & 0 & 0 & 0 & 1 & 0 & 0&0\\0& 0& 0 & 0 &0 & 1& 0 & 0\\0 & 0 & 0 & 0 &0& 0 & -1 & 0\\0 & 0 & 0 & 0 &0 & 0 & 0 & 1\end{smallmatrix}\right)\quad(\ee_1\ee_2\ee_7)\bu_\lrcorner(\cdot) \mapsto\left(\begin{smallmatrix}0 & 0 & 0& 0 & -1& 0 & 0 & 0\\ 0 & 0 &0 & 1 & 0 & 0 &0 & 0\\0 &0 & 0 & 0 & 0 & -1& 0& 0\\0 & 1 & 0 & 0 & 0 & 0& 0 & 0\\-1 & 0 & 0 & 0 & 0 & 0 & 0&0\\0& 0& -1 & 0 &0 & 0& 0 & 0\\0 & 0 & 0 & 0 &0& 0 & 0 & -1\\0 & 0 & 0 & 0 &0 & 0 & -1 & 0\end{smallmatrix}\right)\nonumber\eeq
\beq(\ee_1\ee_3\ee_4)\bu_\lrcorner(\cdot) \mapsto\left(\begin{smallmatrix}-1 & 0 & 0& 0 & 0& 0 & 0 & 0\\ 0 & -1 &0 & 0 & 0 & 0 &0 & 0\\0 &0 & 1 & 0 & 0 & 0& 0& 0\\0 & 0 & 0 & -1 & 0 & 0& 0 & 0\\0 & 0 & 0 & 0 & -1 & 0 & 0&0\\0& 0& 0 & 0 &0 & 1& 0 & 0\\0 & 0 & 0 & 0 &0& 0 & 1 & 0\\0 & 0 & 0 & 0 &0 & 0 & 0 & 1\end{smallmatrix}\right)\quad(\ee_1\ee_3\ee_5)\bu_\lrcorner(\cdot) \mapsto\left(\begin{smallmatrix}0 & 0 & -1& 0 & 0& 0 & 0 & 0\\ 0 & 0 &0 & 0 & 0 & 0 &-1 & 0\\-1 &0 & 0 & 0 & 0 & 0& 0& 0\\0 & 0 & 0 & 0 & 0 & 0& 0 & 1\\0 & 0 & 0 & 0 & 0 & -1 & 0&0\\0& 0& 0 & 0 &-1 & 0& 0 & 0\\0 & -1 & 0 & 0 &0& 0 & 0 & 0\\0 & 0 & 0 & 1 &0 & 0 & 0 & 0\end{smallmatrix}\right)\nonumber\eeq
\beq(\ee_1\ee_3\ee_6)\bu_\lrcorner(\cdot) \mapsto\left(\begin{smallmatrix}0 & 0 & 0& 0 & 0& 0 & 0 & -1\\ 0 & 0 &0 & 0 & 0 & 1 &0 & 0\\0 &0 & 0 & -1 & 0 & 0& 0& 0\\0 & 0 & -1 & 0 & 0 & 0& 0 & 0\\0 & 0 & 0 & 0 & 0 & 0 & 1&0\\0& 1& 0 & 0 &0 & 0& 0 & 0\\0 & 0 & 0 & 0 &-1& 0 & 0 & 0\\-1 & 0 & 0 & 0 &0 & 0 & 0 & 0\end{smallmatrix}\right)\quad(\ee_1\ee_3\ee_7)\bu_\lrcorner(\cdot) \mapsto\left(\begin{smallmatrix}0 & 0 & 0& 0 & 0& 0 & 1 & 0\\ 0 & 0 &-1 & 0 & 0 & 0 &0 & 0\\0 &-1 & 0 & 0 & 0 & 0& 0& 0\\0 & 0 & 0 & 0 & 0 & -1& 0 & 0\\0 & 0 & 0 & 0 & 0 & 0 & 0&-1\\0& 0& 0 & -1 &0 & 0& 0 & 0\\1 & 0 & 0 & 0 &0& 0 & 0 & 0\\0 & 0 & 0 & 0 &-1 & 0 & 0 & 0\end{smallmatrix}\right)\nonumber\eeq
\beq(\ee_1\ee_4\ee_5)\bu_\lrcorner(\cdot) \mapsto\left(\begin{smallmatrix}0 & 0 & 0& 0 & 0& 0 & 1 & 0\\ 0 & 0 &-1 & 0 & 0 & 0 &0 & 0\\0 &-1 & 0 & 0 & 0 & 0& 0& 0\\0 & 0 & 0 & 0 & 0 & 1& 0 & 0\\0 & 0 & 0 & 0 & 0 & 0 & 0&1\\0& 0& 0 & 1 &0 & 0& 0 & 0\\1 & 0 & 0 & 0 &0& 0 & 0 & 0\\0 & 0 & 0 & 0 &1 & 0 & 0 & 0\end{smallmatrix}\right)\quad(\ee_1\ee_4\ee_6)\bu_\lrcorner(\cdot) \mapsto\left(\begin{smallmatrix}0 & 0 & 0& 0 & 0& -1 & 0 & 0\\ 0 & 0 &0 & 0 & 0 & 0 &0 & -1\\0 &0 & 0 & 0 & -1 & 0& 0& 0\\0 & 0 & 0 & 0 & 0 & 0& 1 & 0\\0 & 0 & -1 & 0 & 0 & 0 & 0&0\\-1& 0& 0 & 0 &0 & 0& 0 & 0\\0 & 0 & 0 & 1 &0& 0 & 0 & 0\\0 & -1 & 0 & 0 &0 & 0 & 0 & 0\end{smallmatrix}\right)\nonumber\eeq
\beq(\ee_1\ee_4\ee_7)\bu_\lrcorner(\cdot) \mapsto\left(\begin{smallmatrix}0 & 0 & 1& 0 & 0& 0 & 0 & 0\\ 0 & 0 &0 & 0 & 0 & 0 &1& 0\\1 &0 & 0 & 0 & 0 & 0& 0& 0\\0 & 0 & 0 & 0 & 0 & 0& 0 & 1\\0 & 0 & 0 & 0 & 0 & -1 & 0&0\\0& 0& 0 & 0 &-1& 0& 0 & 0\\0 & 1 & 0 & 0 &0& 0 & 0 & 0\\0 & 0 & 0 & 1 &0 & 0 & 0 & 0\end{smallmatrix}\right)\quad(\ee_1\ee_5\ee_6)\bu_\lrcorner(\cdot) \mapsto\left(\begin{smallmatrix}0 & 0 & 0& 0 & 1& 0 & 0 & 0\\ 0 & 0 &0 & -1 & 0 & 0 &0& 0\\0 &0 & 0 & 0 & 0 & -1& 0& 0\\0 & -1 & 0 & 0 & 0 & 0& 0 & 0\\1 & 0 & 0 & 0 & 0 & 0 & 0&0\\0& 0& -1 & 0 &0& 0& 0 & 0\\0 & 0 & 0 & 0 &0& 0 & 0 & -1\\0 & 0 & 0 & 0 &0 & 0 & -1 & 0\end{smallmatrix}\right)\nonumber\eeq
\beq(\ee_1\ee_5\ee_7)\bu_\lrcorner(\cdot) \mapsto\left(\begin{smallmatrix}-1 & 0 & 0& 0 & 0& 0 & 0 & 0\\ 0 & -1 &0 & 0 & 0 & 0 &0& 0\\0 &0 & 1 & 0 & 0 & 0& 0& 0\\0 & 0 & 0 & 1 & 0 & 0& 0 & 0\\0 & 0 & 0 & 0 & 1 & 0 & 0&0\\0& 0& 0 & 0 &0& -1& 0 & 0\\0 & 0 & 0 & 0 &0& 0 & 1 & 0\\0 & 0 & 0 & 0 &0 & 0 & 0 & -1\end{smallmatrix}\right)\quad(\ee_1\ee_6\ee_7)\bu_\lrcorner(\cdot) \mapsto\left(\begin{smallmatrix}0 & 0 & 0& -1 & 0& 0 & 0 & 0\\ 0 & 0 &0 & 0 & -1 & 0 &0& 0\\0 &0 & 0 & 0 & 0 & 0& 0& 1\\-1 & 0 & 0 & 0 & 0 & 0& 0 & 0\\0 & -1 & 0 & 0 & 0 & 0 & 0&0\\0& 0& 0 & 0 &0& 0& -1 & 0\\0 & 0 & 0 & 0 &0& -1 & 0 & 0\\0 & 0 & 1 & 0 &0 & 0 & 0 & 0\end{smallmatrix}\right)\nonumber\eeq
\beq(\ee_2\ee_3\ee_4)\bu_\lrcorner(\cdot) \mapsto\left(\begin{smallmatrix}0 & 0 & 0&0 & 0& 0 & -1 & 0\\ 0 & 0 &-1 & 0 & 0 & 0 &0& 0\\0 &-1 & 0 & 0 & 0 & 0& 0& 0\\0 & 0 & 0 & 0 & 0 & 1& 0 & 0\\0 & 0 & 0 & 0 & 0 & 0 & 0&-1\\0& 0& 0 & 1 &0& 0& 0 & 0\\-1 & 0 & 0 & 0 &0& 0 & 0 & 0\\0 & 0 & 0 & 0 &-1 & 0 & 0 & 0\end{smallmatrix}\right)\quad(\ee_2\ee_3\ee_5)&\bu_\lrcorner&(\cdot) \mapsto\left(\begin{smallmatrix}0 & 1 & 0&0 & 0& 0 & 0 & 0\\ 1 & 0 &0 & 0 & 0 & 0 &0& 0\\0 &0 & 0 & 0 & 0 & 0& -1& 0\\0 & 0 & 0 & 0 & -1 & 0& 0 & 0\\0 & 0 & 0 & -1 & 0 & 0 & 0&0\\0& 0& 0 & 0 &0& 0& 0 & -1\\0 & 0 & -1 & 0 &0& 0 & 0 & 0\\0 & 0 & 0 & 0 &0 & -1 & 0 & 0\end{smallmatrix}\right)\nonumber\eeq
\beq(\ee_2\ee_3\ee_6)\bu_\lrcorner(\cdot) \mapsto\left(\begin{smallmatrix}0 & 0 & 0&0 & 1& 0 & 0 & 0\\ 0 & 0 &0 & 1 & 0 & 0 &0& 0\\0 &0 & 0 & 0 & 0 & 1& 0& 0\\0 & 1 & 0 & 0 & 0 & 0& 0 & 0\\1 & 0 & 0 & 0 & 0 & 0 & 0&0\\0& 0& 1 & 0 &0& 0& 0 & 0\\0 & 0 & 0 & 0 &0& 0 & 0 & -1\\0 & 0 & 0 & 0 &0 & 0 & -1 & 0\end{smallmatrix}\right)\quad(\ee_2\ee_3\ee_7)\bu_\lrcorner(\cdot) \mapsto\left(\begin{smallmatrix}-1& 0 & 0&0 & 0& 0 & 0 & 0\\ 0 & 1 &0 & 0 & 0 & 0 &0& 0\\0 &0 & -1 & 0 & 0 & 0& 0& 0\\0 & 0 & 0 & -1 & 0 & 0& 0 & 0\\0 & 0 & 0 & 0 & 1 & 0 & 0&0\\0& 0& 0 & 0 &0& 1& 0 & 0\\0 & 0 & 0 & 0 &0& 0 & 1 & 0\\0 & 0 & 0 & 0 &0 & 0 & 0 & -1\end{smallmatrix}\right)\nonumber\eeq
\beq(\ee_2\ee_4\ee_5)\bu_\lrcorner(\cdot) \mapsto\left(\begin{smallmatrix}-1& 0 & 0&0 & 0& 0 & 0 & 0\\ 0 & 1 &0 & 0 & 0 & 0 &0& 0\\0 &0 & -1 & 0 & 0 & 0& 0& 0\\0 & 0 & 0 & 1 & 0 & 0& 0 & 0\\0 & 0 & 0 & 0 & -1 & 0 & 0&0\\0& 0& 0 & 0 &0& -1& 0 & 0\\0 & 0 & 0 & 0 &0& 0 & 1 & 0\\0 & 0 & 0 & 0 &0 & 0 & 0 & 1\end{smallmatrix}\right)\quad(\ee_2\ee_4\ee_6)\bu_\lrcorner(\cdot) \mapsto\left(\begin{smallmatrix}0& 0 & 0&-1 & 0& 0 & 0 & 0\\ 0 & 0 &0 & 0 & 1 & 0 &0& 0\\0 &0 & 0 & 0 & 0 & 0& 0& -1\\-1 & 0 & 0 & 1 & 0 & 0& 0 & 0\\0 & 1 & 0 & 0 & 0 & 0 & 0&0\\0& 0& 0 & 0 &0& 0& -1 & 0\\0 & 0 & 0 & 0 &0& -1 & 0 & 0\\0 & 0 & -1 & 0 &0 & 0 & 0 & 0\end{smallmatrix}\right)\nonumber\eeq
\beq(\ee_2\ee_4\ee_7)\bu_\lrcorner(\cdot) \mapsto\left(\begin{smallmatrix}0& -1 & 0&0 & 0& 0 & 0 & 0\\ -1 & 0 &0 & 0 & 0 & 0 &0& 0\\0 &0 & 0 & 0 & 0 & 0& 1& 0\\0 & 0 & 0 & 0 & -1 & 0& 0 & 0\\0 & 0 & 0 & -1 & 0 & 0 & 0&0\\0& 0& 0 & 0 &0& 0& 0 & -1\\0 & 0 & 1 & 0 &0& 0 & 0 & 0\\0 & 0 & 0 & 0 &0 & -1 & 0 & 0\end{smallmatrix}\right)\quad(\ee_2\ee_5\ee_6)\bu_\lrcorner(\cdot) \mapsto\left(\begin{smallmatrix}0& 0 & 0&0 & 0& 0 & 0 & 1\\ 0 & 0 &0 & 0 & 0 & 1 &0& 0\\0 &0 & 0 & -1 & 0 & 0& 0& 0\\0 & 0 & -1 & 0 & 0 & 0& 0 & 0\\0 & 0 & 0 & 0 & 0 & 0 & 1&0\\0& 1& 0 & 0 &0& 0& 0 & 0\\0 & 0 & 0 & 0 &1& 0 & 0 & 0\\1 & 0 & 0 & 0 &0 & 0 & 0 & 0\end{smallmatrix}\right)\nonumber\eeq
\beq(\ee_2\ee_5\ee_7)\bu_\lrcorner(\cdot) \mapsto\left(\begin{smallmatrix}0& 0 & 0&0 & 0& 0 & -1 & 0\\ 0 & 0 &-1 & 0 & 0 & 0 &0& 0\\0 &-1 & 0 & 0 & 0 & 0& 0& 0\\0 & 0 & 0 & 0 & 0 & -1& 0 & 0\\0 & 0 & 0 & 0 & 0 & 0 & 0&1\\0& 0& 0 & -1 &0& 0& 0 & 0\\-1 & 0 & 0 & 0 &0& 0 & 0 & 0\\0 & 0 & 0 & 0 &1 & 0 & 0 & 0\end{smallmatrix}\right)\quad(\ee_2\ee_6\ee_7)\bu_\lrcorner(\cdot) \mapsto\left(\begin{smallmatrix}0& 0 & 0&0 & 0& 1 & 0 & 0\\ 0 & 0 &0 & 0 & 0 & 0 &0& -1\\0 &0 & 0 & 0 & -1 & 0& 0& 0\\0 & 0 & 0 & 0 & 0 & 0& -1 & 0\\0 & 0 & -1 & 0 & 0 & 0 & 0&0\\1& 0& 0 & 0 &0& 0& 0 & 0\\0 & 0 & 0 & -1 &0& 0 & 0 & 0\\0 & -1 & 0 & 0 &0 & 0 & 0 & 0\end{smallmatrix}\right)\nonumber\eeq
\beq(\ee_3\ee_4\ee_5)\bu_\lrcorner(\cdot) \mapsto\left(\begin{smallmatrix}0& 0 & 0&0 & 0& 0 & 0 & -1\\ 0 & 0 &0 & 0 & 0 & -1 &0& 0\\0 &0 & 0 & -1 & 0 & 0& 0& 0\\0 & 0 & -1 & 0 & 0 & 0& 0 & 0\\0 & 0 & 0 & 0 & 0 & 0 & 1&0\\0& -1& 0 & 0 &0& 0& 0 & 0\\0 & 0 & 0 & 0 &1& 0 & 0 & 0\\-1 & 0 & 0 & 0 &0 & 0 & 0 & 0\end{smallmatrix}\right)\quad(\ee_3\ee_4\ee_6)\bu_\lrcorner(\cdot) \mapsto\left(\begin{smallmatrix}0& 0 & 1&0 & 0& 0 & 0 & 0\\ 0 & 0 &0 & 0 & 0 & 0 &-1& 0\\1 &0 & 0 & 0 & 0 & 0& 0& 0\\0 & 0 & 0 & 0 & 0 & 0& 0 & -1\\0 & 0 & 0 & 0 & 0 & -1 & 0&0\\0& 0& 0 & 0 &-1& 0& 0 & 0\\0 & -1 & 0 & 0 &0& 0 & 0 & 0\\0 & 0 & 0 & -1 &0 & 0 & 0 & 0\end{smallmatrix}\right)\nonumber\eeq
\beq(\ee_3\ee_4\ee_7)\bu_\lrcorner(\cdot) \mapsto\left(\begin{smallmatrix}0& 0 & 0&0 & 0& 1 & 0 & 0\\ 0 & 0 &0 & 0 & 0 & 0 &0& -1\\0 &0 & 0 & 0 & 1 & 0& 0& 0\\0 & 0 & 0 & 0 & 0 & 0& 1 & 0\\0 & 0 & 1 & 0 & 0 & 0 & 0&0\\1& 0& 0 & 0 &0& 0& 0 & 0\\0 & 0 & 0 & 1 &0& 0 & 0 & 0\\0 & -1 & 0 & 0 &0 & 0 & 0 & 0\end{smallmatrix}\right)\quad(\ee_3\ee_5\ee_6)\bu_\lrcorner(\cdot) \mapsto\left(\begin{smallmatrix}-1& 0 & 0&0 & 0& 0 & 0 & 0\\ 0 & 1 &0 & 0 & 0 & 0 &0& 0\\0 &0 & 1 & 0 & 0 & 0& 0& 0\\0 & 0 & 0 & -1 & 0 & 0& 0 & 0\\0 & 0 & 0 & 0 & 1& 0 & 0&0\\0& 0& 0 & 0 &0& -1& 0 & 0\\0 & 0 & 0 & 0 &0& 0 & -1 & 0\\0 & 0 & 0 & 0 &0 & 0 & 0 & 1\end{smallmatrix}\right)\nonumber\eeq
\beq(\ee_3\ee_5\ee_7)\bu_\lrcorner(\cdot) \mapsto\left(\begin{smallmatrix}0& 0 & 0&0 & -1& 0 & 0 & 0\\ 0 & 0 &0 & -1 & 0 & 0 &0& 0\\0 &0 & 0 & 0 & 0 & 1& 0& 0\\0 & -1 & 0 & 0 & 0 & 0& 0 & 0\\-1 & 0 & 0 & 0 & 0& 0 & 0&0\\0& 0& 1 & 0 &0& 0& 0 & 0\\0 & 0 & 0 & 0 &0& 0 & 0 & -1\\0 & 0 & 0 & 0 &0 & 0 & -1 & 0\end{smallmatrix}\right)\quad(\ee_3\ee_6\ee_7)\bu_\lrcorner(\cdot)\mapsto\left(\begin{smallmatrix}0& 1 & 0&0 & 0& 0 & 0 & 0\\ 1 & 0 &0 & 0 & 0 & 0 &0& 0\\0 &0 & 0 & 0 & 0 & 0& 1& 0\\0 & 0 & 0 & 0 & -1 & 0& 0 & 0\\0 & 0 & 0 & -1 & 0& 0 & 0&0\\0& 0& 0 & 0 &0& 0& 0 & 1\\0 & 0 & 1 & 0 &0& 0 & 0 & 0\\0 & 0 & 0 & 0 &0 & 1 & 0 & 0\end{smallmatrix}\right)\nonumber\eeq
\beq(\ee_4\ee_5\ee_6)\bu_\lrcorner(\cdot) \mapsto\left(\begin{smallmatrix}0& -1 & 0&0 & 0& 0 & 0 & 0\\ -1 & 0 &0 & 0 & 0 & 0 &0& 0\\0 &0 & 0 & 0 & 0 & 0& -1& 0\\0 & 0 & 0 & 0 & -1 & 0& 0 & 0\\0 & 0 & 0 & -1 & 0& 0 & 0&0\\0& 0& 0 & 0 &0& 0& 0 & 1\\0 & 0 & -1 & 0 &0& 0 & 0 & 0\\0 & 0 & 0 & 0 &0 & 1 & 0 & 0\end{smallmatrix}\right)\quad(\ee_4\ee_5\ee_7)\bu_\lrcorner(\cdot) \mapsto\left(\begin{smallmatrix}0& 0 & 0&1 & 0& 0 & 0 & 0\\ 0 & 0 &0 & 0 & -1 & 0 &0& 0\\0 &0 & 0 & 0 & 0 & 0& 0& -1\\1 & 0 & 0 & 0 & 0 & 0& 0 & 0\\0 & -1 & 0 & 0 & 0& 0 & 0&0\\0& 0& 0 & 0 &0& 0& -1 & 0\\0 & 0 & 0 & 0 &0& -1 & 0 & 0\\0 & 0 & -1 & 0 &0 & 0 & 0 & 0\end{smallmatrix}\right)\nonumber\eeq
\beq(\ee_4\ee_6\ee_7)\bu_\lrcorner(\cdot) \mapsto\left(\begin{smallmatrix}-1& 0 & 0&0 & 0& 0 & 0 & 0\\ 0 & 1 &0 & 0 & 0 & 0 &0& 0\\0 &0 & 1 & 0 & 0 & 0& 0& 0\\0 & 0 & 0 & 1 & 0 & 0& 0 & 0\\0 & 0 & 0 & 0 & -1 & 0 & 0 & 0\\0& 0& 0 & 0 &0& 1& 0 & 0\\0 & 0 & 0 & 0 &0& 0 & -1 & 0\\0 & 0 & 0 & 0 &0 & 0 & 0 & -1\end{smallmatrix}\right)\quad(\ee_5\ee_6\ee_7)\bu_\lrcorner(\cdot) \mapsto\left(\begin{smallmatrix}0& 0 & -1&0 & 0& 0 & 0 & 0\\ 0 & 0 &0 & 0 & 0 & 0 &1& 0\\-1 &0 & 0 & 0 & 0 & 0& 0& 0\\0 & 0 & 0 & 0 & 0 & 0& 0 & -1\\0 & 0 & 0 & 0 & 0& -1 & 0&0\\0& 0& 0 & 0 &-1& 0& 0 & 0\\0 & 1 & 0 & 0 &0& 0 & 0 & 0\\0 & 0 & 0 & -1 &0 & 0 & 0 & 0\end{smallmatrix}\right)\nonumber\eeq

\end{document}